\documentstyle[aps]{revtex}

\def\lg{{\mathchoice{~\raise.58ex\hbox{$<$}\mkern-14.8mu\lower.52ex\hbox{$>$}~}
	            {~\raise.58ex\hbox{$<$}\mkern-14.8mu\lower.52ex\hbox{$>$}~}
		    {\raise.59ex\hbox{{$\scriptscriptstyle <$}}\mkern-12.8mu%
		     \lower.01ex\hbox{{$\scriptscriptstyle >$}}}   {} 	}} 
\def\gl{{\mathchoice{~\raise.58ex\hbox{$>$}\mkern-12.8mu\lower.52ex\hbox{$<$}~}
                    {~\raise.58ex\hbox{$>$}\mkern-12.8mu\lower.52ex\hbox{$<$}~}
		    {\raise.62ex\hbox{{$\scriptscriptstyle >$}}\mkern-12.0mu%
		     \lower.05ex\hbox{{$\scriptscriptstyle <$}}}  {} 	}}   
\def\ra{{\mathchoice{~\raise.58ex\hbox{$ret$}\mkern-14.8mu\lower.52ex\hbox{$ret$}~}
	            {~\raise.58ex\hbox{$ret$}\mkern-14.8mu\lower.52ex\hbox{$ret$}~}
		    {\raise.68ex\hbox{{$\scriptstyle ret$}} \mkern-22.0mu%
		     \lower.63ex\hbox{{$\scriptstyle adv$}}}   {} 	}} 
\def\ar{{\mathchoice{~\raise.58ex\hbox{$adv$}\mkern-12.8mu\lower.52ex\hbox{$adv$}~}
                    {~\raise.58ex\hbox{$adv$}\mkern-12.8mu\lower.52ex\hbox{$adv$}~}
		    {\raise.68ex\hbox{{$\scriptstyle adv$}}\mkern-22.0mu%
		     \lower.63ex\hbox{{$\scriptstyle ret$}}}  {} 	}}

\date{}

\input{epsf}

\begin{document}

\setcounter{page}{0}
\thispagestyle{empty}

\begin{flushright}
BNL-63632\\
hep-ph/9611400
\end{flushright}
\vspace{3.5cm}

\begin{center}
{\LARGE
{\bf NON-EQUILIBRIUM QCD:}
}
\bigskip

{\LARGE
{\bf Interplay of hard and soft dynamics in high-energy multi-gluon beams}
}

\end{center}
\bigskip

\begin{center}

{\Large
{\bf  Klaus Geiger}
}
\medskip

{\it Brookhaven National Laboratory\\
Physics Department 510-A\\
Upton, N.Y. 11973, U.S.A.\\
e-mail: klaus@bnl.gov}
\end{center}
\vspace{2.5cm}

\begin{center}
{\large {\bf Abstract}}
\end{center}
\bigskip

\noindent
A quantum-kinetic formulation of the dynamical evolution
of a high-energy 
non-equilibrium gluon system at finite density is 
developed, to study the interplay between
quantum fluctuations of high-momentum (hard) gluons
and the low-momentum (soft) mean color-field that is
induced by the collective motion of the hard particles.
From the exact field-equations of motion of QCD, a
self-consistent set of 
approximate quantum-kinetic equations are derived
by separating hard and soft dynamics and choosing
a convenient axial-type gauge.
This set of  master equations describes the
momentum space evolution of the individual hard quanta,
the space-time development of the ensemble of hard
gluons, and the generation  of the  soft mean-field
by the current of the hard particles.
The quantum-kinetic equations are  approximately solved
to order $g^2 (1+g\overline{A})$ for a specific example,
namely the  scenario 
of a high-energy gluon beam along the lightcone,
demonstrating the practical applicability of the approach. 
\smallskip

\noindent
{\small

}

\vspace{0.5cm}


\newpage

\section{INTRODUCTION AND SUMMARY}
\label{sec:section1}
\bigskip

The physics of high-density QCD becomes an increasingly popular
object of research, both from the experimental, phenomenological
interest, and from the theoretical, fundamental point of view.
Presently, and in the near future, the collider facilities HERA
($ep$, $eA$?), Tevatron ($p\bar{p}$, $pA$), RHIC and LHC ($p\bar{p}$,
$AA$) are able to probe new regimes of dense quark-gluon matter 
at very small Bjorken-$x$ or/and at large $A$, 
with rather different dynamical properties.
The common feature of high-density
QCD matter that can be produced in these experiments, is an expected
novel exhibition
of the interplay between the high-momentum (short-distance) perturbative
regime and the low-momentum (long-wavelength) non-perturbative physics.
For example, with HERA and Tevatron experiments,
one hopes to gain insight into problems concerning
the saturation of the strong rise
of the proton structure functions at small Bjorken-$x$,
possibly due to color-screening effects that
are associated with the overlappping of a large number of small-$x$
partons.
Another example is the anticipated formation of a quark-gluon plasma in
RHIC and LHC heavy ion collisions, 
where multiple parton rescattering and cascading may generate a 
high-density environment, in which the collective 
motion of the quanta can give rise to
non-abelian long-wavelength excitations and screening of color charges. 

In any case, the study of coherent low-momentum excitations
in QCD,
that are generated by, and interacting with, the high-momentum
partonic color charges, is of fundamental interest in
several respects:
Firstly, it provides insight into 
the basic features of non-abelian multiparticle dynamics and
a step towards a rigorous decription of parton transport properties
in a dense environment.
Secondly, it may help to resolve current problems encountered 
in perturbative QCD, for instance the absence of static magnetic 
color-screening \cite{magscreen},
the problem of infrared renormalons \cite{renormalons} 
connected with the resummation of perturbation
theory in the small-$x$ regime, or, the problem of confinement associated
with collective `glue'-behavior of non-perturbative gluons \cite{confinement}.
Interesting progress in these areas is continously being
made, and consistent
schemes have emerged to perform calculations of the parton evolution at
very small-$x$ \cite{smallx},  at very large density 
\cite{raju,doketal}
and for high-temperature QCD of a quark-gluon plasma \cite{BI1}.

Most progress in the context of bulk multi-parton dynamics at high density
has been made by studying `hot QCD' with a thermally 
equilibrated quark-gluon system at very high temperature $T$.
`Hot QCD' has the attractive advantage that the parton density is homogenous
and isotropic in momentum, and its exact form $\propto T^3$ is known,
since $T\gg\Lambda\approx 200$ MeV is the only energy scale in the problem.
For this academic scenario, inconsistencies of 
former perturbative calculations have been resolved by gauge-invariant
resummation techniques \cite{hotQCD1} as studied in
various applications \cite{hotQCD2}, and moreover,
a self-consistent kinetic theory has been formulated \cite{BI2}.
\medskip

The present paper, extending previous work of Ref. \cite{ms39}, 
is to be viewed in this very context: it takes
the `hot QCD' developments as inspirational
guideline, but aims to describe the opposite physics extreme, namely a
highly non-equilibrium
\footnote{
The term `non-equilibrium' is used in the sense of statistical
many-body physics, describing a quantum system far off the
state of maximum entropy and thermal equilibrium. Such a 
non-equilibrium system may in general be 
spatially inhomogenous and anisotropic in momentum,
in contrast to a homogenous, thermal ensemble, or 
translation invariant system in vacuum.},
non-uniform and non-isotropic parton system.
Specifically, the attempt is made to derive from first principles
a self-consistent kinetic description 
for a
{\it non-quilibrium scenario of a gluon beam directed along
the lightcone}, that is,
a high-density system of gluons, moving with very large energies
$k_0 \simeq k_z \gg k_\perp \gg\Lambda$ along a 
beam direction (the $k_z$-axis), as it would be typical for 
the initial stage of a high-energy collider experiment
(an extreme example is a collision of two heavy nuclei at the LHC,
involving many thousands of gluons coming down the beam pipe).
For simplicity the quark degrees of freedom are ignored, but 
are straightforward to include.

As illustrated schematically in Fig. 1, the initial
multi-gluon state is imagined as a highly Lorentz contracted
sheet of  bare gluons, characterized by a very large momentum scale
$Q$ (e.g. in an ultra-relativistic nuclear collision, the typical
momentum transfer of hard scatterings that materialize the gluons
out of the colliding beam nuclei).
Hence the typical energy and longitudinal momentum of the initial gluons
is $\sim Q$.
The subsequent evolution of these bare quanta is, 
at leading order $\alpha_s$, well known to 
lead to a rapid multiplication and diffusion
of gluons through real and virtual 
radiation, corresponding to brems-strahlung and Coulomb-field
regeneration, respectively \cite{dokbook}.
As a consequence, the typical gluon momenta both in longitudinal and transverse
direction, decrease (see Fig. 1 a).
As long as the average transverse momentum is sufficiently large,
$k_\perp \ge \mu \sim 1-2$ GeV,
$\alpha_s(\mu^2) \ll 1$,  a perturbative description 
of the evolution of the gluon density ${\tt G}$ is appropriate,
but when
$k_\perp < \mu$, non-perturbative dynamics is expected
to take over, governed by the collective
infrared behavior of a large number of long-wavelength gluons.
If the number density of low-mometum gluons below $\mu$ is large, their
dynamics may approximately be described classically \cite{raju,bmcg}
in terms
of a coherent mean field $\overline{A}$ (see Fig. 1b). 

\begin{figure}
\epsfxsize=250pt
\centerline{ \epsfbox{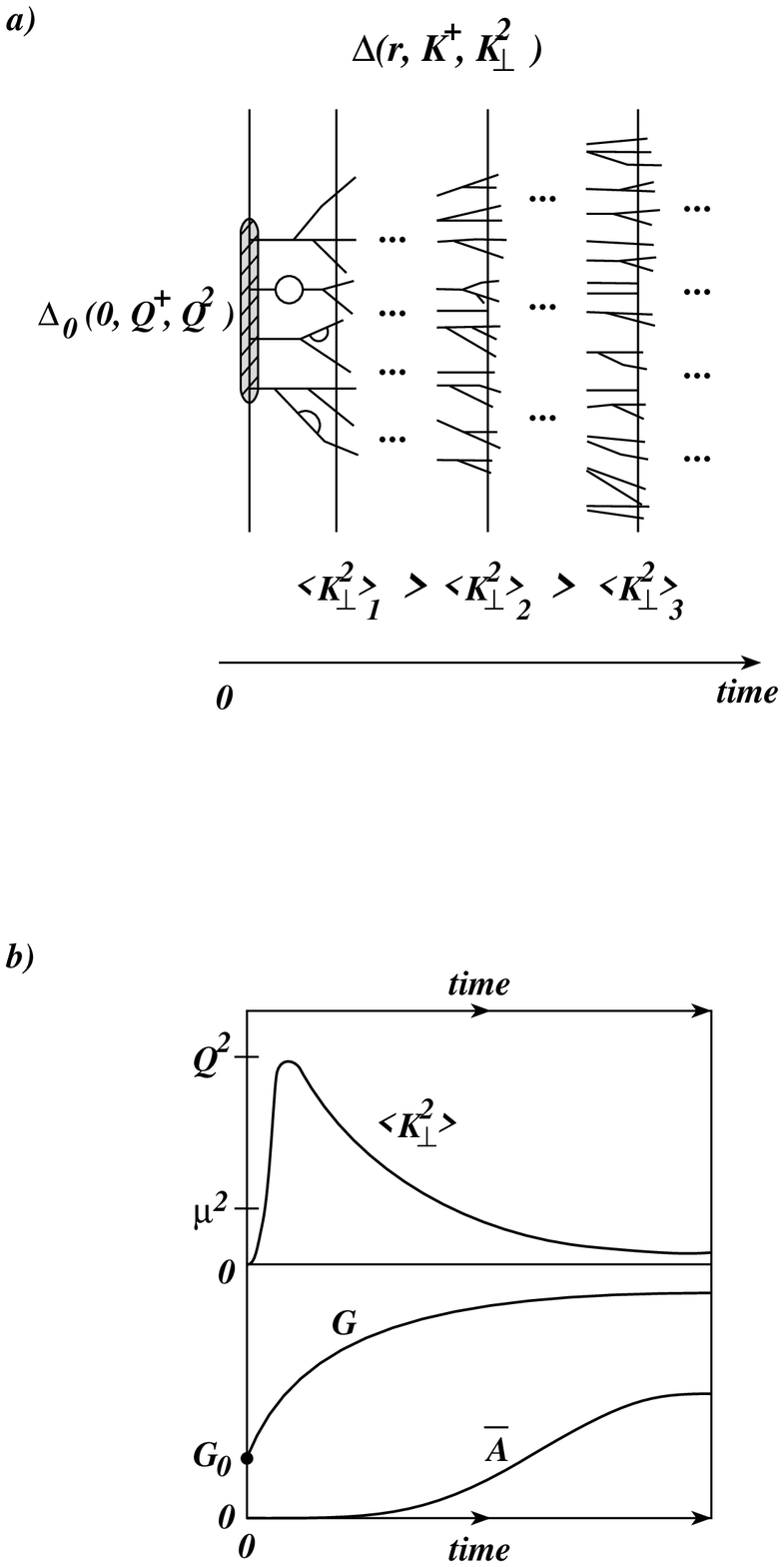} }
\bigskip
\bigskip
\bigskip
\caption{
Non-equilibrium scenario of gluon beam along the lightcone:
{\bf a)}
The initial multi-gluon state,
prepared at time $t_0 =0$ at the hard scale $Q$ with initial
condition $\Delta_0(0,Q^+,Q^2)$, develops forward in time
which is described by the evolution of the gluon propagator
$\widehat{\Delta}(r,K^+,K_\perp^2)$ being a function
of both space-time $r^\mu=(t,\vec{r})$ and momentum $K^\mu=(E,\vec{K})$.
The gluons, propagating with large $K^+=E+K_z \gg K_\perp$
along the $z$-axis, are accompanied by real and virtual radiation which
causes a diffusion in both transverse
direction $r_\perp$ and transverse momentum $K_\perp$ as time goes on:
The emission of gluons increases the multiplicity and decreases 
the average transverse momenta $\langle K_\perp^2\rangle$
at given lightcone time $r^- =t-z$ and position
lightcone position $r^+ = t+z$.
{\bf b)}
{\it Top}: Qualitative picture of time evolution of the
typical transverse momentum $\langle K_\perp^2\rangle$ of hard gluons,
where the earliest emitted daughter gluons have the largest 
$K_\perp^2 \,\lower3pt\hbox{$\buildrel < \over\sim$}\,Q^2$
and later produced gluons have much smaller $K_\perp^2$.
Eventually modes with $K_\perp^2< \mu^2$ will
be populated significantly.
{\it Bottom}: Corresponding time development of the 
number density ${\tt G}$ of hard gluons from initial value ${\tt G}_0$ and of 
the average soft field $\overline{A}$ that is induced
by the population of gluons with $K_\perp^2< \mu^2$ , starting
from zero initial value.
Speculatively, one would expect a saturation at 
asymptotic times due to screening of further gluon emission by the presence of
the soft mean field. 
\label{fig:fig1}
}
\end{figure}
\newpage

Given this heuristic picture,
the near-at-hand rationale is therefore to subdivide
the dynamical development of the gluon ensemble into a perturbative
quantum evolution in the short-distance regime $Q^2 \ge k_\perp^2 \ge \mu^2$,
and a non-perturbative, but classical, mean-field 
in the long-wavelength regime $k_\perp^2 < \mu^2$.
The corresponding degrees of freedom 
are referred to as {\it hard gluons}
for $k_\perp \ge \mu$, whereas excitations with
$k_\perp < \mu$ represent the {\it soft mean field}.

Because the hard gluons have small transverse extent
$\lambda \sim 1/k_\perp\le 0.2$ fm (for $\mu = 1$GeV), they can be considered,
locally in space-time, as incoherent self-interacting quanta,
if the interparticle distance is significantly larger than $\lambda$.
On the other hand, when the 
typical transverse momenta drop below $\mu$, 
the gluons begin to act coherently, and collectivity arises,
because the motion taking place over a distance scale
$1/\mu$  or larger, involves coherently a large number
of hard particles, which gives rise to an average soft color field.
The crucial point of this {\it hard-soft separation} is that the
over long distances  $\lambda > 1/\mu$, the soft mean field 
represents the average gluon motion, but  at short-distances 
$\lambda \le 1/\mu$ the hard gluons may be described
approximately as in free space.
Certainly, such a rigid division of hard and soft physics in terms 
of a single parameter $\mu$, is at his point an arbitrary
and idealizing definition.
However, the arbitrariness can in principle be removed by considering
the variation with respect to $\mu$, as in
the usual renormalization-group framework.
This interesting task is beyond the scope of this paper, and remains
to be addressed in the future.
\bigskip

The non-quilibrium scenario of a lightcone beam of gluons
along the lightcone has two major advantages over the opposite
thermal equilibrium extreme, the isotropic quark-gluon plasma:
First, it favors the 2-scale separation between hard and soft physics.
Second, it allows to choose an axial-type gauge which eliminates
to large extent the problems of non-linearities and of ghost degrees
of freedom that are encountered in usual covariant gauges.
The 2-scale separation arises naturally here, because Lorentz contraction
and time dilation along the beam direction plus the limited
transverse momenta, force the hard gluon fluctuations and 
self-interactions to be highly localized to short-distances,
and separates their
quantum motion from the low-momentum mean-field dynamics over 
comparably long distances.
On the other hand,
the choice of a non-covariant axial gauge, characterized by
a directed four-vector $n$ along a fixed axis, is very suggestive, 
because the geometry and kinematics allows to choose $n$
paralell to the gluon momentum $k_z$, in which case
perturbative QCD calculations formally reduce in many respects to 
the abelian QED counter parts.
This is not possible for an isotropic thermal system, where
all possible directions of gluon motion  are equally probable.
Given these premises, 
the {\it quantum dynamics} is dominated by the self-interactions 
of the hard gluons, which make them fluctuate localized around the lightcone,
whereas the {\it kinetic dynamics} can well be desribed 
statistical-mechanically in terms of mutual interactions
among them and in the presence of their generated soft mean field.
As elaborated in Ref. \cite{ms39}, these notions are the keys to 
formulating a quantum-kinetic description, by combining 
standard techniques of parton evolution and renormalization group,
with  relativistic many-body transport theory.
\smallskip

The {\it main result of this study} within the outlined physics framework,
is a {\it set of three master equations},
which couple
the quantum evolution of short-distance fluctuations of the individual
hard gluons, the space-time development of the gluon system as-a-whole,
and the generation of the soft mean field:
\begin{description}
\item{(i)}
an {\it evolution equation} for the spectral density
$\widehat{\rho}$
of each individual hard gluon, which determines the
intrinsic gluon distribution of a hard gluon in accord with
mass- and coupling-constant renormalization,
and which dresses up the bare initial gluons
to renormalized `quasi-particles'.
\item{(ii)}
a {\it transport equation} for the space-time development of the
whole ensemble of these renormalized gluons with respect 
to their propagation in the self-generated soft mean field,
as well as due to their scatterings off each other,
which determines the physical gluon phase-space density ${\tt G}$.
\item{(iii)}
a {\it Yang-Mills equation} for the generation of the
soft mean field $\overline{A}$, which is induced by the effective 
color current of the hard, renormalized  gluons, where the current
is obtained from the momentum-weighted gluon phase-space density.
\end{description}
Although this set of equations appears at first sight
to be of impractical complexity, it allows in fact for
a practical applicable calculation scheme, as will be demonstrated
with an explicit sample calculation.

To arrive at the above master equations, 
three essential aspects of the problem
have to be merged:
first, the physics-dictated aspect of space-time, 
kinematics and geometry, second,
the quantum field aspect of gluon excitations and self-interactions,
and third, the statistical aspect of multi-particle interactions
in the presence of the mean field.
The non-trivial interconnection of these aspects require to 
{\it work directly at the level of equations of motion}, 
rather than on the level of Feynman diagrams,
because the relative proportions and interactions 
of hard and soft quanta must can only be calculated
self-consistently from the equations of motion.

The strategy for deriving the above three-some of master equations
follows closely the previous work of 
Refs. \cite{ms39}.
The path-integral representation of the Yang-Mills action
gives an infinite set of equations of motion for the
non-equilibrium $n$-point Green functions, which is the
well known analogue of the BBGKY hierarchy \cite{BBGKY}.
This hierarchy, which represents the exact theory, is truncated 
to a system
of equations involving only the 1- and 2-point functions,
by arguing that higher-order correlators $n\ge 3$ are comparably small.
To achieve self-consistency of the truncated set of equations
at the $n=2$ level, the $n\ge3$ functions must be implicitely
lumped into the 1- and 2-point functions.
After  separating hard and soft field modes, as alluded before, 
the 1-point function is identified with the
soft average field 
$\overline{A}_\mu= \langle A_\mu(x) \rangle$ and the
2-point function is given by the hard
gluon correlator
$i\widehat{\Delta}_{\mu\nu}= \langle a_\mu(x)\,a_\nu(y)  \rangle_P$,
where $A_\mu$ and $a_\mu$ represent the soft and hard modes, repectively.
The truncated set of equations of motion then involves
the non-equilibrium version of the Dyson-Schwinger equation
for $\widehat{\Delta}$ and the classical Yang-Mills equation
for the soft mean-field $\overline{A}$.
The two field-equations of motion 
for $\widehat{\Delta}$ and $\overline{A}$ can be cast into
much simpler quantum-kinetic equations with the help of 
the  Wigner-function technique and gradient expansion, 
and the assumption of 2-scale separation implying that the
long-wavelength $\overline{A}$-field is slowly varying on the 
short-distance scale of the hard quantum fluctuations.
The result is then the above set of master equations.
\bigskip

A powerful theoretical framework 
to derive from the exact field equations of motions the above
approximate quantum kinetic equations, is the so-called 
{\it Closed-Time-Path} formalism (CTP).
The CTP formalism is a  general tool
for treating initial value problems of irreversible
multi-particle dynamics in quantum field theory.
It therefore provides an appropriate language to describe
the  problem  of non-equilibrium gluon dynamics within a well-established
theoretical framework.
Originally
introduced by Schwinger \cite{schwinger} and Keldysh \cite{keldysh} 
the CTP formalism and its diverse applications is documented in great
detail in the literature \cite{baym,lifshitz,chou,calzetta,jordan,rammer,stan}.
In particular, I refer
to Ref. \cite{ms39}, where the CTP method
is applied to high-energy QCD, and
to Appendices B and C.

\begin{figure}
\epsfxsize=450pt
\centerline{ \epsfbox{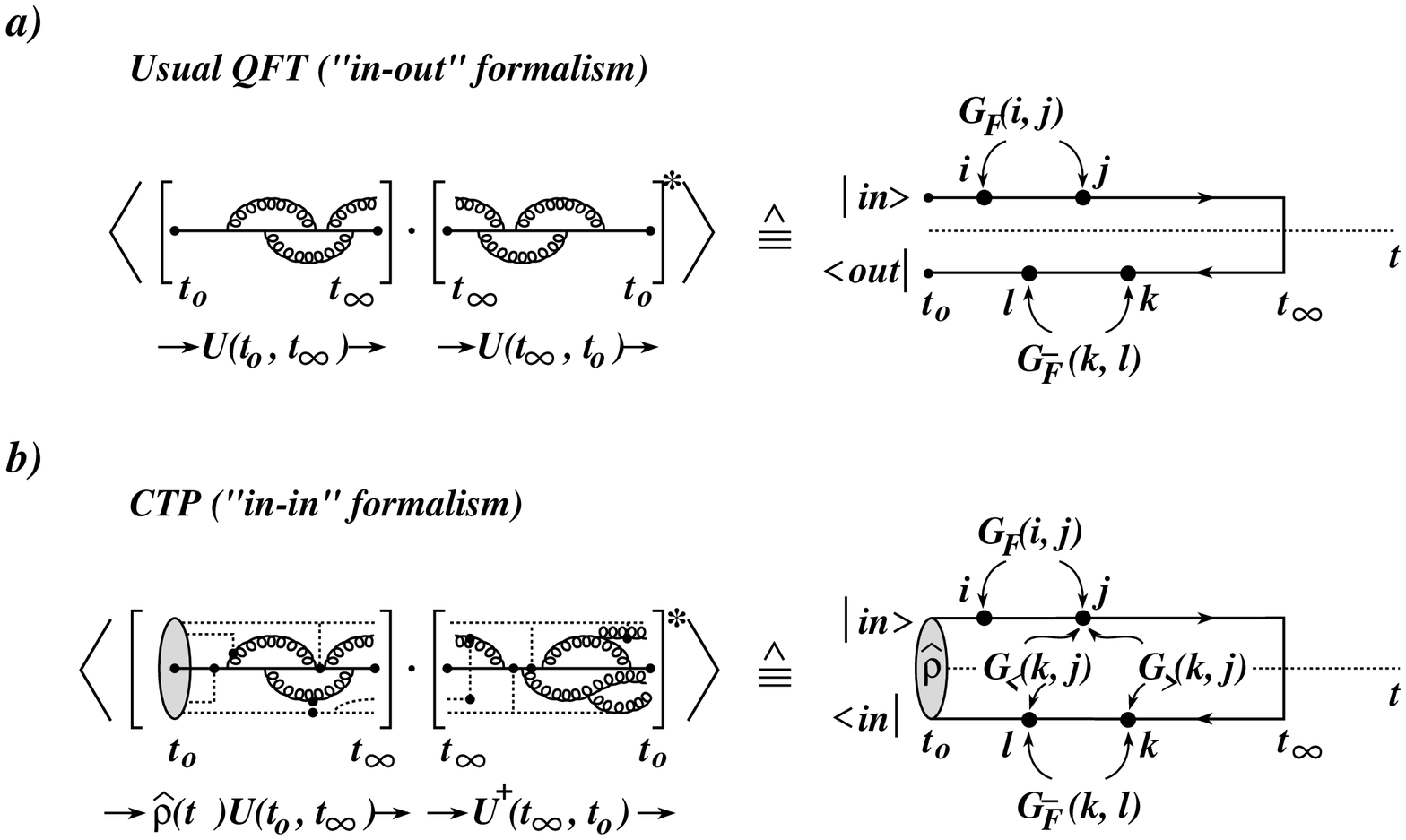} }
\bigskip
\bigskip
\caption{
Difference between {\it in-out} formalism of usual quantum field theory
(in free space or `vacuum')
and the {\it in-in} formalism of the CTP formulation (in the presence
of surrounding particles or `medium').
{\bf a)}
The {\it in-out amplitude}  describes by the evolution
of an asymptotic $|in\rangle$ state at $t_0 \rightarrow -\infty$ to an 
asymptotic $|out\rangle$ state at $t_\infty \rightarrow \infty$
by means of the time evolution operator $U(t_0,t_\infty)$.
Because $U(t_0,t_\infty)=U^\dagger(t_\infty,t_0)$, forward and backward
evolution are identical, and there is 
no correlation between the two time branches.
Consequently, the Feynman propagator
$G_F = G_{\overline{F}}^\ast$ contains the full dynamics of  
the 2-point correlations.
{\bf b)}
The {\it in-in amplitude} starts at $t_0$ with a non-trivial
initial multi-particle state described  by the density
matrix $\widehat{\rho}(t_0)$ and evolves again by means of
time evolution operator $U(t_0,t_\infty)$. Due to statistical
interactions among the many evolving particles acting as a medium,
in this case $\widehat{\rho}(t_0)U(t_0,t_\infty)\ne
U^\dagger(t_\infty,t_0)$.
Consequently, the statistical correlation between
the two time branches have the effect that
$G_F \ne G_{\overline{F}}^\ast$, and moreover require
the introduction of additional correlation functions $G_>$ and $G_<$ to account 
for the cross-talk between upper and lower time branches.
\label{fig:fig2}
}
\end{figure}
\bigskip

The fundamental starting point of non-equilibrium
field theory in the CTP formalism  is to  write 
down the {\it in-in amplitude} $Z_P$ for the evolution of
the initial quantum state $\vert in\rangle$ forward in time
into the remote future.
As reviewed in Appendix B,
this generalizes the  usual quantum field theory approach based on
the vacuum-vacuum transition amplitude, or {\it in-out} amplitude,
to account for the
a-priori-presence of medium particles described by the density 
matrix $\hat{\rho}(t_0)$ and to evolve this
non-trivial initial state 
in the presence of the medium from $t_0$
to $t_\infty$ in the future, 
The {\it in-in} amplitude
$Z_P$ is graphically depicted in Fig. 2, and formally it is given by 
$
Z_P[{\cal J},\hat{\rho}] 
=
\langle \; in \;\vert \;in\;\rangle _{{\cal J},\hat{\rho}}
$,
where 
${\cal J} = ({\cal J}^+,{\cal J}^-)$ is an external source with 
components on the upper $+$ and
lower $-$ time branch, and $\hat{\rho}(t_0)$ denotes the the initial state  density matrix.
From the path-integral representation of $Z_P$ one obtains then the 
non-equilibrium Green functions.
The convenient feature of this Green function formalism on the 
closed-time path $P$ 
is that it is formally  analogous to standard quantum field theory,
based on the vacuum-vacuum, or {\it in-out} amplitude 
$
Z[{\cal J}] 
=
\langle \;out \;\vert \;in\;\rangle _{{\cal J}}
=
\langle 0\vert 0\rangle _{{\cal J}}
$,
except for the fact that in the CTP formalism, the fields have contributions from both
time branches.
For details I refer to
Appendix B, where the basics of the CTP formalism are summarized, in particular,
how to obtain the path-integral for $Z_P$ that generates the
Green functions on the closed-time-path $P$.

The interpretation of this formal apparatus for the
evolution along the closed-time path $P$ is rather simple:
If the initial state is the vacuum itself, that is, 
the absence of a medium generated
by other particles, 
then the density matrix $\hat{\rho}$
is diagonal and
one has  $\vert in\rangle \rightarrow \vert 0 \rangle$. 
In this case the evolution along the $+$ branch is identical to the anti-time ordered
evolution along the $-$ branch (modulo an irrelevant phase),
and space-time points on different branches cannot cross-talk.
In the presence of a medium however, the density matrix contains off-diagonal elements,
and there are statistical correlations between the quantum system and the
medium particles (e.g. scatterings) that lead to correlations between space-time 
points on the $+$ branch with space-time points on the $-$ branch.
Hence, when addressing the evolution of a multi-particle system, both the
deterministic self-interaction of the quanta, i.e. 
the time- (anti-time-) ordered evolution along the $+$ ($-$) branch,
{\it and} the statistical mutual interaction with each other,
i.e. the non-time-ordered cross-talk between $+$ and $-$ branch,  
must be included in a self-constistent manner. 
The CTP method achieves this through the time integtation along the contour $P$.
Although for physical observables the time values are on the $+$ branch, 
both $+$ and $-$ branches will come into play at intermediate steps in 
a self-consistent calculation.
\bigskip

The {\it outline of the paper} is as follows:
In Section 2 the  field equations of motion
for the hard gluon propagator and the soft mean field
are derived from the path-integral representation of the
{\it in-in} amplitude $Z_P$ for non-covariant gauges. 
After separating hard and soft degrees of freedom, 
two key approximations are made, that allow to cast
the infinite hierarchy of exact equations
of motion 
in terms of a truncated system of only two approximate 
equations, namely a Dyson-Schwinger equation for the hard
gluon propagator, and a Yang-Mills equation for the soft field.
In Section 3 the transition to a quantum kinetic description
is worked out. This requires one further key approximation
in conjunction with a clear definition of
quantum and kinetic space-time regimes, such that the aforementioned
2-scale separation is guaranteed. This also defines
the limits for the applicability of the quantum kinetic approximation.
Provided that the separability condition is satisfied,
one finally arrives at the set of master equations discussed above,
for which a systematic calculation scheme is proposed.
In Section 4 an explicit calculation to solve the master equations
is presented for the physics scenario
depicted in Fig. 1. I consider the evolution of an initial
incoherent  ensemble of bare gluons moving colinearly along
the lightcone as it proceeds in its  momentum- and space-time
development and generates its soft mean field.
To avoid overkill of too many technical details, each Section
is accompanied by Appendices.
Appendix A defines the notation and conventions used throughout the paper.
Appendix B reviews the basics of the CTP formalism.
Appendix C discusses the application of the CTP method to QCD for
non-covariant gauges.
Appendix D shows the advantagous absence of ghosts in non-covariant gauges.
Appendix E gives details on how to obtain from the {\it in-in} amplitude ${\cal Z}_P$
an approximate effective action functional from which the 
of motion for the hard gluon propagator and the soft field are derived.
Appendix F summarizes some basic analycity properties 
of the free-field  propagators in the CTP formalism and
discusses their relation to the gluon phase-space density.
\bigskip

\noindent
Finally some remarks on most closely {\it related work} in the literature
(for an extended discussion, see introduction of Ref. \cite{ms39}):

Blaizot and Iancu \cite{BI2}
have in a series of papers developed 
a kinetic theory for `hot QCD', i.e. the case of a high-temperature
quark-gluon plasma.
One of the key elements of their approach is the formulation
of well-defined and consistent approximation scheme.
I adopted many features of this approach to the present,
rather different physics context. New is here 
the inclusion  of the aspect of quantum evolution and renormalization.

McLerran, Venugopalan, {\it et al.} \cite{lmcl}, as well as Makhlin 
\cite{makhlin},
have developed different approaches to calculate
the quantum evolution of parton systems with lightcone dominance, i.e.
in a beam-type scenario as considered in the present work.
The McLerran-Venugopalan model also gives a predictive estimate
for the feedback effect of the coherent mean field on the
hard gluon evolution that generates this field.
In several respects I
follow a similar route. New in the present work is
that it embodies in addition
the aspect of space-time development of the evolution.

Boyanovski {\it et al.} have intensly studied the non-equilibrium
evolution in scalar field theory, using extensively
the techniques of the CTP formalism in conjunction with a large-$N$
expansion. Although the focus on this paper is rather different,
many of the concepts and results in their papers concerning
the first-principle time evolution of quantum system
with associated particle production, dissipation, mean-field dynamics, etc.,
may serve as a scalar toy model for QCD.
\bigskip
\bigskip

\section{INTERPLAY OF `HARD' AND `SOFT' GLUON DYNAMICS}
\label{sec:section2}

\subsection{The {\it in-in} amplitude for QCD in non-covariant gauges
and the concept of approximation}

The {\it in-in} amplitude $Z_P$ introduced in Sec. 1
admits a path-integral representation which
is the generating functional for the non-equilibrium Green functions
defined on closed-time-path $P$, as discussed in Appendices B and C:
\begin{equation}
Z_P[{\cal K}] \;=\;
\int \,{\cal D}{\cal A} \; 
\exp \left\{i \left( \frac{}{} I \left[ {\cal A}, {\cal K}\right]
\right)\right\}
\label{ZZ3}
\;,
\end{equation}
where ${\cal A}_\mu^a = ({\cal A}^{a\;+}_\mu,{\cal A}^{a\;-}_\mu)$ 
has two components, living on the upper ($+$) and lower ($-$) 
time branches of Fig. 2, with 
${\cal D}{\cal A}= 
\prod_{\mu, a}{\cal D}{\cal A}^{a\;+}_\mu {\cal  D}{\cal A}^{a\;-}_\mu$, 
and ${\cal K}$ represents the presence of external sources.
I consider here the
{\it class of non-covariant gauges} defined by
\cite{gaugereview,gaugebook},
\begin{equation}
\langle \;n^\mu\,{\cal A}_\mu^a(x)\;\rangle \;=\;0
\label{gauge100}
\;,
\end{equation}
where
$n^\mu$ is a constant 4-vector, being either space-like ($n^2 < 0$), 
time-like ($n^2 > 0$), or light-like ($n^2=0$).
The particular choice of 
the vector $n^\mu$ is usually dictated by
the physics or computational convenience, and distinguishes
{\it axial gauge} ($n^2 < 0$),
{\it temporal gauge} ($n^2 >  0$), and
{\it lightcone gauge} ($n^2=  0$).
Referring to Appendix D, 
the great advantage of these gauges is that the Fadeev-Popov ghosts
decouple, so that in practical calculations the ghost degrees of freedom
can be ignored, just as in abelian gauge theories.

Then the action $I$ in the exponential of (\ref{ZZ3}) is given by (c.f. Appendix C),
\begin{equation}
I \left[ {\cal A}, {\cal K}\right]
\;\equiv\;
I_{YM}\left[{\cal A}\right] 
\;+\;
I_{GF}\left[n\cdot{\cal A}\right]
\;+\; 
{\cal K}[{\cal A}]
\;,
\label{Ieff}
\end{equation}
containing the Yang-Mills action $I_{YM}$, the gauge fixing term $I_{GF}$,
and the initial state source term ${\cal K}$,
containing multi-point correlations concentrated
at $t=t_0$:
\begin{eqnarray}
I_{YM}\left[{\cal A}\right]
& =&
-\frac{1}{4} \int_P d^4x \,{\cal F}_{\mu\nu}^a(x) {\cal F}^{\mu\nu, \,a}(x) 
\;,\;\;\;\;\;\;\;\;\;\;\;\;\;\;\;
I_{GF}\left[n\cdot{\cal A}\right]
\;=\;
- \frac{1}{2\alpha}
\int_P d^4x \, \left[n\cdot {\cal A}^a(x)\right]^2
\;,
\\
{\cal K}[{\cal A}]
&=&
{\cal K}^{(0)} \;+\;
\int_P d^4x \;{\cal K}^{(1)\;a}_{\;\;\;\,\mu}(x) \;{\cal A}^{\mu, \,a}(x) 
\;+\;
\frac{1}{2}
\int_P d^4xd^4y \;{\cal K}^{(2) \;ab}_{\;\;\;\,\mu\nu}(x,y)\;{\cal A}^{\mu, \,a} (x)
\,{\cal A}^{\nu, \,b}(y) 
\;\;+\;\;
\ldots
\;.
\end{eqnarray}

The exact knowledge of 
the {\it in-in} amplitude $Z_P$ from (\ref{ZZ3}), would require
to the calculation of all Green functions up to infinite order, and
would correspond to the full solution of QCD in non-equilibrium media.
Rather than that, the realistic goal is to
formulate a practical calculation scheme for the kinetic evolution
of a multi-guon system.
In  order to make progress, one needs 
to make reasonable approximations that are
consistent with the specific physical problem under study,
and truncate the infinite hierarchy of Green functions.

In this Section a closed set of
approximate equations is derived that are in principle solvable, 
given a suitable physics scenario.  The  basic idea is to describe an
evolving gluon system in terms of  two distinct components, namely,
{\it hard, short-range quantum fluctuations} and 
{\it soft, long-wavelength collective excitations}, 
which I assume to be separable by a 
characteristic space-time distance. 
It is clear that the relative proportions and interactions 
of hard and soft degrees of freedom must be calculated
self-consistently from the equations of motion.
\medskip

Starting from the {\it in-in} amplitude (\ref{ZZ3}), the strategy of procedure 
is the following:
\begin{description}
\item[1.]
The exact expression of the {\it in-in} amplitude
$Z_P \equiv \exp(i W_P)$ is rewritten in terms
of soft and hard field modes by splitting the gauge field
${\cal A}_\mu = A_\mu + a_\mu$. Therefrom one obtains an infnite set of
coupled equations for the Green functions. In order to reduce
this to a finite system, I make 
\begin{itemize}
\item{}
{\bf approximation 1:}
The functional $W_P= -i \ln Z_P$ is expressed 
in terms of connected 1- and 2-point functions
${\cal G}^{(1)}$, ${\cal G}^{(2)}$ alone by eliminating
${\cal G}^{(n)}$ for $n\ge 3$ as dynamical variables.
Then the expectation values of
${\cal G}^{(1)}$ and  ${\cal G}^{(2)}$ describe the induced soft 
mean field $\overline{A}_\mu$ and the
hard (soft) correlation functions $\widehat{\Delta}_{\mu\nu}$ ($\widehat{D}_{\mu\nu}$).
\end{itemize}

\item[2.]
From the truncated functional $W_P$ the corresponding
effective action $\Gamma_P$ is obtained, which generates the desired 
self-consistent equations of motion for 
$\overline{A}_\mu$,
$\widehat{\Delta}_{\mu\nu}$ and $\widehat{D}_{\mu\nu}$.
Here I make 
\begin{itemize}
\item{}
{\bf approximation 2:}
It is assumed that
the soft field dynamics can  be treated classically by the
non-propagating average field $\overline{A}_\mu$, and that the long-range
propagation of soft modes, described by $\widehat{D}_{\mu\nu}$ 
may be ignored at this level, i.e.
$\widehat{D}_{\mu\nu}\ll \overline{A}_\mu\overline{A}_\nu$.
This assumption is motivated by
the widely studied \cite{lmcl,bmcg} observation that a classical treatment of the 
long-distance  dynamics of bosonic quantum fields at high density,
obeying the classical field equations, should provide a good approximation,
if the soft modes are sufficiently occupied.
\end{itemize}
\end{description}
\smallskip

The original infinite equation system can then be reduced to 
a Yang-Mills equation for the classical, soft field $\overline{A}_\mu$, as it is
induced by the current of hard quanta, and a Dyson-Schwinger equation for the
hard propagator $\widehat{\Delta}_{\mu\nu}$ subject to the presence
of the soft mean field and to quantum fluctuations.
These field equations of motion are still of very intractable  non-linear character. 
They are further simplified to quantum-kinetic equations in Sec. III.
\medskip

\subsection{Separating soft and hard dynamics}

The first step in the strategy is the separation of
soft and hard physics in the
path-integral formalism with Green functions of
both the soft and hard quanta in the presence of the soft classical field
that is induced by and feeding back to the quantum dynamics.
A frequently used method
for separate treatment of quantum and classical dynamics in field theory is the
so-called `background field method'  
\cite{bgm1} which has been studied,
e.g., in the context of dynamical symmetry breaking, 
vacuum structure, confinement and gravity,
or for hot plasmas in finite temperature QCD.
Within the background field method, one would split up the gauge field
appearing in the classical action 
into an external  classical background field  and a quantum field 
which remains the sole dynamical variable in the path integral.
I will however not follow this path, and
rather prefer to {\it treat soft and hard physics on equal footing}, 
that is, to separate the gauge field into a soft classical field plus  
its soft quantum excitations, and a hard quantum field. Then both
soft and hard fields can quantized and remain as dynamical variables a priori.

The gauge field
${\cal A}_\mu$ appearing in the classical action 
$I_{YM}\left[{\cal A}\right]$ is split up into a soft (
long-range) part $A_\mu$,  and a hard (short-range) quantum field $a_\mu$:
\begin{equation}
{\cal A}_\mu^a(x) \;=\;
\int \frac{d^4k}{(2\pi)^4}\,e^{+i k \cdot x} 
{\cal A}_\mu^a(k) \;\, \theta(\mu^2 - k_\perp^2) 
\;\,+\;\,
\int \frac{d^4k}{(2\pi)^4}\,e^{+i k \cdot x} 
{\cal A}_\mu^a(k) \,\; \theta(k_\perp^2 -\mu^2)
\;\;\equiv\;\;
A_\mu^a(x) \;+\; a_\mu^a(x)
\label{Aa}
\;.
\end{equation}
This is the formal definition of the terms `soft' and `hard',
as used in this paper.
The soft and hard physics are separated by the 
momentum scale $\mu$ which is at this point arbitrary. 
However, this arbitrariness can in principle be
overcome by considering $\mu(x)$ as a dynamical {\it variable}
depending on the space-time point $x$, 
rather than a fixed parameter,
and determining it self-consistently from the local
stability condition
$d {\cal A}_\nu(x)/d\mu^2(x)=0$.
From (\ref{Aa}) it is obvious, that
the corresponding scale in space-time, $\lambda \equiv 1/\mu$,
divides soft and hard regimes in terms of 
the transverse wavelength of field modes,
so that one may associate the soft field $A_\mu$ being responsible 
for long range color collective effects, 
and the hard field $a_\mu$ embodying the short-range quantum dynamics.
Consequently, the field strength tensor receives a soft part,
a hard part, and a mixed contribution,
\begin{equation}
{\cal F}_{\mu\nu}^{a}(x) \;\equiv\;
\left(\frac{}{}
F_{\mu\nu}^{a}[A] \;+\; f_{\mu\nu}^{a}[a] \;+\; \phi_{\mu\nu}^{a}[A,a]
\right) (x)
\label{Ffphi}
\;.
\end{equation}
\smallskip

When quantizing this decomposed theory by writing down the
appropriate {\it in-in}-amplitude $Z_P$, one must be
consistent with the gauge field decomposition (\ref{Aa}) into soft and 
hard components and with the classical character of the former.
Substituting the soft-hard mode decomposition (\ref{Aa}) 
into (\ref{ZZ3}), 
the functional integral of the {\it in-in} amplitude (\ref{Z3}) becomes:
\begin{equation}
Z_P[ {\cal  K}] \;=\;
\int \,{\cal D} A \,{\cal D} a \; 
\exp \left\{i \left( \frac{}{} I \left[ A\right]
\;+\; I \left[ a\right] \;+\; I \left[ A, a\right]
\right)\right\}
\label{Z4}
\;,
\end{equation}
with the soft, hard, and mixed contribution, respectively,
\begin{eqnarray}
I \left[ A\right]
&=&
\int d^4x \left(
\frac{}{}
-\frac{1}{4} F_{\mu\nu}^{a} F^{\mu\nu,\;a} -
\frac{1}{2\alpha}\,(n\cdot A^a)^2 
\right)
\nonumber \\
& &
\;\;\;\;\;\;\;\;\;\;\;\;\;\;\;\;\;\;\;\;\;\;\;\;\;
\;\;\;\;\;\;\;\;\;\;\;\;\;\;\;\;
\;+\;
\int d^4x \;{\cal K}_{\;\;\;\mu}^{(1)\;a} A^{\mu,\;a} 
\;+\;
\int d^4xd^4y\; A^{\mu,\;a} \,{\cal K}_{\;\;\;\mu\nu}^{(2)\;ab}\, A^{\nu,\;b} 
\;\;+\;\;\ldots
\label{Isoft}
\;,
\\
I \left[ a\right]
&=&
\int d^4x \left(
\frac{}{}
-\frac{1}{4} f_{\mu\nu}^{a} f^{\mu\nu,\;a} -
\frac{1}{2\alpha}\,(n\cdot a^a)^2 
\right)
\nonumber \\
& &
\;\;\;\;\;\;\;\;\;\;\;\;\;\;\;\;\;\;\;\;\;\;\;\;\;
\;\;\;\;\;\;\;\;\;\;\;\;\;\;\;\;
\;+\;
\int d^4x \;{\cal K}_{\;\;\;\mu}^{(1)\;a} a^{\mu,\;a} 
\;+\;
\int d^4xd^4y\; a^{\mu,\;a} \,{\cal K}_{\;\;\;\mu\nu}^{(2)\;ab}\, a^{\nu,\;b} 
\;\;+\;\;\ldots
\label{Ihard}
\;,
\\
I \left[ A, a\right]
&=&
\int d^4x \left(
\frac{}{}
-\frac{1}{4} \phi_{\mu\nu}^{a} \phi^{\mu\nu,\;a}
-\frac{1}{2} \left\{
\phi_{\mu\nu}^{a} F^{\mu\nu,\;a}
\;+\;
\phi_{\mu\nu}^{a} f^{\mu\nu,\;a}
\;+\;
F_{\mu\nu}^{a} f^{\mu\nu,\;a}
\right\}
\right)
\nonumber \\
&=&
\int d^4x \left\{\frac{}{}
-g f^{abc}
\left[
\frac{}{}
\;
(\partial_\mu^x A_\nu^a) \;\left( a^{\mu,\;b} a^{\nu,\;c} +
A^{\mu,\;b} a^{\nu,\;c} + a^{\mu,\;b} A^{\nu,\;c} \right)
\right.
\right.
\nonumber \\
& &
\left.
\frac{}{}
\;\;\;\;\;\;\;\;\;\;\;\;\;\;\;\;\;\;\;\;\;\;\;\;\;\;\;\;\;\;\;\;\;\;\;\;
\;+\;
(\partial_\mu^x a_\nu^a) \;\left( A^{\mu,\;b} A^{\nu,\;c} +
a^{\mu,\;b} A^{\nu,\;c} + A^{\mu,\;b} a^{\nu,\;c} \right)
\right]
\nonumber \\
& &
\left.
\;\;\;\;\;\;\;\;\;\;\;
\;-\;
g^2 f^{ace} f^{bde}
\left[
\frac{}{}
2\, A_\mu^a  A_\nu^b  a^{\mu,\;c} a^{\nu,\;d} \;+\;
A_\mu^a  A_\nu^b  A^{\mu,\;c} a^{\nu,\;d} \;+\;
a_\mu^a  a_\nu^b  a^{\mu,\;c} A^{\nu,\;d}
\right]
\;
\right\}
\label{Imixed}
\;.
\end{eqnarray}
Note that in (\ref{Ihard}) and (\ref{Imixed}) terms involving 2-products
$\propto a_\mu\,A_\nu$ do not contribute   to $Z_P$, because their
expectation value vanishes due to
the soft-hard separation (\ref{Aa}) which defines $a_\mu$ and $A_\nu$ 
as complimentary.
\medskip

At this point I make  {\it approximation 1} from above.
It is assumed that initial state can be represented
as an ensemble of incoherent hard gluons, 
each of which has very small spatial extent
$\Delta r_\perp\ll \lambda = 1/\mu$, 
corresponding to transverse momenta $k_\perp^2 \gg \mu^2$.
By definition of $\mu$, the short-range character of
these quantum fluctuations implies that the expectation value $\langle a_\mu\rangle$
vanishes at all times. However, the long-range correlations of the eventually
populated soft modes with small momenta $k_\perp^2\le\mu^2$ may lead
to a collective mean field with non-vanishing $\langle A_\mu\rangle$.
Accordingly, I impose the following condition
on the expectation values of the fields:
\begin{equation}
\langle \; A_\mu^a(x)\;\rangle \;
\left\{
\begin{array}{ll}
\;=\;0 \;\;\;\mbox{for} \;t\,\le\,t_0 \\
\;\ge\;0 \;\;\;\mbox{for} \;t\,>\,t_0
\end{array}
\right.
\;\;\;\;\;\;\;\;\;\;\;\;\;\;\;\;\;\;\;\;
\langle \; a_\mu^a(x)\;\rangle \;\stackrel{!}{=}\; 0
\;\;\;\mbox{for all} \;t
\;.
\label{MFconstraint1}
\end{equation}
Now I make {\it approximation 2}, that is,
the quantum fluctuations of the soft field are ignored, assuming
any multi-point correlations of soft fields to be small,
\begin{equation}
\langle \; A_{\mu_1}^{a_1}(x_1)\; \ldots A_{\mu_n}^{a_n}(x_n) \;\rangle
\ll\;\langle \;  A_{\mu_1}^{a_1}(x_1)\;\rangle
\;\ldots \;\langle \;  A_{\mu_1}^{a_n}(x_n)\;\rangle
\;\;\;\;\;\;\mbox{for all $n\ge 2$}
\;,
\end{equation}
i.e. take $A_\mu$ as a non-propagating and non-fluctuating, 
classical field. In particular,
\begin{equation}
iD_{\mu\nu}^{ab}(x,y) \;\equiv\;
\langle \; A_{\mu}^{a}(x) A_{\nu}^{b}(y) \;\rangle
\ll\;\langle \;  A_{\mu}^{a}(x)\;\rangle
\;\langle \;  A_{\nu}^{b}(y)\;\rangle
\;,
\label{MFconstraint1a}
\end{equation}
so that the limit "$D_{\mu\nu} \rightarrow 0$" can be considered.

As explained in more detail in Appendix E, 
the generating functional for the 
{\it connected} Green functions,
\begin{equation}
W_P\left[{\cal K}\right] \;\;=\;\;
- i \,\ln\, Z_P\left[ {\cal K}\right]
\;,
\end{equation}
which generates 
the infinite set of connected $n$-point 
Green functions ${\cal G}^{(n)}$ via
\begin{equation}
(-i)\, {\cal G}_{\;\;\;\;\mu_1\ldots \mu_n}^{(n)\;a_1\ldots a_n}(x_1,\ldots, x_n)
\;\equiv\;
\left.
\frac{\delta}{i \,\delta{\cal K}^{(n)}}
W_P[{\cal K}]\right|_{{\cal K}=0}
\;\;=\;\;
\langle\; a_{\mu_1}^{a_1}(x_1)\ldots a_{\mu_k}^{a_k}(x_k)\,
A_{\mu_{k+1}}^{a_{k+1}}(x_{k+1})\ldots A_{\mu_n}^{a_n}(x_n)\;\rangle_P^{(c)}
\label{Green2a}
\;,
\end{equation}
is truncated at level $n\ge 3$ on the basis of approximation
(\ref{MFconstraint1}) and (\ref{MFconstraint1a}). As a result,
$W_P$ becomes a functional of
the 1-point function (soft mean field $\overline{A}$) 
and  the 2-point function (hard propagator $\widehat{\Delta}$) {\it only}:
\begin{eqnarray}
{\cal G}_{\;\;\;\;\mu}^{(1)\;a}(x)
\;=\;\langle\;A_\mu^a(x)\;\rangle_P^{(c)}
\;\;\equiv\; \;\overline{A}_\mu^a(x)
\;\;\;\;\;\;\;\;\;
{\cal G}_{\;\;\;\;\mu\nu}^{(2)\;ab}(x,y)
\;=\;
\langle \;a_\mu^a(x) a_\nu^b(y)\;\rangle_P^{(c)}
\;\;\equiv\;\; 
i \widehat{\Delta}_{\mu\nu}^{ab}(x,y)
\;.
\end{eqnarray}
These relations 
define the soft, classical  mean field $\overline{A}$, and the hard  
quantum propagator $\widehat{\Delta}$ in terms of expectation values
of soft and hard field operators $A_\mu$ and $a_\mu$, respectively.
One now readily 
one obtains the {\it effective action} $\Gamma_P$ (or proper vertex functional) via
Legendre transformation (c.f. Appendix E):
\begin{equation}
\Gamma_P\left[{\cal G}\right]
\;\;\approx\;\;
\Gamma_P\left[\overline{A}, \widehat{\Delta}\right]
\;\;=\;\;
W_P\left[{\cal K}^{(1)},{\cal K}^{(2)}\right] 
\;-\;{\cal K}^{(1)}\circ\overline{A}
\;-\;\frac{1}{2} 
\;{\cal K}^{(2)}\circ \left(\,i\widehat{\Delta}\;+\;\overline{A}\;\overline{A}
\right)
\label{Gamma000}
\;,
\end{equation}
which is a functional of {\it only} the soft field $\overline{A}$ and the
hard propagator $\widehat{\Delta}$ as independent dynamical degrees of freedom.
\smallskip

The  {\it equations of motion} for the 
mean field $\overline{A}$ and for the hard propagator $\widehat{\Delta}$ 
in the presence of  sources, follow now by differentiaition of 
(\ref{Gamma000}) with respect to $\overline{A}$ and $\widehat{\Delta}$
(c.f. Appendix E): 
\begin{eqnarray}
\frac{\delta \Gamma_P}{\delta \overline{A}_\mu^a(x)} &=&
- {\cal K}^{(1) \;\mu,a}(x) \;-\; 
\int_P d^4y \,K^{(2)\;\mu\nu,ab}(x,y) \;\overline{A}^{\nu,\,b}(y)
\;,
\label{eom0A}
\\
\frac{\delta \Gamma_P}{\delta \widehat{\Delta}_{\mu\nu}^{ab}(x,y)} 
&=& \frac{1}{2i}\;{\cal K}^{(2)\;\mu\nu,ab}(x,y)
\label{eom0B}
\;.
\end{eqnarray}
The 
self-consistent equations of motion of the dynamically evolving system are then
obtained from (\ref{eom0A}) and (\ref{eom0B}) by 
(i) imposing initial conditions in terms
of the ${\cal K}$ kernels at $t=t_0$, and (ii) by obtaining an explicit formula
for $\Gamma_P$ in terms of $\overline{A}_\mu$ and $\widehat{\Delta}_{\mu\nu}$.
Concerning the initial conditions,
I restrict  to  non-equilibrium initial states of
Gaussian form (i.e. quadratic in the hard modes) and
do not consider possible linear force terms. That is, I set
\begin{equation}
\left. {\cal K}^{(1)}(x)\right|_{x^0=t_0} \;=\; 0
\;,\;\;\;\;\;\;
\left. {\cal K}^{(2)}(x,y)\right|_{x^0=y^0=t_0} \;\ge\; 0
\;.
\label{inicond}
\end{equation}
\smallskip
To obtain an explicit expression for $\Gamma_P$, the
formal loop expansion of (\ref{Gamma000})
results in the well known
Cornwall-Jackiw-Tomboulis formula \cite{cornwall}:
\begin{equation}
\Gamma_P\left[\overline{A}, \widehat{\Delta}\right]\;=\;
\;\overline{I}_{eff}[\overline{A},a] 
\;-\; 
\frac{i}{2}\,\mbox{Sp}\left[\;\ln(\Delta_{0}^{-1} \widehat{\Delta})\,-\, 
\overline{\Delta}_{0}^{\;-1} \widehat{\Delta} \,+\, 1 \;\right]
\;\;+\;\; \Gamma_P^{(2)}\left[\overline{A}, \widehat{\Delta}\right]
\;,
\label{Gamma2}
\end{equation}
where 
$\mbox{Sp}[AB\ldots]\equiv \mbox{Tr} \int_P d^4x_1d^4x_2 \ldots A(x_1)B(x_2)\ldots$ 
stands for both the trace over color and Lorentz indices, as well as the 
integration over all space-time positions, hence giving the
expectation value $\langle AB\ldots \rangle_P$ as defined in Appendix C.
The {\it physical interpretation} of the various terms in this expression for $\Gamma_P$ 
is the following \cite{ms39}:
\begin{description}
\item[(i)]
The first term is of order $\hbar^0$ and is given by 
the classical action (\ref{Isoft})-(\ref{Ihard}) at $A=\overline{A}$ and
switched-off sources ${\cal K}$: 
\begin{equation}
\overline{I}_{eff}[\overline{A},a] 
\;\equiv\;
\left[ \frac{}{}
I \left[ A\right] \;+\; I \left[ a\right] \;+\; I \left[ A, a\right]
\right]_{A = \overline{A},\,{\cal K}=0}
\;.
\label{I3}
\end{equation}
Notice that in the limit $a=0$, this 
reduces to the classical action 
for the soft mean field,
$\overline{I}_{eff}[\overline{A},0] 
=I_{YM}[\overline{A}] + I_{GF}[n\cdot\overline{A}]$.
\item[(ii)]
The second  term in (\ref{Gamma2}) is of order $\hbar^1$ and  contains the
the contributions of the coupling between the soft mean field $\overline{A}$
and the hard quantum propagator $\widehat{\Delta}$.
The {\it free} propagator (see Fig. 3a) is given by
$
[\delta^2 \overline{I}_{eff}[\overline{A},a]
/\delta a(x) \delta a(y) ]_{A=0;\,a = 0} 
$
with $\overline{A}$ switched off,
which yields
\begin{equation}
(\Delta_{0}^{-1})_{\mu\nu}^{ab}(x,y) 
\;\;=\;\;
-\;
\delta_{ab} \,
\delta_P^4(x,y)  \;\; d_{\mu\nu}(\partial_x) \;\;\partial_x^2
\;,
\label{D0}
\end{equation}
where it is understood that the space-time arguments 
$x$ and $y$ in $\Delta_{0}$ satisfy $(x-y)_\perp^2 < 1/\mu^2$, and
\begin{equation}
d_{\mu\nu}(\partial_x) 
\;\equiv\;
g^{\mu\nu} \;-\;
\frac{n^\mu\partial_x^\nu+ n^\nu\partial_x^\mu}{n\cdot\partial_x}
\;+\; \left(n^2 + \alpha^{-1} \partial_x^2\right)\,
\frac{\partial^\mu_x\partial^\nu_x}{(n\cdot \partial_x)^2}
\;,
\label{Box}
\end{equation}
\newpage

\begin{figure}
\epsfxsize=550pt
\centerline{ \epsfbox{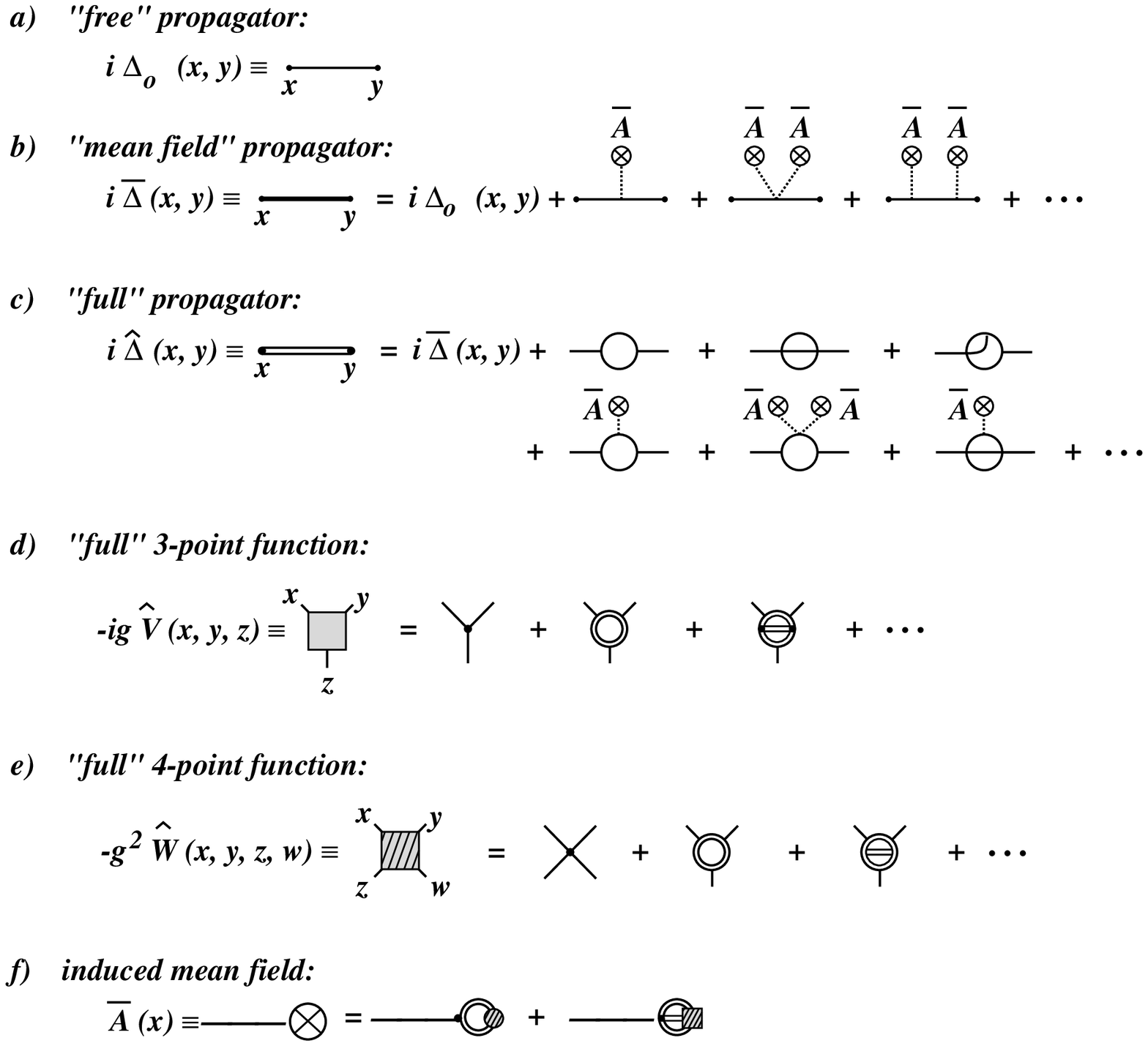} }
\bigskip
\bigskip
\bigskip
\caption{
Diagrammatics of the various terms used for the
$n$-point functions appearing in the text.
The 2-point function ${\cal G}^{(2)}= i\Delta$ is the hard gluon
propagator with the {\it free-field} propagator
$\Delta_0$ (no interactions), the {\it mean-field} propagator 
$\overline{\Delta}$ (including the 
interactions with the classical soft field
$\overline{A}$), and the {\it full} propagator $\widehat{\Delta}$
(including both mean-field and quantum (loop) interactions).
Similarly, the connected 3-point function 
${\cal G}^{(3)}= -ig\widehat{V}$ and the 4-point function
${\cal G}^{(4)}= -g^2\widehat{W}$ contain
soft mean-field plus hard quantum contributions with internal
full propagator $\widehat{\Delta}$.
Finally, the 1-point function is the soft mean field $\overline{A}$
that is generated by the hard gluons through the coupling to the full 
3-point and 4-point functions $\widehat{V}$ and $\widehat{W}$.
\label{fig:fig3} 
}
\end{figure}
\newpage

Even in the absence of quantum fluctuations,
these contributions amount to a modification of the free propagator,
such that the free propagator $\Delta_{0}$ becomes an effective
propagator  $\overline{\Delta}$ in the mean field,
{\it dressed} up by the presence of  $\overline{A}$.
This {\it mean field propagator} (see Fig. 3b), 
denoted by $\overline{\Delta}$,
is obtained from
$
[\delta^2 \overline{I}_{eff}[\overline{A},a]
/\delta a(x) \delta a(y) ]_{A=\overline{A};\,a = 0} 
$
with finite $\overline{A}\ne 0$,
which results in
\begin{equation}
(\overline{\Delta}^{\;-1})_{\mu\nu}^{ab}(x,y) 
\;\;=\;\;
(\Delta_{0}^{-1})_{\mu\nu}^{ab}(x,y) 
\;\,-\;\,
\overline{\Pi}_{\mu\nu}^{ab}(x,y) 
\label{DMF}
\;,
\end{equation}
where $\overline{\Pi}$ denotes the self-energy contribution associated with
the presence of the mean-field $\overline{A}\ne 0$. 
Its explicit expression is given below in (\ref{PiMF}).
In other words,
the effect of the mean field 
is to shift the pole in the free propagator $\Delta_{0}$ of (\ref{D0})
by a dynamically induced `mass' term $\propto\overline{\Pi}$, which
can produce screening and damping effects.
Note that 
$\Delta_{0}^{-1} = \overline{\Delta}^{\;-1}\vert_{\overline{A}=0}$.
It is important to realize that this mean field effect is still on the 
classical tree-level, and does not involve quantum fluctuations
associated with radiative  self-interactions among the hard gluons.

\item[(iii)]
The  last term $\Gamma_P^{(2)}$ in (\ref{Gamma2})
represents the sum of all two-particle irreducible graphs of order 
$\hbar^2,\hbar^3,\ldots$ \cite{cornwall},
with the {\it full} propagator 
$\widehat{\Delta}$,
dressed by both the soft mean field and the quantum self-interactions 
(see Fig. 3c),
\begin{equation}
\widehat{\Delta}_{\mu\nu}^{ab}(x,y)
\;\equiv \;
\widehat{\Delta}_{[\overline{0}]\;\mu\nu}^{ab}(x,y)
\;+\;
\delta\widehat{\Delta}_{[\overline{A}]\;\mu\nu}^{ab}(x,y)
\;,
\label{DELTA}
\end{equation}
where
the dependence of the full propagator
on the soft mean field $\overline{A}$
is indicated by an explicit  subscript, and
\begin{equation}
\widehat{\Delta}_{[\overline{0}]\;\mu\nu}^{ab}
\;=\;
\left. 
\widehat{\Delta}_{[\overline{A}]\;\mu\nu}^{ab} \right|_{\overline{A}=0}
\;\;\;\;\;\;\;\;\;\;\;
\left.
\delta\widehat{\Delta}_{[\overline{A}]\;\mu\nu}^{ab} \right|_{\overline{A}=0}
\;=\;0
\;.
\end{equation}
Note that 
$\widehat{\Delta}_{[\overline{0}]} \ne \Delta_{0}$,
that is, 
$\widehat{\Delta}_{[\overline{0}]}$ denotes 
the {\it full} propagator 
for $\overline{A}=0$,
whereas $\Delta_{0}$ is the {\it free} propagator (\ref{D0}).
The real (dispersive) part of $\Gamma_P^{(2)}$ contains
the virtual loop corrections associated with the gluon self-interactions, 
whereas the imaginary (dissipative) 
part contains the  emission, absorption, and 
scattering processes of hard gluons.
In other words, $\Gamma_P^{(2)}$ embodies all the interesting quantum dynamics
that is connected with renormalization group, 
entropy generation, dissipation, etc..
The explicit form of $\Gamma_P^{(2)}$ is diagramatically shown in Fig. 4,
with the vertices and lines defined by Fig. 3.
Suppressing color and Lorentz indices and employing
a condensed notation, e.g., $\widehat{\Delta}(x_1,x_2) \equiv 
\widehat{\Delta}_{\mu\nu}^{ab}(x_1,x_2)$, 
the corresponding formula is,
\begin{equation}
\Gamma_P^{(2)}\left[\overline{A}, \widehat{\Delta} \right]
\;=\;
\Gamma_{(1)} \;+\; \Gamma_{(2)} \;+\; \Gamma_{(3)} \;+\; 
\Gamma_{(4)}
\;,
\label{Gamma20}
\end{equation}
with the following contributions, 
\begin{eqnarray}
\Gamma_{(1)}
&=&
\;\frac{1}{8} \,g^2\;
\;\int_P d^4x d^4y \int_P d^4x_1d^4y_1 
W_{0}(x,y,x_1,y_1)\;\widehat{\Delta}(y_1,x_1)\;\widehat{\Delta}(y,x)
\label{Gamma2a}
\\
\Gamma_{(2)}
&=&
\;\frac{i}{12} \,g^2\;
\;\int_P d^4x d^4y \int_P\prod_{i=1}^{2} d^4x_id^4y_i 
\;\;V_{0}(x,x_1,x_2)\;\;\widehat{\Delta}(x_1,y_1)\;\widehat{\Delta}(x_2,y_2)\; 
\widehat{V}(y_2,y_1,y) \;\widehat{\Delta}(y,x)
\label{Gamma2b}
\\
\Gamma_{(3)}
&=&
\;\frac{1}{48} \,g^4\;
\;\int_P d^4x d^4y \int_P\prod_{i=1}^{3} d^4x_id^4y_i 
\;\;W_{0}(x,x_1,x_2,x_3)\;\;\widehat{\Delta}(x_1,y_1)\;
\widehat{\Delta}(x_2,y_2)\; \widehat{\Delta}(x_3,y_3)\;
\nonumber \\
& &
\;\;\;\;\;\;\;\;\;\;\;\;\;\;\;\;\;\;\;\;\;\;\;\;\;
\;\;\;\;\;\;\;\;\;\;\;\;\;\;\;\;\;\;\;\;\;\;\;\;\;
\;\;\;\;\;
\times\;
\widehat{W}(y_3,y_2,y_1,y) \;\;\widehat{\Delta}(y,x)
\label{Gamma2c}
\\
\Gamma_{(4)}
&=&
\;\frac{i}{96} \,g^4\;
\;\int_P d^4x d^4y \int_P\prod_{i=1}^{2} d^4x_id^4y_id^4z_i 
\;\;W_{0}(x,x_1,x_2,x_3)\;\;\widehat{\Delta}(x_2,z_2)\;\widehat{\Delta}(x_3,z_3)\; 
\nonumber \\
& &
\;\;\;\;\;\;\;\;\;\;\;\;\;\;\;\;\;\;\;\;\;\;\;\;\;
\;\;\;\;\;\;\;\;\;\;\;\;\;\;\;\;\;\;\;\;\;\;\;\;\;
\;\;\;\;\;\;
\times\;
\widehat{V}(z_3,z_2,z_1) \;\;\widehat{\Delta}(z_1,y_1) \;
\widehat{\Delta}(x_1,y_2) \;\;\widehat{V}(y_1,y_2,y)
\;\;\widehat{\Delta}(y,x)
\label{Gamma2d}
\;.
\end{eqnarray}
The functions $\widehat{V}$ and $\widehat{W}$ 
are the {\it full} proper vertex functions for the 3-gluon and 
4-gluon coupling, respectively. Their diagramatic representation is shown in 
Fig.  3d and 3e, and formally they are given by 
the functional derivatives of $\Gamma_P$ at $\overline{A}\ne 0$, namely,
$
\left[
\delta^n \Gamma_P / 
\delta a(x_1) \ldots \delta a(x_i)\delta A(x_{i+1})\ldots\delta A(x_n)
\right]_{A=\overline{A};\,a = 0} 
$
for $n=3$ and $n=4$, respectively:
\begin{eqnarray}
& \;\;&
\;\;\;\;-i g \;\widehat{V}^{abc}_{\lambda\mu\nu}(x,y,z)
\;\;=\;\;
-i g \;V^{\;\;abc}_{0\;\lambda\mu\nu}(x,y,z)
\;\;+\;\; O(g^3)
\nonumber \\
&\;\;&
-g^2\; \widehat{W}^{abcd}_{\lambda\mu\nu\sigma}(x,y,z,w)
\;\;=\;\;
-g^2\;W^{\;\;abcd}_{0\;\lambda\mu\nu\sigma}(x,y,z,w)
\;\;+\;\; O(g^4)
\;,
\label{fullvertices}
\end{eqnarray}
which, to lowest order in the coupling constant, reduce to the {\it bare} 
3- and 4-gluon vertices $V_{0}$ and $W_{0}$, respectively:
\begin{eqnarray}
V_{0\;\lambda\mu\nu}^{\;\;abc}(x,y,z)
& =&
f^{abc} \;
\left\{
\frac{}{}
g_{\lambda\mu} (\partial_y - \partial_x)_\nu \,\delta^4_P(x,z) \delta^4_P(y,z)
\;+\;
g_{\mu\nu} (\partial_z - \partial_y)_\lambda \,\delta^4_P(y,x) \delta^4_P(z,x)
\right.
\nonumber \\
& &
\left.
\frac{}{}
\;\;\;\;\;\;\;\;\;\;\;
\;\;\;\;\;\;\;\;\;\;\;
\;+\;
g_{\nu\lambda} (\partial_x - \partial_z)_\mu \,\delta^4_P(x,y) \delta^4_P(z,y)
\right\}
\label{3vertex}
\\
W_{0\;\lambda\mu\nu\sigma}^{\;\;abcd}(x,y,z,w)
&=&
-\;
\left\{
\frac{}{}
\left( f^{ace}f^{bde} -  f^{ade}f^{cbe} \right) \, g_{\lambda\mu} g_{\nu\sigma}
\;+\;
\left( f^{abe}f^{cde} -  f^{ade}f^{bce} \right) \, g_{\lambda\nu} g_{\mu\sigma}
\right.
\nonumber\\
& &
\left.
\frac{}{}
\;\;\;\;\;\;\;\;\;\;\;
\;+\;
\left( f^{ace}f^{dbe} -  f^{abe}f^{cde} \right) \, g_{\lambda\sigma} g_{\nu\mu}
\right\}
\;
\delta^4_P(x,y) \delta^4_P(z,w) \delta^4_P(y,z)
\label{4vertex}
\;.
\end{eqnarray}
\end{description}
\bigskip

\begin{figure}
\epsfxsize=400pt
\centerline{ \epsfbox{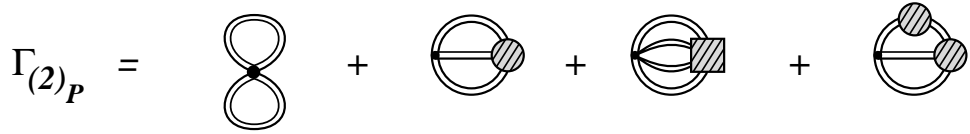} }
\bigskip
\caption{
The 2-loop 
contribution $\Gamma^{(2)}_P$ , eqs. (\ref{Gamma20})-(\ref{Gamma2d}),
to the effective action $\Gamma_P$ of eq. (\ref{Gamma2}),
in the diagrammatic representation of Fig. 3.
Formally, $\Gamma^{(2)}_P$ is the sum of 
all two-particle irreduclible graphs with internal
lines representing the full gluon propagators $\widehat{\Delta}$ 
and full 3- and 4-gluon vertices $\widehat{V}$ and $\widehat{W}$.
\label{fig:fig4}
}
\end{figure}

\bigskip
\bigskip

\subsection{Equations of motion}

As scetched above and discussed in more detail in Appendix E,
the equations of motion (\ref{eom0A}) and (\ref{eom0B}) result
from approximating the exact theory by truncation of the infinite hierarchy of
equations for the $n$-point Green functions
to the 1-point function (the soft mean field $\overline{A}(x)$) and the
2-point function (the hard propagator $i\widehat{\Delta}(x,y)$),
with all higher-point functions being combinations of these and connected
by the 3-gluon and 4-gluon vertices
$-ig\widehat{V}(x,y,z)$ and  $-g^2\widehat{W}(x,y,z,w)$, respectively.
Before writing down the explicit form of the resulting
equations of motion, it is useful to summarize the terminology
introduced in the course of the above discussion:
\begin{itemize}
\item{}
{\it mean field} $\overline{A}$ (Fig. 3f):  denotes
the classical soft field as the expectation value of the gauge field
$A$, which is induced by the abundance of emitted
hard gluons and their collective motion.
\item{}
{\it free} propagator $i\Delta_{0}$ (Fig. 3a):  
refers to the free propagation of hard gluons in the absence
of interactions, i.e. vanishing coupling $g=0$.
\item{}
{\it mean-field} propagator $i\overline{\Delta}$ (Fig. 3b):  
denotes to the tree-level propagator
without quantum corrections, i.e. the free propagator
with an arbitrary number $\ge 0$ of attached external legs coupling
to the soft mean field,
but without closed loops that correspond to quantum self-interactions.
\item{}
{\it full} propagator $i\widehat{\Delta}$ (Fig. 3c):  
terms the dressed  propagator of the hard quanta,
that is renormalized by both the interactions with the soft mean field 
{\it and} the self-interactions among the hard quanta.
\item{}
{\it full} vertex functions $-ig \widehat{V},\; -g^2 \widehat{W}$
(Fig. 3d, 3e): 
represent the 3-gluon and 4-gluon vertices
with the internal lines being the full hard propagator including
mean-field and quantum interactions.
\end{itemize}
\medskip

\subsubsection{Yang-Mills equation for the soft mean field}

The equation of motion for the soft field $\overline{A}_\mu^a(x)$,
is given by  (\ref{eom0A}), i.e.,  
$
\delta \Gamma_P
/
\delta \overline{A}
=
-{\cal K}^{(1)}
- {\cal K}^{(2)} \circ\overline{A}
$,
from which one obtains,
upon taking into account the initial condition (\ref{inicond}),
${\cal K}^{(1)} = 0$, the {\it Yang-Mills equation for} $\overline{A}$:
\begin{equation}
\left[\frac{}{}
\overline{D}^{\lambda,\;ab}  , \; \overline{F}_{\lambda \mu}^{b}
\right](x)
\;=\;
- \,\widehat{j}_{\mu}^{a}(x) 
\;-\; \left(\frac{}{}
{\cal K}_{\;\;\;\mu\lambda}^{(2)\,ab}
\,\circ\,\overline{A}^{\lambda,\,b}\right)(x)
\label{YME2}
\;,
\end{equation}
where $[\overline{D}, \overline{F}] =\overline{D}\, \overline{F} -
\overline{F}\, \overline{D}$
with the covariant derivative defined as
$\overline{D}^\lambda \equiv D^\lambda[\overline{A}] = 
\partial_x^\lambda - ig \overline{A}^\lambda(x)$,
and $\overline{F}_{\lambda \mu}\equiv F_{\lambda \mu}[\overline{A}]
= \left[ \overline{D}_{\lambda}\,,\, \overline{D}_{\mu}\right]/(-ig)$.
The second term on the right side is the
initial state contribution to the current,
according to the condition (\ref{inicond}),
${\cal K}^{(2)}\circ\overline{A}
=
\int_P d^4y \,{\cal K}_{\;\;\;\mu\lambda}^{(2)\,ab}(x,y)
\,\overline{A}^{\lambda,\,b}(y)$.

Rewriting the left hand side of (\ref{YME2}) as
\begin{equation}
\left[\frac{}{}
\overline{D}^{\lambda,\;ab}  , \; \overline{F}_{\lambda \mu}^{b}
\right](x)
\;=\;
{\cal D}_{0\;\mu\lambda}^{-1\;\;ab} \;\overline{A}^{\lambda,\,b}(x)
\;\;+\;\;\overline{\Xi}_{\mu}^{a}(x)
\;\;\;\;\;\;\;\;\;\;\;\;\;\;\;
{\cal D}_{0\;\mu\lambda}^{-1\;\;ab} 
\;\equiv\;\delta^{ab} \,\left( g_{\mu\lambda} \partial_x^{2} - 
\partial^x_\mu\partial^x_\lambda 
- n_\mu n_\lambda \right)
\;,
\label{DF1}
\end{equation}
where, upon taking into account the gauge constraint (\ref{gauge100}),
the $-n_\mu n_\lambda \overline{A}^\lambda$ 
in ${\cal D}_{0\;\mu\lambda}^{-1} \;\overline{A}^{\lambda}$
does not contribute, because
$0=\langle n\cdot A\rangle = n^\nu \overline{A}_\nu$,
eq. (\ref{YME2}) may be expressed in the alternative form (see Fig. 5a):
\begin{equation}
\left\{
\frac{}{}
\left( {\cal D}_{0}^{-1} \;+\;{\cal K}^{(2)}
\,\right) \,\overline{A}\right\}_\mu^a(x)
\;\;+\;\;\overline{\Xi}_{\mu}^{a}(x)
\;\;+\;\;\widehat{j}_{\mu}^{a}(x)
\;\;=\;\;0
\label{YME2a}
\;.
\end{equation}
Here the function 
$\overline{\Xi}$ contains the soft-field self-coupling,
\begin{equation}
\overline{\Xi}_{\mu}^{a}(x)
\;\;=\;\;
\overline{\Xi}_{(1)\;\mu}^{a}(x)\;\;+\;\; \overline{\Xi}_{(2)\;\mu}^{a}(x)
\label{DF2} 
\end{equation}
\begin{eqnarray}
\overline{\Xi}_{(1)\;\mu}^{a}(x)
&= &
\;-\; \frac{g}{2} \,
\int_P\prod_{i=1}^{2} d^4x_i \;
V_{0\;\mu\nu\lambda}^{\;\;\;\;abc}(x,x_1,x_2) 
\;\overline{A}^{\nu,\,b}(x_1) \overline{A}^{\lambda,\,c}(x_2)
\label{DF3} \\
\overline{\Xi}_{(2)\;\mu}^{a}(x)
&= &
\;+\;
\frac{i\,g^2}{6} \,
\int_P\prod_{i=1}^{3} d^4x_i \;
W_{0\;\mu\nu\lambda\sigma}^{\;\;\;\;abcd}(x,x_1,x_2,x_3) 
\;\overline{A}^{\nu,\,b}(x_1) \overline{A}^{\lambda,\,c}(x_2)\overline{A}^{\sigma,\,d}(x_3)
\label{DF4}
\;,
\end{eqnarray}
and the current $\widehat{j}$ is the
{\it induced current} due to the hard quantum dynamics
in the presence of the soft field $\overline{A}$:
\begin{equation}
\widehat{j}_{\mu}^{a}(x) 
\;\;=\;\;
\widehat{j}_{(1)\;\mu}^{a}(x) \;\;+\;\;\widehat{j}_{(2)\;\mu}^{a}(x) 
\;\;+\;\;\widehat{j}_{(3)\;\mu}^{a}(x)
\label{J}
\end{equation}
\begin{eqnarray}
\widehat{j}_{(1)\;\mu}^{a}(x) 
& = &
-\; \frac{i\,g}{2} \,
\int_P\prod_{i=1}^{2} d^4x_i \;
V_{0\;\mu\nu\lambda}^{\;\;\;\;abc}(x,x_1,x_2) 
\;\widehat{\Delta}^{\nu\lambda,\,bc}(x_1,x_2)
\label{J1}
 \\
\widehat{j}_{(2)\;\mu}^{a}(x) 
&= &
-\;
\frac{g^2}{2} \,
\; \int_P\prod_{i=1}^{3} d^4x_i
\;\;W_{0\;\mu\nu\lambda\sigma}^{\;\;\;\;abcd}(x,x_1,x_2,x_3)\;
\overline{A}^{\nu,\,b}(x_1)
\;\widehat{\Delta}^{\lambda\sigma,\,cd}(x_2,x_3)
\label{J2}
\\
\widehat{j}_{(3)\;\mu}^{a}(x) 
&= &
-\;
\frac{i g^3}{6} \,
\; \int_P\prod_{i=1}^{3} d^4x_i d^4y_i
\;\;W_{0\;\mu\nu\lambda\sigma}^{\;\;\;\;abcd}(x,x_1,x_2,x_3)\;
\widehat{\Delta}^{\nu\nu',\,bb'}(x_1,y_1) \;
\widehat{\Delta}^{\lambda\lambda',\,cc'}(x_2,y_2) \;
\nonumber \\
& &
\;\;\;\;\;\;\;\;\;\;\;\;\;\;\;\;\;\;\;\;
\;\;\;\;\;\;\;\;\;\;\;\;\;\;\;\;\;\;\;\;
\times \;
\widehat{\Delta}^{\sigma\sigma',\,dd'}(x_3,y_3) \;
\;V_{0\;\mu'\nu'\lambda'\sigma'}^{\;\;\;\;abcd}(y_1,y_2,y_3)
\label{J3}
\;.
\end{eqnarray}
\smallskip

It should be remarked
that the function $\overline{\Xi}$ on the left hand side of (\ref{YME2})
contains the non-linear self-coupling of the soft field $\overline{A}$
alone, whereas the induced current $\widehat{j}$ on the right hand side
is  determined by the hard propagator $\widehat{\Delta}$,
thereby generating the soft field.
\medskip

\begin{figure}
\epsfxsize=450pt
\centerline{ \epsfbox{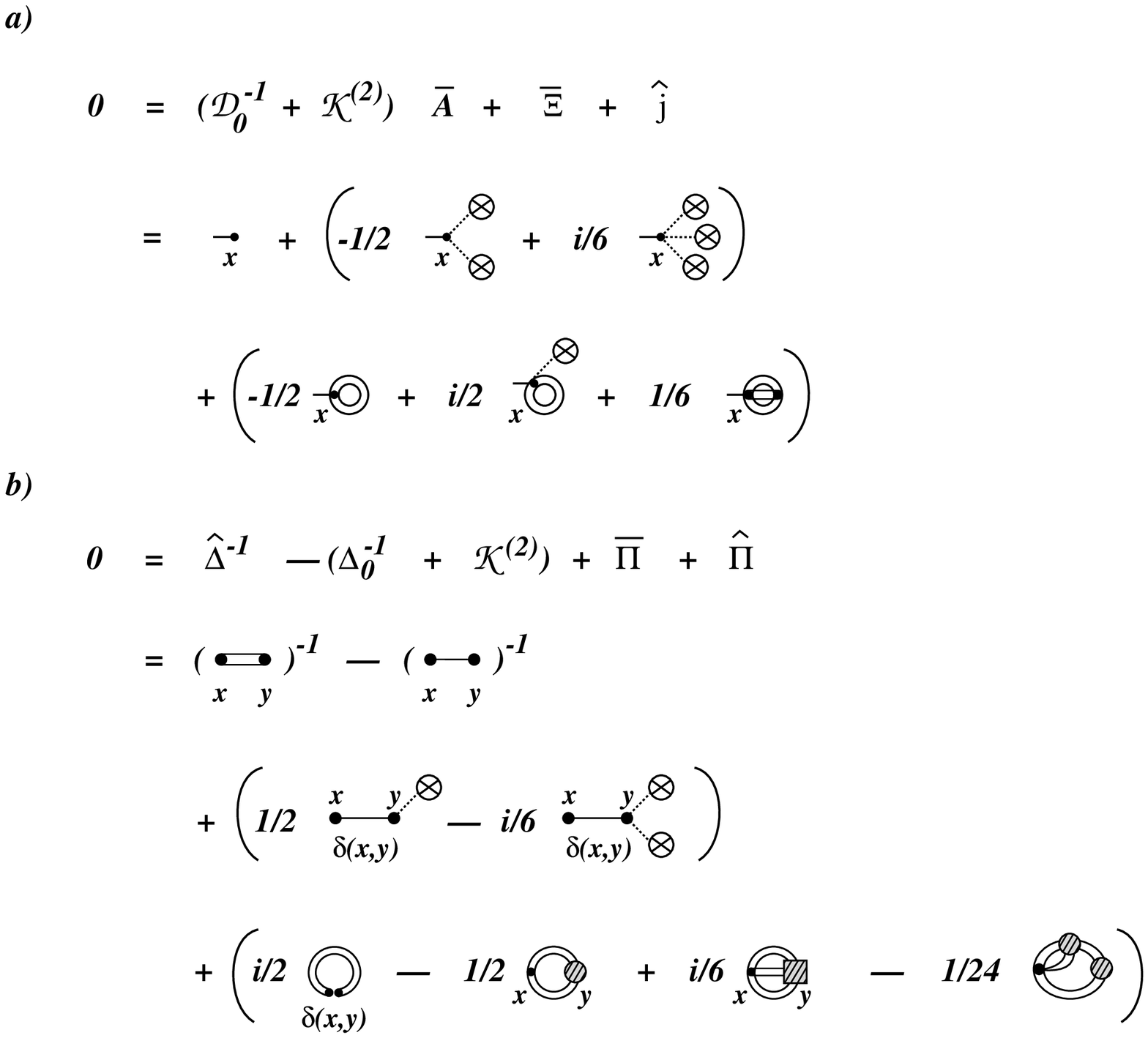} }
\bigskip
\bigskip
\caption{
Diagrammatic representation in terms of the rules of Fig. 3,
of the equations of motion:
{\bf a)}
The Yang-Mills equation (\ref{YME2}), (\ref{DF1}) for the soft field
$\overline{A}$ with the self-coupling  contribution $\overline{\Xi}$, 
eqs. (\ref{DF2})-(\ref{DF4}) 
and the generating hard gluon current $\widehat{j}$,
eqs. (\ref{j1})-(\ref{j3}).
{\bf a)}
The Dyson-Schwinger equation (\ref{DSE2}) for the hard propagator
$\widehat{\Delta}$
with the mean-field polarization tensor $\overline{\Pi}$, eqs.
(\ref{PiMF})-(\ref{PiMF2})
and the quantum contribution $\widehat{\Pi}$, 
eqs. (\ref{PiQU})-(\ref{Pid}).
\label{fig:fig5}
}
\end{figure}

\subsubsection{Dyson-Schwinger equation for the hard gluon propagator}

From the  equation of motion (\ref{eom0B}) for the hard propagator, 
$\widehat{\Delta}_{\mu\nu}^{ab}(x,y)$,  that is, 
$
\delta \Gamma_P
/
\delta  \widehat{\Delta} 
=
{\cal K}^{(2)}/(2i)
$,
one finds after incorporating  condition (\ref{inicond}), ${\cal K}^{(1)}=0$,
the {\it Dyson-Schwinger equation for} $\widehat{\Delta}$
(see Fig. 5b):
\begin{equation}
\left\{\frac{}{}
\;\left(\widehat{\Delta}\right)^{-1} \;\;-\;\; 
\left(\Delta_{0}\right)^{-1} 
\;\;-\;\;
{\cal K}^{(2)}
\;\;+\;\;
\overline{\Pi} \;\;+\;\;\widehat{\Pi} 
\; \right\}_{\mu\nu}^{ab}(x,y)
\;\;=\;\;0
\label{DSE2}
\;,
\end{equation}
where $\widehat{\Delta}$
is the {\it fully dressed propagator} of
the hard quantum fluctuations in the presence of the soft mean field,
defined by  (\ref{DELTA}), 
whereas $\Delta_{0}$ is the {\it free propagator}, given by (\ref{D0}).
The polarization tensor $\Pi$ has been decomposed in two parts,
a mean-field part $\overline{\Pi}$  and a quantum fluctuation part
$\widehat{\Pi}$.
The {\it mean-field polarization tensor} $\overline{\Pi}$ incorporates
the local interaction between the hard quanta and the soft mean field, 
\begin{equation}
\overline{\Pi}_{\mu\nu}^{ab}(x,y) \;\;=\;\;
\overline{\Pi}_{(1)\;\mu\nu}^{\;\;\;\;ab}(x,y) \;\;+\;\;
\overline{\Pi}_{(2)\;\mu\nu}^{\;\;\;\;ab}(x,y)
\label{PiMF}
\end{equation}
\begin{eqnarray}
\overline{\Pi}_{(1)\;\mu\nu}^{\;\;\;\;ab}(x,y)
&=&
\frac{i g}{2}
\;\delta_P^4(x,y)\;
\, \int_P d^4z \,V_{0\;\mu\nu\lambda}^{abc}(x,y,z)
\,\overline{A}^{\lambda ,\,c}(z)
\label{PiMF1}
\\
\overline{\Pi}_{(2)\;\mu\nu}^{\;\;\;\;ab}(x,y)
&=&
\frac{g^2}{6}
\;\delta_P^4(x,y)\;
\, \int_P d^4z d^4 w \,W_{0\;\mu\nu\lambda\sigma}^{abcd}(x,y,z,w)
\,\overline{A}^{\lambda ,\,c}(z) \,\overline{A}^{\sigma ,\,d}(w)
\label{PiMF2}
\;.
\end{eqnarray}
plus terms of order $g^3 \overline{A}^3$ which one may safely ignore
within the  present approximation scheme.
The {\it fluctuation polarization tensor} $\widehat{\Pi}$ contains
the quantum self-interaction among the hard quanta in the presence of 
$\overline{A}$. It is given by the 
variation 
$2i\delta \Gamma_P^{(2)} / \delta\widehat{\Delta}$
of the 2-loop part $\Gamma_P^{(2)}$,
eq. (\ref{Gamma20}), of the effective action $\Gamma_P$,
\begin{equation}
\widehat{\Pi}^{ab}_{\mu\nu}(x,y)
\;\;=\;\;
\widehat{\Pi}^{\;\;\;\;ab}_{(1)\;\mu\nu}(x,y)
\;\;+\;\;
\widehat{\Pi}^{\;\;\;\;ab}_{(2)\;\mu\nu}(x,y)
\;\;+\;\;
\widehat{\Pi}^{\;\;\;\;ab}_{(3)\;\mu\nu}(x,y)
\;\;+\;\;
\widehat{\Pi}^{\;\;\;\;ab}_{(4)\;\mu\nu}(x,y)
\;,
\label{PiQU}
\end{equation}
\begin{eqnarray}
\widehat{\Pi}^{\;\;\;\;ab}_{(1)\;\mu\nu}(x,y)
&=&
-\, \frac{g^2}{2}
\; \int_P d^4x_1 d^4y_1
\;\;W_{0\;\mu\nu\lambda\sigma}^{\;\;\;\;abcd}(x,y,x_1,y_1)\;
\widehat{\Delta}^{\lambda\sigma\,,cd}(y_1,x_1) \;
\\
\widehat{\Pi}^{\;\;\;\;ab}_{(2)\;\mu\nu}(x,y)
&=&
-\,\frac{i\,g^2}{2}
\; \int_P\prod_{i=1}^{2} d^4x_id^4y_i 
\;\;V_{0\;\mu\lambda\sigma}^{\;\;\;\;acd}(x,x_1,x_2)\;\;
\widehat{\Delta}^{\lambda\lambda',\,cc'}(x_1,y_1)\;
\widehat{\Delta}^{\sigma\sigma',\,dd'}(x_2,y_2)\;
\widehat{V}_{\sigma'\lambda'\nu}^{d'c'b}(y_2,y_1,y)
\label{Pib}
\\
\widehat{\Pi}^{\;\;\;\;ab}_{(3)\;\mu\nu}(x,y)
&=&
-\;\frac{g^4}{6}
\; \int_P\prod_{i=1}^{3} d^4x_id^4y_i 
\;\;W_{0\;\mu\lambda\sigma\tau}^{\;\;\;\;acde}(x,x_1,x_2,x_3)\;\;
\widehat{\Delta}^{\lambda\lambda',\,cc'}(x_1,y_1)\;
\widehat{\Delta}^{\sigma\sigma',\,dd'}(x_2,y_2)\;
\nonumber \\
& &
\;\;\;\;\;\;\;\;\;\;\;\;\;\;\;\;\;\;\;\;\;\;\;\;\;
\;\;\;\;\;\;\;\;\;\;\;\;\;\;\;\;\;\;\;\;\;\;\;\;\;
\;\;\;\;\;\;
\times\;
\widehat{\Delta}^{\tau\tau',\,ee'}(x_3,y_3)\;
\widehat{W}_{\tau'\sigma'\lambda'\nu}^{e'd'c'b}(y_3,y_2,y_1,y) 
\label{Pic}
\\
\widehat{\Pi}^{\;\;\;\;ab}_{(4)\;\mu\nu}(x,y)
&=&
-\;\frac{i\,g^4}{24}
\;\int_P\prod_{i=1}^{2} d^4x_id^4y_id^4z_i 
\;\;W_{0\;\mu\lambda\sigma\tau}^{\;\;\;\;acde}(x,x_1,x_2,x_3)\;\;
\widehat{\Delta}^{\sigma\rho',\,df'}(x_2,z_2)\;
\widehat{\Delta}^{\tau\rho'',\,ef''}(x_3,z_3)\;
\nonumber \\
& &
\;\;\;\;\;\;\;\;\;\;\;\;\;\;\;\;\;\;\;\;\;\;\;\;\;
\;\;\;\;\;\;
\times\;
\widehat{V}_{\rho''\rho'\rho}^{f''f'f}(z_3,z_2,z_1) \;\;
\widehat{\Delta}^{\rho\lambda',\,fc'}(z_1,y_1)\;
\widehat{\Delta}^{\lambda\sigma',\,cd'}(x_1,y_2)\;
\;\;\widehat{V}_{\lambda'\sigma'}^{c'd'}(y_1,y_2,y)
\label{Pid}
\;.
\end{eqnarray}
Note that the usual Dyson-Schwinger equation
in {\it vacuum} is contained in (\ref{DSE2}) -(\ref{Pid}) as the special case 
when the mean field vanishes, $\overline{A}(x)= 0$,
and initial state correlations are absent, ${\cal K}^{(2)}(x,y)=0$. In this case,
the propagator becomes the usual vacuum propagator, 
since the mean-field contribution $\overline{\Pi}$ is identically zero, 
and the quantum part $\widehat{\Pi}$ reduces to the vacuum contribution.
\bigskip
\bigskip

\section{TRANSITION TO QUANTUM KINETICS}
\label{sec:section4}
\bigskip

The equations of motion (\ref{YME2}) or (\ref{YME2a}) for
$\overline{F}_{\mu\nu}$ or $\overline{A}_\mu$,  and (\ref{DSE2}) 
for $\widehat{\Delta}_{\mu\nu}$, are 
non-linear integro-differential equations and clearly not solvable in
all their generality.
However, 
the field-equations of motion (\ref{YME2}) or (\ref{DSE2})
can be cast into
much simpler quantum-kinetic equations with the help of 
the  Wigner-function technique and gradient expansion, 
and the assumption of 2-scale separation.
As a result
one obtains finally the three master equations mentionened in Sec. 1:
a simplified Yang-Mills equation decribing the space-time change of 
$\overline{A}$, and two equations for the 
gluon propagator $\widehat{\Delta}$, namely,
first, an {\it evolution equation} for the QCD evolution in momentum space,
and second, a {\it transport equation} for the space-time development in the
presence of $\overline{A}$.
In order to achieve this result, one needs to make a third
key approximation (in addition to the two approximations of Sec. II A),
namely,
\begin{itemize}
\item{}
{\bf approximation 3:}
It is assumed that the induced soft field $\overline{A}_\mu$ is slowly
varying on the scale of the short-range, hard quantum fluctuations,
that is, the gradient of the soft field is small compared to the
Compton wavelength of the hard quanta.
Then one can treat the quantum fluctuations 
of $\widehat{\Delta}(r,k)$ at short distances 
separately from the collective effects represented by to the soft field 
$\overline{A}(r)$ with long wavelength.
\end{itemize}
\medskip

\subsection{Quantum and kinetic space-time regimes}

The  key to derive from (\ref{YME2}) or (\ref{DSE2})
the corresponding approximate quantum-kinetic equations is
the separability of hard and soft dynamics in terms 
of the space-time scale $\lambda\equiv 1/\mu$, where $\mu$ is the 
parametric momentum scale introduced in (\ref{Aa}).
This implies that one 
may characterize the dynamical evolution of the gluon system
by a short-range {\it quantum scale} $r_{qua}\ll \lambda$, and a comparably
long-range {\it kinetic scale} 
$r_{kin}\,\lower3pt\hbox{$\buildrel > \over\sim$}\,\lambda$.
Low-momentum collective excitations
that may develop at the particular momentum scale $g\mu$ are
thus well separated from the typical hard gluon momenta $k_\perp \ge \mu$,
if $g\ll 1$. Therefore, collectivity can arise, because the wavelength of the
soft oscillations $\sim 1/g\mu$ is much larger than the typical
extention of the hard quantum fluctuations $\sim 1/\mu$.
I emphasize that this notion of two characteristic scales
is not just an academic 
construction, but rather is a typical property of quantum field theory.
A simple example is a freely propagating electron:
In this case, the quantum scale is given its the Compton wavelength 
$\sim 1/m_e$ in the restframe of the charge, and measures the size of
the radiative vacuum polarization cloud around the bare charge.
The kinetic scale, on the other hand, is determined by the mean-free-path 
of the charge, which is infinite in vacuum, and in medium is
inversely proportional to the 
local charge density times the interaction cross-section,
$\sim 1/(n_e \,\sigma_{int})$.
Adopting this notion
to the present case of gluon  dynamics, I define
$r_{qua}$ and $r_{kin}$ as follows:

\begin{itemize}
\item{}
The {\bf quantum scale {\boldmath $r_{qua}$}}
measures the spatial extension of quantum fluctuations associated with virtual
and real radiative emission and re-absorption
off a given hard gluon, described by the hard propagator $\widehat{\Delta}$. 
It can thus be interpreted as
its Compton wavelength, corresponding
to the typical transverse extension of the fluctuations and
thus inversely proportional to the average transverse momentum,
\begin{equation}
r_{qua} \;\,\equiv\;\,\widehat{\lambda}
\;\;\simeq\;\;\frac{1}{\langle \;k_\perp\;\rangle}
\;,
\;\;\;\;\;\;\;\;\;\;\;\;\;\;\;
\langle \;k_\perp\;\rangle \;\ge \;\mu
\;,
\label{rqua}
\end{equation}
where the second relation is imposed by means of the definition (\ref{Aa})
of hard and soft modes.
In general, $\widehat{\lambda}$ can be  space-time dependent quantity, because
the magnitude of $\langle k_\perp \rangle$ is determined by both  the
radiative self-interactions of the hard gluons and ther interactions
with the soft field.
\item{}
The {\bf kinetic scale {\boldmath $r_{kin}$}}
measures the range of the long-wavelength correlations, described by
the soft mean-field $\overline{A}$, and may be parametrized in terms of the
average transverse wavelength of soft modes $\langle q_\perp \rangle$, 
such that
\begin{equation}
r_{kin} \;\,\equiv\;\,\overline{\lambda}
\;\;\simeq\;\;\frac{1}{\langle \;q_\perp\;\rangle}
\;,
\;\;\;\;\;\;\;\;\;\;\;\;\;\;\;
\langle \;q_\perp\;\rangle \; \,\lower3pt\hbox{$\buildrel < \over\sim$}\,
\;g\,\mu
\;,
\label{rkin}
\end{equation}
where $\overline{\lambda}$ may vary from one space-time point to
another, because the population of soft modes $\overline{A}(q)$ is determined 
locally by the hard current $\widehat{j}$ with dominant contribution
from gluons with transverse momentum $\simeq \mu$. 
\end{itemize}

The above classification of quantum- (kinetic-)  scales
specifies in space-time the relevant regime for the hard (soft)
dynamics, so that the separability of the two scales 
$r_{qua}$ and  $r_{kin}$
imposes the following condition on the relation between space-time and
momentum: 
\begin{equation}
\widehat{\lambda} \;\;\ll\;\;\overline{\lambda}
\;,\;\;\;\;\;\;\;\;\;\;\;\;\;
\mbox{or}
\;\;\;\;\;\;\;\;\;\;\;\;\;\;
\langle\; k_\perp \;\rangle
\; \;\;\ge
\;\mu\;\;\gg\;\; g\,\mu \;\;
\approx \;\;\;
\langle\; q_\perp \;\rangle
\label{kqcond}
\;.
\end{equation}
The physical interpretation of (\ref{kqcond})
is simple:
At short distances $r_{qua} \ll 1/(g \mu)$ a hard gluon can be considered 
as an {\it incoherent quantum} which emits and partly reabsorbs 
daughter gluons, corresponding to the combination of real
brems-strahlung and virtual radiative fluctuatiuons.
Only a hard probe with a short wavelength $\widehat{\lambda}\le r_{qua}$
can resolve this quantum dynamics.
On the other hand,
at larger distances $r_{kin} \approx 1/(g \mu)$, a gluon appears
as a {\it coherent quasi-particle}, that is, as an extended
object with a changing transverse size corresponding
to the extent of its intrinic quantum fluctuations. This dynamical
substructure is however not resolvable by long-wavelength modes
$\overline{\lambda}\ge r_{kin}$ of the soft field $\overline{A}$.
\medskip

Accordingly, one may classify the quantum and kinetic regimes, respectively,
by associating with two distinct space-time points $x^\mu$ and $y^\mu$ the
following characteristic scales:
\begin{eqnarray}
s^\mu 
\;&\equiv&
\;\;\;\;\;\;
 x^\mu\;-\;y^\mu \;\;
\;\;\;\sim 
\;\widehat{\lambda} \;=\;\frac{1}{g\mu}
\;,\;\;\;\;\;\;\;\;\;\;\;\;\;
\partial_s^\mu 
\;=\;\frac{1}{2}\;\left(\partial_x^\mu\;-\;\partial_y^\mu \right)
\;\;\sim\;\;g\,\mu
\nonumber \\
r^\mu 
\;&\equiv& 
\frac{1}{2}\;\left(x^\mu\;+\;y^\mu\right) 
\;\;\;\;\sim 
\;\overline{\lambda} \;=\;\frac{1}{\mu}
\;,\;\;\;\;\;\;\;\;\;\;\;\;\;\;\;
\partial_r^\mu 
\;=\; \;\;\;\;\partial_x^\mu\;+\;\partial_y^\mu
\;\;\;\;\sim\;\;\;\mu
\label{sr}
\;.
\end{eqnarray}
On the {\it kinetic scale}
the effect of the soft field modes of $\overline{A}$ 
on the hard quanta involves the coupling $g \overline{A}$
to the hard propapgator 
and is of the order of the soft wavelength $\overline{\lambda} = 1/(g\mu)$, 
so that one may characterize the soft field strength by
\begin{equation}
g \overline{A}_\mu(r)  \;\;\sim \;\; g \mu
\;,\;\;\;\;\;\;\;\;\;\;\;\;
g \overline{F}_{\mu\nu}(r)  \;\;\sim \;\;g^2\,\mu^2
\label{AF2}
\;,
\end{equation}
plus corrections of order $g^2\mu^2$ and $g^3\mu^3$, respectively, 
which are assumed to be small.
\noindent

On the {\it quantum scale}, on the other hand,
\begin{equation}
\widehat{\Delta}_{\mu\nu}^{-1}
\;\;\sim k_\perp^2 \;
\,\lower3pt\hbox{$\buildrel > \over\sim$}\,
\;\mu^2
\;\;\gg\;\;g^2\mu^2 \;\; \sim \;\; g\,\overline{F}_{\mu\nu}
\;,
\end{equation}
and one expects that
that the short-distance fluctuations corresponding to
emission and reabsorption of gluons with momenta $k_\perp \ge \mu$,
are little affected by the long-range, soft mean field, because the
color force $\sim g\overline {F}$ acting on a gluon with momentum
$k_\perp \sim \mu$ produces only a very small change  in its momentum.
\smallskip


\subsection{The kinetic approximation}

The realization of the two space-time scales, short-distance quantum and 
quasi-classical kinetic, allows to reformulate the quantum field-theoretical
problem as a relativistic many-body problem within kinetic theory.
The key element is to establish the connection between the preceding
description in terms of  Green functions
and a probabilistic kinetic description in terms of
of so-called Wigner functions \cite{wigner}.
Whereas the 2-point functions, 
such as the propagator or the polarization tensor,
depend on two separate  space-time points $x$ and $y$, 
their Wigner transforms 
utilizes a mixed space-time/momentum representation, which is  
particularly convenient
for implementing  the assumption of separated quantum and
kinetic scales, i.e., that the long-wavelength
field $\overline{A}$ is slowly varying in space-time on the scale
of short-range quantum fluctuations. 
Moreover, the trace of the Wigner-transformed propagator 
is the quantum analogue 
of  the single particle phase-space distribution of gluons, and
therefore provides the basic quantity to make contact with 
kinetic theory of  multi-particle dynamics \cite{lifshitz}.
\smallskip

In terms of the center-of-mass coordinate,
$r = \frac{1}{2}(x+y)$, and relative coordinate $s=  x-y$,
of two space-time points $x$ and $y$, eq. (\ref{sr}),
one can express any 2-point function 
${\cal G}(x,y)$, such as $\widehat{\Delta},\Pi$, in terms of these coordinates,  
\begin{equation}
{\cal G}_{\mu\nu}^{ab}(x,y)\; =\; {\cal G}_{\mu\nu}^{ab}\left(r+\frac{s}{2}, r-\frac{s}{2}\right)
\;\;\equiv\;\; {\cal G}_{\mu\nu}^{ab}\left( r,s\right)
\;,\;\;\;\;\;\;\;\;\;\;\;\;\;
\;.
\end{equation}
The {\it Wigner transform} ${\cal G}(r,k)$
is then defined as the  Fourier transform with respect to the relative 
coordinate $s$, being the canonical conjugate to the momentum $k$.
In general, the necessary preservation of local gauge symmetry
requires a careful definition that obeys the gauge transformation 
properties \cite{BI1}, but for the specific choice
of  gauge (\ref{gauge100}), the Wigner transform is
simply \cite{remark1}: 
\begin{equation}
{\cal G}(r,s) \;=\;
\int \frac{d^4k}{(2\pi)^4} \, e^{-i\,k\,\cdot\, s}\;\, 
{\cal G}\left( r,k\right)
\;\;,\;\;\;\;\;\;\;\;\;\;\;\;\;
{\cal G}\left( r,k\right) \;=\;
\int d^4 s \, e^{i\,k\,\cdot\, s}\;\, {\cal G}\left( r,s\right)
\;.
\label{W}
\end{equation}
The Wigner representation (\ref{W}) will facilitate a systematic identification
of the dominant contributions of the soft field $\overline{A}$ to the hard
propagator $\widehat{\Delta}$, a concept that has been developed
by Blaizot and Iancu \cite{BI1}:
First, one expands both $\overline{A}$ and 
$\widehat{\Delta}= \widehat{\Delta}_{[\overline{0}]}+
\delta\widehat{\Delta}_{[\overline{A}]}$
in terms of gradients of the long-range variation with the kinetic 
scale $r$, and second,
one makes an additional expansion in powers of the soft field $\overline{A}$
and of the induced perturbation
$\delta\widehat{\Delta}_{[\overline{A}]} 
\sim g\widehat{\Delta}_{[\overline{0}]}$.

\subsubsection{Gradient expansion}

To proceed, recall that
the coordinate $r^\mu$ describes
the kinetic space-time dependence $\sim r_{kin}$,
whereas $s$ measures the quantum space-time distance $\sim r_{qua}$.
In translational invariant situations, e.g.
in vacuum or thermal equilibrium,  ${\cal G}(r,s)$ 
in (\ref{W}) is independent of $r^\mu$ and
sharply peaked about $s^\mu =0$. Here the range of the variation is fixed
by $\lambda = 1/\mu$, eq. (\ref{rqua}), corresponding to
the confinement length $const. \times 1/\Lambda$ in the case of vacuum, 
or to the thermal wavelength $const. \times 1/T$ in equilibrium.
On the other hand, in the presence of a slowly varying
soft field $\overline{A}$ with a wavelength  $\overline{\lambda} = 1/(g\mu)$,
eq. (\ref{rkin}), the $s^\mu$ dependence is little affected, 
while the acquired $r^\mu$ dependence 
will have a long-wavelength variation.
In view of the estimates (\ref{sr}), one may
therefore  neglect the derivatives
of ${\cal G}(r,k)$  with respect to 
$r^\mu$ which are of order $g\mu$, relative to those with respect to
$s^\mu$ which are of order $\mu$.

Hence one can perform the so-called 
{\it gradient expansion} of the soft field and the hard propagator and
polarization tensor in terms of gradients $(s\cdot\partial_r)^n$,  
and keep only terms up to first order $n=1$,
i.e.,
\begin{equation}
\overline{A}_\mu(x) 
\;=\;
\overline{A}_\mu\left(r+\frac{s}{2}\right) \;\simeq \;
\overline{A}_\mu(r)+ 
\frac{s}{2}\cdot \partial_r\overline{A}_\mu(r)
\label{gradexp1}
\;,
\end{equation}
and similarly for
$\overline{A}_\mu(y) 
=\overline{A}_\mu(r-s/2)$,
as well as
\begin{eqnarray}
\widehat{\Delta}_{\mu\nu}\left(x,y\right)
&=&
\widehat{\Delta}_{\mu\nu}\left(r,s\right) \;\simeq \;
\widehat{\Delta}_{\mu\nu}\left(0,s\right) \;+ \;
s\,\cdot\, \partial_r\, \widehat{\Delta}_{\mu\nu}\left(r,s\right)
\label{gradexp2}
\\
& &\nonumber \\
\widehat{\Pi}_{\mu\nu}\left(x,y\right)
&=&
\widehat{\Pi}_{\mu\nu}\left(r,s\right) \;\simeq \;
\widehat{\Pi}_{\mu\nu}\left(0,s\right) \;+ \;
s\,\cdot\, \partial_r\, \widehat{\Pi}_{\mu\nu}\left(r,s\right)
\label{gradexp3}
\;,
\end{eqnarray}
Then,
by using  the following conversion rules \cite{ms39,stan} to
carry out the Wigner transformations,
\begin{eqnarray}
\int d^4x' f(x,x') \,g(x',y)
\;\Rightarrow \;
f(r,k) \,g(r,k)
\,&+& \,
\frac{i}{2}\;\left[ (\partial_k f)\cdot(\partial_r g)
\;-\; (\partial_r f)\cdot(\partial_k g) \right]
\label{rules1}
\\
h(x) \,g(x,y)
\;\Rightarrow \;
h(r) \,g(r,k)
\;-\;
\frac{i}{2}\, (\partial_r h)\cdot(\partial_k g)
\;\;\;& & \;\;\;
h(y) \,g(x,y)
\;\Rightarrow\;
h(r) \,g(r,k)
\;+\;
\frac{i}{2}\, (\partial_r h)\cdot(\partial_k g)
\\
\partial_x^\mu f(x,y)
\;\Rightarrow\;
(-i k^\mu \,+\,\frac{1}{2} \partial_r^\mu)  \,f(r,k)
\;\;\;& & \;\;\;
\partial_y^\mu f(x,y)
\;\Rightarrow\;
(+i k^\mu \,+\,\frac{1}{2} \partial_r^\mu)  \,f(r,k)
\label{rules4}
\;,
\end{eqnarray}
the transformed  polarization tensor $\Pi(r,k)$ 
is obtained from $\Pi(x,y)$, eqs. (\ref{PiMF}) and (\ref{PiQU}), with
\begin{equation}
\Pi_{\mu\nu}(r,k) =
\overline{\Pi}_{\mu\nu}(r,k) +
\widehat{\Pi}_{\mu\nu}(r,k)
\end{equation}
\smallskip

\noindent
where the soft {\it mean-field contribution} [c.f. (\ref{PiMF1}), (\ref{PiMF2})]
is
\begin{equation}
\overline{\Pi}_{\mu\nu}^{ab}(r,k)
\;\;=\;\;
\left(\frac{}{}
\overline{\Pi}_{(1)}
\;+\;
\overline{\Pi}_{(2)}
\right)_{\mu\nu}^{ab}(r,k)
\label{RPiMF}
\end{equation}
\begin{eqnarray}
\overline{\Pi}_{(1)\;\mu\nu}^{\;\;\;\;ab}(r,k)
&=&
\frac{i g}{2}
\,V_{0\;\mu\nu\lambda}^{abc}(k,0,-k)
\,\overline{A}^{\lambda ,\,c}(r)
\label{PiMF5}
\\
\overline{\Pi}_{(2)\;\mu\nu}^{\;\;\;\;ab}(r,k)
&=&
\;-\;
\frac{i g}{6}
\,W_{0\;\mu\nu\lambda\sigma}^{abcd}(k,0,0,-k)
\,\overline{A}^{\lambda ,\,c}(r) \,\overline{A}^{\sigma ,\,d}(r)
\label{PiMF6}
\;,
\end{eqnarray}
\smallskip

\noindent
and the {\it quantum contribution} [c.f. (\ref{PiQU})-(\ref{Pid})]
is
\begin{equation}
\widehat{\Pi}_{\mu\nu}^{ab}(r,k)
\;\;=\;\;
\left(\frac{}{}
\widehat{\Pi}_{(1)}
\;+\;
\widehat{\Pi}_{(2)}
\;+\;
\widehat{\Pi}_{(3)}
\;+\;
\widehat{\Pi}_{(4)}
\right)_{\mu\nu}^{ab}(r,k)
\label{RPiQU}
\end{equation}
\begin{eqnarray}
\widehat{\Pi}^{\;\;\;\;ab}_{(1)\;\mu\nu}(r,k)
&=&
+\, \frac{i g^2}{2}
\; \int \frac{d^4q}{(2\pi)^4\;i}
\;\;W_{0\;\mu\nu\lambda\sigma}^{\;\;\;\;abcd}(k,q,-q,-k)\;
\widehat{\Delta}^{\lambda\sigma\,,cd}(r,k) \;
\\
\widehat{\Pi}^{\;\;\;\;ab}_{(2)\;\mu\nu}(r,k)
&=&
+\,\frac{g^2}{2}
\; \int \frac{d^4q}{(2\pi)^4\;i}
\;\;V_{0\;\mu\lambda\sigma}^{\;\;\;\;acd}(k,-q,-q')\;\;
\widehat{\Delta}^{\lambda\lambda',\,cc'}(r,q)\;
\widehat{\Delta}^{\sigma\sigma',\,dd'}(r,q')\;
\widehat{V}_{\sigma'\lambda'\nu}^{d'c'b}(r; q', q, -k)
\label{RPib}
\\
\widehat{\Pi}^{\;\;\;\;ab}_{(3)\;\mu\nu}(r,k)
&=&
-\;\frac{g^4}{6}
\; \int \frac{d^4q}{(2\pi)^4\;i} \frac{d^4p}{(2\pi)^4\;i}
\;\;W_{0\;\mu\lambda\sigma\tau}^{\;\;\;\;acde}(k,-q,-q',-p)\;\;
\widehat{\Delta}^{\lambda\lambda',\,cc'}(r,q)\;
\widehat{\Delta}^{\sigma\sigma',\,dd'}(r,q')\;
\nonumber \\
& &
\;\;\;\;\;\;\;\;\;\;\;\;\;\;\;\;\;\;\;\;\;\;\;\;\;
\;\;\;\;\;\;\;\;\;\;\;\;\;\;\;\;\;\;\;\;\;\;\;\;\;
\;\;\;\;\;\;
\times\;
\widehat{\Delta}^{\tau\tau',\,ee'}(r,p)\;
\widehat{W}_{\tau'\sigma'\lambda'\nu}^{e'd'c'b}(r; q, q', p,-k)
\label{RPic}
\\
\widehat{\Pi}^{\;\;\;\;ab}_{(4)\;\mu\nu}(r,k)
&=&
-\;\frac{g^4}{24}
\; \int \frac{d^4q}{(2\pi)^4\;i} \frac{d^4p}{(2\pi)^4\;i}
\;\;W_{0\;\mu\lambda\sigma\tau}^{\;\;\;\;acde}(k,-q,-q',-p')\;\;
\widehat{\Delta}^{\sigma\rho',\,df'}(r,q)\;
\widehat{\Delta}^{\tau\rho'',\,ef''}(r,q')\;
\nonumber \\
& &
\;\;\;\;\;\;\;\;\;\;\;\;\;\;\;\;\;\;\;\;\;\;\;\;\;
\;\;\;\;\;\;
\times\;
\widehat{V}_{\rho''\rho'\rho}^{f''f'f}(r; q, q', -p) \;\;
\widehat{\Delta}^{\rho\lambda',\,fc'}(r,p)\;
\widehat{\Delta}^{\lambda\sigma',\,cd'}(r,p')\;
\;\;\widehat{V}_{\lambda'\sigma'}^{c'd'}(r; p, p', -k)
\label{RPid}
\;.
\end{eqnarray}
Here
the 3- and 4-gluon vertex functions from
(\ref{fullvertices}), (\ref{3vertex}) and (\ref{4vertex})
depend explicitly on $r$, 
\begin{equation}
\widehat{V}(r;k_i) = V_{0}(k_i)+ O\left(g^4 f(r,k_i)\right)
\;\;\;\;\;\;\;\;\;\;\;\;\;\;\;\;
\widehat{W}(r;k_i) = W_{0}(k_i)+ O\left(g^4 f(r,k_i)\right)
\label{kfullvertices}
\end{equation}
with the {\it bare} point-like
vertices $V_{0}$, $W_{0}$ being $r$-independent and given by
\begin{eqnarray}
V_{0\;\lambda\mu\nu}^{abc}(k_1,k_2,k_3)
& =&
-\,i\,f^{abc} \,
\left\{
\frac{}{}
g_{\lambda\mu} (k_1 - k_2)_\nu \;+\;
g_{\mu\nu} (k_2 - k_3)_\lambda \;+\;
g_{\nu\lambda} (k_3 - k_2)_\mu \right\}
\label{3kvertex}
\\
W_{0\;\lambda\mu\nu\sigma}^{abcd}(k_1,k_2,k_3,k_4)
& =&
-\;
\left\{
\frac{}{}
\left( f^{ace}f^{bde} -  f^{ade}f^{cbe} \right) \, g_{\lambda\mu} g_{\nu\sigma}
\;+\;
\left( f^{abe}f^{cde} -  f^{ade}f^{bce} \right) \, g_{\lambda\nu} g_{\mu\sigma}
\right.
\nonumber \\
& &
\left.
\frac{}{}
\;\;\;\;\;\;\;\;\;\;\;\;\;\;\;\;\;\;\;\;\;\;
\;+\;
\left( f^{ace}f^{dbe} -  f^{abe}f^{cde} \right) \, g_{\lambda\sigma} g_{\nu\mu}
\right\}
\label{4kvertex}
\;.
\end{eqnarray}
\smallskip

With the above formulae, one can now convert both the 
Yang-Mills equation (\ref{YME2}) and the 
Dyson-Schwinger equation (\ref{DSE2}) into a set of much simpler equations.
For the  Dyson-Schwinger equation,
the Wigner transformation together with the gradient expansion
yields 
{\it two} distinct equations 
for the hard propagator $\widehat{\Delta}^{\mu\nu} (r,k)$, 
namely,  (i) an {\it evolution equation}
\footnote{
In Ref. \cite{ms39} the `evolution' equation was termed 
`renormalization' equation,
a term that may be misleading. In order to avoid
confusion with the `renormalization group' equation, the name evolution equation
appears more suitable.},
and (ii) a {\it transport equation}.
They are obtained \cite{ms39,stan} by taking the sum and difference 
of Wigner-transform of (\ref{DSE2}) and its adjoint, using the rules 
(\ref{rules1})-(\ref{rules4}),

\noindent
(i) {\it evolution equation}:
\begin{eqnarray}
(ii) & &
\;\;\;
\left(
k^2 -\frac{1}{4} \partial_r^2 
\right)
\;
\widehat{\Delta}^{\mu\nu} (r,k) 
\;\;-\;
\frac{1}{2} \;
\left\{\overline{\Pi}^\mu_\sigma\,,\, \widehat{\Delta}^{\sigma\nu} \right\} 
(r,k)
\;+\; 
\frac{i}{4}\;
\left[ \partial^\lambda_r \overline{\Pi}^\mu_{\sigma} \, , 
\partial_\lambda^k \widehat{\Delta}^{\sigma\nu}\right]
(r,k)
\nonumber \\
& & \;\;\;\;\;\;\;\;\;\;\;\;\;\;\;\;\;\;\;\;
\;\;=\;\;
\,d^{\mu\nu}(k)\,\,\hat 1_P
\;+\;
\frac{1}{2} \;
\left\{\widehat{\Pi}^\mu_\sigma\,,\, \widehat{\Delta}^{\sigma\nu} \right\} 
(r,k)
\;+\; 
\frac{i}{4}\;
\left[ \partial^\lambda_k \widehat{\Pi}^\mu_{\sigma} \, , 
\partial_\lambda^r \widehat{\Delta}^{\sigma\nu}\right]
(r,k)
\;-\;
\frac{i}{4}\;
\left[ \partial^\lambda_r \widehat{\Pi}^\mu_{\sigma} \, , 
\partial_\lambda^k \widehat{\Delta}^{\sigma\nu}\right]
(r,k)
\label{R}
\end{eqnarray}

\noindent
(ii)  {\it transport equation}: 
\begin{eqnarray}
& &
\;\;\;\;
\;\;
\left(
k\cdot\partial_r \right)
\;\;  \widehat{\Delta}^{\mu\nu} (r,k) 
\;\;
+\;\frac{i}{2} \;\left[\overline{\Pi}^\mu_\sigma\,,\, 
\widehat{\Delta}^{\sigma\nu} \right] 
(r,k)
\;-\;\frac{1}{4}\;
\left\{ \partial^\lambda_r \overline{\Pi}^\mu_{\sigma} \, , 
\partial_\lambda^k \widehat{\Delta}^{\sigma\nu}\right\}
(r,k)
\nonumber \\
& & \;\;\;\;\;\;\;\;\;\;\;\;\;\;\;\;\;\;\;\;
\;\;=\;\;
-\;\frac{i}{2} \;\left[\widehat{\Pi}^\mu_\sigma\,,\, 
\widehat{\Delta}^{\sigma\nu} \right] 
(r,k)
\;+\;\frac{1}{4}\;
\left\{ \partial^\lambda_k \widehat{\Pi}^\mu_{\sigma} \, , 
\partial_\lambda^r \widehat{\Delta}^{\sigma\nu}\right\}
(r,k)
\;-\;
\frac{1}{4}\;
\left\{ \partial^\lambda_r \widehat{\Pi}^\mu_{\sigma} \, , 
\partial_\lambda^k \widehat{\Delta}^{\sigma\nu}\right\}
(r,k)
\label{T}
\;,
\end{eqnarray}
where
$\partial_r^2 \equiv \partial_r\cdot\partial_r$, 
$[A,B] \equiv AB-BA$, $\{A,B\}\equiv AB+BA$.
In (\ref{R}) and (\ref{T}),  
$\hat{1}_P = 1\;(0)$ for $\widehat{\Delta}^F, \widehat{\Delta}^{\overline F}$ 
($\widehat{\Delta}^>, \widehat{\Delta}^<$) 
arises as the transform of 
$\delta^4_P(x,y)$.
The function
$d_{\mu\nu}(k)$ is the  sum over the  gluon polarizations $s$,
(emerging from the Fourier transform of the operator (\ref{Box})),
\begin{equation}
d_{\mu\nu}(k)\;=\;\sum_{s=1,2}\,\varepsilon_\mu(k,s) \cdot \varepsilon_{\nu}^\ast(k,s)
\;=\;
g_{\mu\nu} - \frac{n_\mu k_\nu+n_\nu k_\mu}{n\cdot k}
\;+\; (n^2 \,+\,\alpha\,k^2) \, \frac{k_\mu k_\nu}{(n\cdot k)^2}
\label{dmunu}
\;.
\end{equation}
with the properties
$ d_\mu^\mu(k)= 2 $ 
\footnote{
This property reflects that in the non-covariant gauges (\ref{gauge1a})
only the two physical 
polarization states propagate, i.e. those with with $\varepsilon_\mu k^\mu =0$. 
For comparison, in the covariant Feynman 
gauge, $d^{\mu\nu}=g^{\mu\nu}$, $d_\mu^\mu=4$, and
$k_\mu d^{\mu\nu}=k^\nu \ne 0$.},
$k_\mu d^{\mu\nu}(k) \stackrel{k^2\rightarrow 0}{\longrightarrow}0
$
and
$n_\mu d^{\mu\nu}= 0 =d^{\mu\nu} n_\nu$.
Furthermore, the initial state contribution
${\cal K}^{(2)}$ appearing in (\ref{YME2}) and (\ref{DSE2}),
which contributes only at $r^0 = t_0$, has been absorbed 
into the hard propagator,
\begin{equation}
\widehat{\Delta}_{\mu\nu}^{-1} (r,k) \;\;\equiv\;\;
\widehat{\Delta}_{\mu\nu}^{-1} (r,k) \;\;-\;\;
{\cal K}^{(2)}_{\mu\nu}(r,k) \;\delta(r^0 - t_0)
\;.
\end{equation}

\noindent
For the  Yang-Mills equation (\ref{YME2}), determining 
$\overline{F}^{\mu\nu} (r)$, 
one obtains on the same level of approximation
a compact expression in terms of the hard current $\widehat{j}$
\footnote{
Note that in the kinetic approximation,
the  piece $\widehat{j}_{(3)}$, eq. (\ref{J3}),
does not contribute, because it has
two additional $\widehat{\Delta}$ insertions and is down by a factor
$g/\mu^4$ as compared to $\widehat{j}_{(1)}$ and $\widehat{j}_{(2)}$. 
}:
\begin{eqnarray}
 & &
\;\;\;
\left[\frac{}{}
\overline{D}^{\lambda,\;ab}  , \; \overline{F}_{\lambda \mu}^{b}
\right](r)
\;\;=\;\;
-\;\widehat{j}_\mu^a(r) 
\;\;=\;\;
 -\,g
\,\gamma^{\mu\nu\lambda\sigma} 
\;\int \frac{d^4k}{(2\pi)^2}
\;
\mbox{Tr}\left\{
\;
T^a\,\left(
k_\lambda
\,\widehat{\Delta}_{\nu\sigma}(r,k)
\;+\;
\frac{i}{2}
\left[
\overline{D}_\lambda^r , \widehat{\Delta}_{\nu\sigma}(r,k) 
\right]
\right)
\right\}
\label{YME3}
\;,
\end{eqnarray}
where
$
\gamma_{\mu\nu\lambda\sigma} \;=\;
2\,g_{\mu\nu} \,g_{\lambda\sigma}
\;-\;
g_{\mu\lambda} \,g_{\nu\sigma}
\;-\;
g_{\mu\sigma} \,g_{\nu\lambda}
$
and
$ \overline{D}_\lambda^r  = \partial_\lambda^r -ig \overline{A}_\lambda(r)$.
\smallskip

\subsubsection{Expansion in powers of $g \overline{A}$}

In order to isolate the leading effects of the soft mean field $\overline{A}$
on the hard quantum propagator $\widehat{\Delta}$, 
I follow Ref. \cite{BI1} to separate
the quantum contribution from the mean field contribution
on the basis of the assumption  that the $\overline{A}$ field is
slowly varying on the short-range scale of the quantum fluctuations. 
To do so, recall  eq. (\ref{DELTA}),
\begin{equation}
\widehat{\Delta} (r,k)\;\;\equiv\;\;
\widehat{\Delta}_{[\overline{0}]}(r,k) \;\,+\;\, 
\delta\widehat{\Delta}_{[\overline{A}]}(r,k)
\label{Dsep}
\end{equation}
with the  quantum piece 
$\widehat{\Delta}_{[\overline{0}]}$
and the mean-field part
$\delta\widehat{\Delta}_{[\overline{A}]}$ defined by
\begin{equation}
\widehat{\Delta}_{[\overline{0}]}^{\;-1} 
\;=\; \left. \widehat{\Delta}^{-1} \right|_{\overline{A}=0}
\;=\;
\Delta_{0}^{-1} \;-\; \left.\widehat{\Pi} \right|_{\overline{A}=0}
\;\;\;\;\;\;\;\;\;\;\;\;\;\;\;\;\;\;
\delta\widehat{\Delta}_{[\overline{A}]}^{-1} \;=\;
\overline{\Delta}^{\,-1} \;-\; \Delta_{0}^{-1} 
\;=\; -\; \overline{\Pi}
\label{DA}
\;,
\end{equation}
and the free-field  propagator ${\Delta}_{0}$ and the
mean-field proagator $\overline{\Delta}$ are given by
(\ref{D0}) and (\ref{DMF}), respectively.
Given the ansatz (\ref{Dsep}), with the
feedback of the induced soft field to the hard propagator
being  contained in $\delta\widehat{\Delta}_{[\overline{A}]}$,
the latter is now  be expanded 
in powers of the soft field coupling $g \overline{A}$, and 
it is anticipated that the mean-field induced part 
$\delta\widehat{\Delta}_{[\overline{A}]}$,
is a correction being {\it at most $g$ times} 
the quantum  piece $\widehat{\Delta}_{[\overline{0}]}$,
that is,
\begin{equation}
\delta\widehat{\Delta}_{[\overline{A}]}(r,k) \;=\;
\sum_{n=1,\infty} \frac{1}{n!}\,\left( g\overline{A}(r)\cdot\partial_k\right)^n
\widehat{\Delta}_{[\overline{0}]}(k)
\;\;\simeq\;\;
 g\overline{A}(r)\cdot\partial_k
\widehat{\Delta}_{[\overline{0}]} (r,k)
\label{DA2}
\;,
\end{equation} 
and, to the same order of approximation, 
\begin{equation}
\partial_r^\mu\delta\widehat{\Delta}_{[\overline{A}]\,\mu\nu}(r,k)
\;\simeq\;
g(\partial_r^\mu\overline{A}^\lambda)\partial_k^\lambda 
\widehat{\Delta}_{[\overline{0}]\;\mu\nu}(r,k)
\;,
\end{equation}
where, on the right side, the space-time derivative acts only on $\overline{A}$.
Now
the decomposition (\ref{Dsep}) with the approximation (\ref{DA2})
is inserted into eqs. (\ref{R}), (\ref{T}), (\ref{YME3}),
and  all terms up to order 
$g^2\mu^2 \widehat{\Delta}_{[\overline{0}]}$ are kept.
The resulting equations can be compactly expressed in terms
of the {\it kinetic} momentum $K_\mu$ rather than the {\it canonical} 
momentum $k_\mu$ (as always in the context of interactions with a
gauge field \cite{jackson}), which for the 
class of axial-type gauges (\ref{gauge100}) amounts to the 
replacements 
\begin{equation}
k_\mu \;\longrightarrow \;K_\mu \;=\; k_\mu \;-\;g \overline{A}_\mu(r)
\;\;,\;\;\;\;\;\;\;\;\;\;
\partial^r_\mu \;\longrightarrow \;
\overline{D}_\mu^r \;=\; \partial^r_\mu \;-\; 
g \partial^r_\mu \overline{A}^\nu(r)\,\partial_\nu^k
\;.
\label{kincan}
\end{equation}
Taking into account the approximation 3 of Sec. III A
implying $K^2\widehat{\Delta} \gg  \overline{D}_r^2 \widehat{\Delta}$, 
one finds for the 
the  evolution-, transport-, and  Yang-Mills equation,
eqs. (\ref{R}), (\ref{T}), and (\ref{YME3}),
respectively,
\begin{eqnarray}
\left\{
\; K^2\, 
,
\; \widehat{\Delta}^{\mu\nu}_{[\overline{0}]}
\right\}
(r,K)
&=&
\,d^{\mu\nu}(K)\,\,
\;+\;
\frac{1}{2} \;
\left\{\widehat{\Pi}^\mu_\sigma\,,\, 
\widehat{\Delta}^{\sigma\nu}_{[\overline{0}]} \right\}(r,K)
\label{R2}
\\ & & \nonumber
\\
\left[
K\cdot\overline{D}_r , \;  \widehat{\Delta}^{\mu\nu}
\;\right]
(r,K)
&=&
-\;\frac{i}{2} \;\left[\overline{\Pi}^\mu_{\sigma}\,,\, 
\widehat{\Delta}^{\sigma\nu}_{[\overline{0}]} \right] (r,K)
\;-\;
\frac{i}{2} \;\left[\widehat{\Pi}^\mu_\sigma\,,\, 
\widehat{\Delta}^{\sigma\nu}_{[\overline{0}]} \right] (r,K) 
\label{T2}
\\ & & \nonumber
\\
\left[\frac{}{}
\overline{D}_r^{\lambda}  , \; \overline{F}_{\lambda \mu}
\right](r)
&=&
-\;\widehat{j}_\mu(r) 
\;\;=\;\;
- g
\;\int \frac{d^4K}{(2\pi)^2}
\;
\mbox{Tr}\left\{
\;
- K_\mu \, \widehat{\Delta}_\nu^{\;\;\;\;\nu}(r,K)
\;+\;
\widehat{\Delta}_\mu^{\;\;\;\;\nu}(r,K) \,K_\nu
\right\}
\label{YME5}
\;,
\end{eqnarray}
where the color indices are suppressed, noting that 
$\widehat{\Delta}_{\mu\nu}^{ab} =\widehat{\Delta}_{\mu\nu}$,
$\overline{F}_{\lambda \mu} =T^a \overline{F}_{\lambda \mu}^a$,
$\widehat{j}_\mu = T^a \widehat{j}_\mu^a$,
and $\mbox{Tr}[ \ldots ] = \mbox{Tr}T^a[ \ldots ]$.
\bigskip

\subsubsection{The physical representation}

One sees that the original
Dyson-Schwinger equation (\ref{DSE2}) reduces in the kinetic approximation to
the  set of algebraic equations
(\ref{R2}) and (\ref{T2}).
Now recall  (c.f. Appendix B3) that
in the CTP framework these equations
are still  $2\times 2$ matrix equations which mix the four different
components of  $\widehat{\Delta} = (\widehat{\Delta}^F,\widehat{\Delta}^>,
\widehat{\Delta}^<,\widehat{\Delta}^{\overline{F}})$ and of
$\widehat{\Pi} = (\widehat{\Pi}^F,\widehat{\Pi}^>,
\widehat{\Pi}^<,\widehat{\Pi}^{\overline{F}})$.
For the following it is more convenient to
employ instead 
an equivalent  set of independent functions, namely,
the {\it retarded} and {\it advanced functions} $\widehat{\Delta}^{ret}$,
$\widehat{\Delta}^{adv}$, plus 
the {\it correlation function} $\widehat{\Delta}^{cor}$,
and analogously for $\widehat{\Pi}$.
This latter set is more directly connected with 
physical, observable quantities, and is commonly referred to as 
{\it physical representation} \cite{chou}:
\begin{equation}
\widehat{\Delta}^{ret} \;=\;   \widehat{\Delta}^F \;-\; \widehat{\Delta}^<
\;\;\;\;\;\;\;\;\;\;\;\;\;\;\;
\widehat{\Delta}^{adv} \;=\; \widehat{\Delta}^F \;-\; \widehat{\Delta}^> 
\;\;\;\;\;\;\;\;\;\;\;\;\;\;\;
\widehat{\Delta}^{cor} \;=\; \widehat{\Delta}^< \;+\; \widehat{\Delta}^> 
\label{retadv1}
\end{equation}
Similarly, for the polarization tensor the retarded, advanced
and correlation functions are defined as 
(note the subtle difference to (\ref{retadv1})):
\begin{equation}
\widehat{\Pi}^{ret} \;=\;   \widehat{\Pi}^F \;+\; \widehat{\Pi}^<
\;\;\;\;\;\;\;\;\;\;\;\;\;\;\;
\widehat{\Pi}^{adv} \;=\; \widehat{\Pi}^F \;+\; \widehat{\Pi}^>
\;\;\;\;\;\;\;\;\;\;\;\;\;\;\;
\widehat{\Pi}^{cor} \;=\; -\left(\widehat{\Pi}^> \;-\; \widehat{\Pi}^<\right) 
\label{retadv2}
\end{equation}
Loosely speaking, the retarded and advanced functions characterize
the intrinsic quantum nature of 
a `dressed' gluon, describing its substructural state of 
emitted and reabsorbed gluons, whereas the correlation function describes
the kinetic correlations among different such `dressed' gluons.
The great advantage \cite {chou,rammer} of this  physical representation is that 
in general the dependence 
on the phase-space occupation of gluon states (the local density) 
is essentially carried by the correlation functions $\widehat{\Delta}^{>}$,
$\widehat{\Delta}^<$, 
whereas the dependence of the retarded and advanced functions, 
$\widehat{\Delta}^{ret}$, $\widehat{\Delta}^{adv}$,
on the local density is weak.
More precisely,
the retarded and advanced propagators
and the imaginary parts of the self-energies embody the
renormalization effects and dissipative quantum dynamics that is associated
with short-distance emission and absorption of quantum fluctuations,
whereas the correlation function contains both the effect of interactions
with the soft mean  field and of statistical binary scatterings among the hard
gluons.
\medskip

In going over to the physical representation,  one finds  then that
eqs. (\ref{R2}) and (\ref{T2}) give  a set of `self-contained'
equations for the retarded and advanced functions alone,
\begin{eqnarray}
\left\{
\;K^2\,, 
\; \widehat{\Delta}^\ra 
\right\}_{\mu\nu}
&=&
d^{\mu\nu}
\;+\;
\frac{1}{2} \;\left(\frac{}{}\Pi^\ra\, \widehat{\Delta}^\ra\;+\; 
\widehat{\Delta}^\ra\, \Pi^\ra\right)_{\mu\nu} 
\label{AGRA}
\\
\left[
K\cdot \overline{D}_r  \, ,\,  \widehat{\Delta}^\ra
\right]_{\mu\nu}
&=&
-\;\frac{i}{2} \;\left(\frac{}{}\Pi^\ra\, \widehat{\Delta}^\ra\;-\; 
\widehat{\Delta}^\ra\, \Pi^\ra\right)_{\mu\nu}
\label{BGRA}
\;,
\end{eqnarray}
plus a set of `mixed' equations for the correlation functions,
\begin{eqnarray}
\left\{
\;K^2\,, 
\; \widehat{\Delta}^\gl
\right\}_{\mu\nu} 
&=&
-\frac{1}{2} \;\left(\frac{}{}\Pi^{\gl}\, \widehat{\Delta}^{adv}\;+\; 
\Pi^{ret}\,\widehat{\Delta}^{\gl}
\;+\; \widehat{\Delta}^\gl\, \Pi^{adv}\;+\; 
\widehat{\Delta}^{ret}\,\Pi^\gl\right)_{\mu\nu}
\label{AGC}
\\
\left[
K\cdot \overline{D}_r  \, ,\,  \widehat{\Delta}^\gl
\right]_{\mu\nu}
&=&
-\;\frac{i}{2} \;\left(\frac{}{}\Pi^{\gl}\, \widehat{\Delta}^{adv}\;+\; 
\Pi^{ret}\,\widehat{\Delta}^{\gl}
\;-\; \widehat{\Delta}^\gl\, \Pi^{adv}\;-\; 
\widehat{\Delta}^{ret}\,\Pi^\gl\right)_{\mu\nu}
\label{BGC}
\end{eqnarray}
\smallskip

The equations (\ref{AGRA})-(\ref{BGC})
may be further manipulated by the following trick.
Let the imaginary and real components of the 
retarded and advanced propagators be denoted by 
\begin{equation}
\widehat{\rho}_{\mu\nu} \;\equiv\; 2\,\mbox{Im} \widehat{\Delta}_{\mu\nu}\;=\; 
i \, \left( \widehat{\Delta}^{ret}-\widehat{\Delta}^{adv}\right)_{\mu\nu}
\;\;\;\;\;\;\;\;\;\;\;\;
\mbox{Re} \widehat{\Delta}_{\mu\nu}\;=\; 
\frac{1}{2}\, \left( \widehat{\Delta}^{ret}+\widehat{\Delta}^{adv}\right)_{\mu\nu}
\;.
\label{rhoxi}
\end{equation}
with 
$\widehat{\Delta}^{ret}=(\widehat{\Delta}^{adv})^\ast$ and
$\theta(K^0)\,\widehat{\Delta}^{ret}=\theta(-k^0)\,\widehat{\Delta}^{adv}$.
The  analogous decomposition 
of the polarization tensor  
in terms  
of its real and imaginary components
defines the quantum part 
$\widehat{\Pi}$
as the sum and difference of the retarded
and advanced contributions, respectively,
\begin{equation}
\widehat{\Gamma}_{\mu\nu} \;\equiv\;
2\,\mbox{Im} \widehat{\Pi}_{\mu\nu}\;=\; 
i \, \left( \widehat{\Pi}^{ret}-\widehat{\Pi}^{adv}\right)_{\mu\nu}
\;\;\;\;\;\;\;\;\;\;\;\;
\mbox{Re} \widehat{\Pi}_{\mu\nu}\;=\; 
\frac{1}{2} \,\left( \widehat{\Pi}^{ret}+\widehat{\Pi}^{adv}\right)_{\mu\nu}
\label{ImRePi}
\;,
\end{equation}
and similarly for the mean-field part $\overline{\Pi}$, associated with 
presence of  soft field.
The imaginary parts $\widehat{\rho}$
and $\widehat{\Gamma}$
are the  {\it spectral density} and {\it spectral width}, respectively,
of the hard gluons.
\smallskip

In terms of this representation one obtains from eqs. 
(\ref{T2})-(\ref{YME5}) and (\ref{AGRA})-(\ref{BGC})
the following  final set of {\it master equations}:
\begin{eqnarray}
\left\{
\;K^2\,, 
\; \widehat{\rho}
\right\}_{\mu\nu}
&=&
\left\{\mbox{Re}\widehat{\Pi}\, , \,  \widehat{\rho}
\right\}_{\mu\nu}
\;+\;
\left\{ \widehat{\Gamma}\, , \,  \mbox{Re}\widehat{\Delta}_{[\overline{0}]}
\right\}_{\mu\nu}
\;+\; 
 g \left(
\overline{F}_\mu^\lambda\,\widehat{\rho}_{\lambda\nu}
+
\widehat{\rho}_{\mu}^{\;\lambda}\, \overline{F}_{\lambda\nu}
\right)
\label{X1}
\\
&& \nonumber \\
\left[
K\cdot \overline{D}_r  \, ,\,  \widehat{\Delta}^{cor}
\right]_{\mu\nu}
&=&
+ \,i
\left[ \widehat{\Pi}^{cor}\, , \,  \mbox{Re} \widehat{\Delta}_{[\overline{0}]}
\right]_{\mu\nu}
\;+\;
i \left[ \mbox{Re}\widehat{\Pi}
\, , \,  \widehat{\Delta}^{cor}_{[\overline{0}]}
\right]_{\mu\nu}
\;-\;
\frac{1}{2} 
\left\{ \widehat{\Pi}^{cor}\, , \,  \widehat{\rho}
\right\}_{\mu\nu}
\;-\;
\frac{1}{2} 
\left\{ \widehat{\Gamma}
\, , \,  \widehat{\Delta}^{cor}_{[\overline{0}]}
\right\}_{\mu\nu}
\;,
\nonumber \\
& &
\;-\;  g \,K_\lambda \overline{F}^{\lambda\sigma} \partial^K_{\sigma}
\widehat{\Delta}^{cor}_{[\overline{0}]\;\mu\nu}
\;-\;  g \left(
\overline{F}_\mu^\lambda\,\widehat{\Delta}^{cor}_{[\overline{0}]\;\lambda\nu}
-
\widehat{\Delta}_{[\overline{0}]\;\mu}^{cor\;\lambda}\, 
\overline{F}_{\lambda\nu}
\right)
\;,
\label{X2}
\\
&& \nonumber \\
\left[\frac{}{}
\overline{D}_r^{\lambda}  , \; \overline{F}_{\lambda \mu}
\right]
&=&
-\;\widehat{j}_\mu
\;\;=\;\;
- g
\;\int \frac{d^4k}{(2\pi)^2}
\;
\mbox{Tr}\left\{
\;
\left(
- K_\mu \,\widehat{\Delta}_{\;\nu}^{cor\;\nu}
\;+\;
\widehat{\Delta}_{\;\mu}^{cor\;\nu} \,K_\nu
\right) 
\right\}
\label{YME6}
\;.
\end{eqnarray}
\noindent
The {\it physical significance} of the (\ref{X1}) and (\ref{X2}) is
the following \cite{ms39}:
Eq. (\ref{X1}) determines, in terms of the spectral density $\widehat{\rho}$,
the state of a single gluon with
respect to its virtual fluctuations and real emission (absorption) processes,
corresponding to the real and imaginary parts of the retarded and advanced
polarization tensor in the presence of the soft field $\overline{F}$.
Eq. (\ref{X2}), on the other hand characterizes, in terms of the
correlation function $\widehat{\Delta}^{cor}$, the correlations
among different such gluon states. 
The polarization tensor appears here in  distinct ways.
The first two terms on the right hand side account for scatterings between 
the single-gluon states.
The next two terms incorporate the renormalization effects 
which result from the fact that the gluons between collisions do not behave as
free particles, but change their dynamical structure due to virtual
fluctuations, as well as real emission and absorption of quanta.
The last two terms account for the soft interaction with the mean
field $\overline{F}$.
Eq. (\ref{YME6}) finally determines the rate of
change of he soft field $\overline{F}$ by the
hard gluon current, which involves the full
correlation function $\widehat{\Delta}^{cor}$.
\smallskip

The  interlinked structure of
eqs. (\ref{X1})-(\ref{YME6}) is 
 very convenient for explicit calculations
(demonstrated in Sec. 4).
It  provides a systematic solution scheme,
as discussed below, to solve for the three quantities of interest,
namely, the spectral density $\widehat{\rho}$, the correlation function
$\widehat{\Delta}^{cor}$, and the mean field $\overline{A}$.
In view of (\ref{X1})-(\ref{YME6}) the natural logic 
is a stepwise determination of 
$\widehat{\rho}\rightarrow
\widehat{\Delta}^{cor}_{[\overline{0}]}\rightarrow
\delta\widehat{\Delta}^{cor}_{[\overline{A}]}\rightarrow
\widehat{\Delta}^{cor}\rightarrow
\widehat{j}\rightarrow \overline{F}$.
\bigskip

\subsection{General solution scheme}

Let me exemplify the above interpretation of (\ref{X1}) and
(\ref{X2}) in more quantitative detail
(see also Refs. \cite{chou,rammer}). 
The {\it formal solution} of (\ref{X1}) for the retarded and advanced functions is
\cite{chou},
\begin{equation}
\widehat{\Delta}^{ret}_{\mu\nu}\;=\;
{\Delta}^{ret}_{0\;\mu\nu}\;+\;
\left( {\Delta}^{ret}_{0}\;{\Pi}^{ret}\;
\widehat{\Delta}^{ret}\right)_{\mu\nu}
\;\;\;\;\;\;\;\;\;\;
\widehat{\Delta}^{adv}_{\mu\nu}\;=\;
{\Delta}^{adv}_{0\;\mu\nu}\;+\;
\left( {\Delta}^{adv}_{0}\;{\Pi}^{adv}\;
\widehat{\Delta}^{adv}\right)_{\mu\nu}
\label{gensol1}
\;,
\end{equation}
where $\Pi^\ra = \widehat{\Pi}^\ra + \overline{\Pi}^\ra$.
This determines 
$\widehat{\rho}_{\mu\nu}$ via (\ref{rhoxi}).
Once  $\widehat{\Delta}^\ra$ is known, the solution of (\ref{X2}) for the
correlation function is given by \cite{chou},
\begin{equation}
\widehat{\Delta}^{cor}_{\mu\nu}\;=\;
- \left(
\widehat{\Delta}^{ret}\; \left.{\Delta}^{cor}_{0}\right.^{-1}\;
\widehat{\Delta}^{adv}\right)_{\mu\nu}
\;+\;
\left(
\widehat{\Delta}^{ret}\; \widehat{\Pi}^{cor}\;
\widehat{\Delta}^{adv}\right)_{\mu\nu}
\label{gensol2}
\;,
\end{equation}
with $\Pi^{cor} = \widehat{\Pi}^{cor} + \overline{\Pi}^{cor}$.
It has the general form \cite{rammer} 
\begin{equation}
\widehat{\Delta}^{cor}_{\mu\nu}(r,K)\;=\;
-i\,\widehat{\rho}_{\mu\nu}(r,K)\; {\tt G}(r,K)
\label{gensol3}
\;,
\end{equation}
i.e., the convolution of the spectral density $\widehat{\rho}_{\mu\nu}$
with the phase-space density of hard gluons 
${\tt G}$,
\begin{equation}
{\tt G}(r,K) \;=\; 1\;+\; 2 {\tt g}(r,K)
\label{Gdist}
\end{equation}
where the 1 comes from the vacuum contribution of a single gluon state,
and the $2{\tt g}$ represents the correlations with other hard gluons
that are close by in phase-space. Note that the function ${\tt g}$ is constrained to be
a real and even function in $K$ (c.f. Appendix F).
From (\ref{gensol3}) it follows that  
the total number of gluons ${\tt N}$ in a space-time element $d^4r$
is
\begin{equation}
{\tt N}(r) \;\equiv\;
\frac{d{\tt N}}{d^4 r} \;=\;
\int\frac{d^4K}{(2\pi)^4} \;\mbox{Tr}\left[
d_{\mu\nu}^{-1}(K)\,i\widehat{\Delta}^{cor}_{\mu\nu}(r,K)\right]
\;=\;
\int\frac{d^4K}{(2\pi)^4} \; \widehat{\rho}(r,K)\;{\tt G}(r,K)
\;,
\label{gensol4}
\end{equation} 
where 
$d_{\mu\nu}(K)$ is the polarization sum given by (\ref{dmunu}),
$d_{\mu\nu}^{-1} = 2 d_{\mu\nu}$, and
$\widehat{\rho} = \frac{1}{2} d_{\mu\nu}\widehat{\rho}_{\mu\nu}$
and an averaging over the transverse polarizations and the 
color degrees of freedom is understood.
\medskip

The above formulae become immediately familiar, 
when considering for illustration
the simplest case of a non-interacting system of gluons, the
{\it free-field} case.
In this case, $\Pi = 0$ and one finds, utilizing the formulae of Appendix F,
for the  free retarded and advanced functions:
\begin{equation}
{\Delta}_{0\;\mu\nu}^{ret}(K)\; =\;
\frac{d_{\mu\nu}(K)}{K^2 + i\epsilon}
\;\;\;\;\;\;\;\;\;\;\;
{\Delta}_{0\;\mu\nu}^{adv}(K)\; =\;
\frac{d_{\mu\nu}(K)}{K^2 - i\epsilon}
\label{freesol1}
\;.
\end{equation}
Hence, the free-field spectral density $\rho_0$ which is the difference between
${\Delta}_{0}^{ret}$ and ${\Delta}_{0}^{adv}$, is 
on-shell, 
\begin{equation}
-i\,{\rho}_{0\;\mu\nu}\; =\;
{\Delta}_{0\;\mu\nu}^{ret} - {\Delta}_{0\;\mu\nu}^{adv} 
\;=\;
2\pi \delta(K^2) d_{\mu\nu}(K)
\label{freesol2}
\;,
\end{equation}
by means of the principal-value formula 
$(K^2\pm i\epsilon)^{-1} = \mbox{PV}(1/K^2)\mp i\pi\delta(K^2)$.
The  free-field correlation function 
${\Delta}^{cor}_{0}$ is then readily determined via (\ref{gensol3}),
\begin{equation}
{\Delta}^{cor}_{0\;\mu\nu}(r,K)\;=\;
-2\pi i\,\delta(K^2)\; {\tt G}_0(r,K)
\;\,d_{\mu\nu}(K)
\label{freesol3}
\;,
\end{equation}
and so, with ${\tt G}_0 = 1 + 2 {\tt g}_0$, the number of 
on-shell gluons per $d^4r$ is
\begin{equation}
{\tt N}_0(r) \;=\;
\int\frac{d^3K}{(2\pi)^3\,2 K^0} \;
{\tt G}_0(r,\vec{K})
\;\;,\;\;\;\;\;\;\;\;\;
{\tt G}_0(r,\vec{K}) \;=\; \left.{\tt G}_0(r,K)\right|_{K^0=|\vec{K} |}
\;=\;
(2\pi)^3 \,2K^0 \frac{d{\tt N}_0}{d^3K}
\label{freesol4}
\end{equation} 
\medskip

The free-field exercise, eqs. (\ref{freesol1})-(\ref{freesol4}),
illustrates the two main properties, which hold also for the general
interacting case, eqs. (\ref{gensol1})-(\ref{gensol4}):
\begin{description}
\item[(i)]
The spectral density $\widehat{\rho}_{\mu\nu}(r,K)$
describes the `dressing' of a {\it single} gluon state with momentum $K$
with respect to its radiative quantum fluctuations, i.e., 
its fluctuating coat of emitted and reabsorbed gluons.
The function $\mbox{Tr}[d_{\mu\nu}^{-1}\widehat{\rho}^{\mu\nu}]$ is
the {\it intrinsic} gluon distribution, that is,  the number of gluons
inside this gluon state. The spectral density is 
a property of the state itself and therefore is nonvanishing even
in vacuum, in the absence of a medium.
For on-shell particles, $\widehat{\rho}_{\mu\nu} \propto \delta(K^2)$
and therefore there are no intrinsic gluons present.
\item[(ii)]
The correlation function 
$\widehat{\Delta}^{cor}_{\mu\nu}(r,K)$ describes an interacting
{\it ensemble} of such fluctuating gluon states,
and is given by  the number density ${\tt G}(r,K)$ of 
those gluons weighted
with their spectral density $\widehat{\rho}_{\mu\nu}$,
containing the intrinsic gluon density of each of them.
For the non-interacting case, it obviously reduces to an ensemble
of on-shell particles with $K^0 = |\vec{K}|$.
\end{description}
\medskip

In closure of this Section, 
a generic {\it solution scheme} may be 
the following iteration recipe (which is exemplified in the next Section):
\begin{description}
\item[1.]
Solve the evolution equation (\ref{X1}) for $\widehat{\Delta}^\ra$
and the associated spectral density 
$\widehat{\rho}$ at starting point $t=t_0$ 
with specified initial condition $\widehat{\rho}(t_0) = \rho_0$
at a large initial momentum or energy scale $Q$.
This can be done just as in free space, except that the 
kinetic momentum $K = k-g\overline{A}$
carries now an implicit dependence on the soft field $\overline{A}$
with specified initial value $\overline{A}(t_0)$.
\item[2.]
Solve the transport equation (\ref{X2}) 
for the correlation function
$\widehat{\Delta}^{cor}=\widehat{\Delta}^{cor}_{[\overline{0}]}+
\delta\widehat{\Delta}^{cor}_{[\overline{A}]}$.
This involves, a) the construction of 
$\widehat{\Delta}^{cor}_{[\overline{0}]}$ with the help of
$\widehat{\rho}$ and $\widehat{\Delta}^\ra$ from step 1, and b)
the calculation of the mean-field induced correction 
$\delta\widehat{\Delta}^{cor}_{[\overline{0}]}$ from
the right side of (\ref{X2}).
The resulting space-time evolution of $\widehat{\Delta}^{cor}$ 
describes then the evolution of the gluon density ${\tt G}$
within a time interval between $t_0$ and 
$t_1\sim 1/\langle K_{\perp 1} \rangle$,
corresponding to the evolution  from
$Q$ down to a mean $K_{\perp 1}$ at $t_1$.
\item[3.]
Insert the solution for the full correlation function
$\widehat{\Delta}^{cor}$ into the current $\widehat{j}_\mu$ 
on the left side of 
(\ref{YME6}) and integrate over all momenta $K$ from
the initial momentum scale $Q$
down to the hard-soft scale $\mu$.
This gives the current induced by the  motion of the
total aggregat of hard gluons during the evolution
between $t_0$ and $t_1$.
Then solve the Yang-Mills equation (\ref{YME6}) 
to determine the  
soft field $\overline{A}$ (equivalently
$\overline{F}$) that is generated at $t_1$ 
as a result of the hard gluon evolution.
\item[4.]
Return to 1., and  proceed with second iteration,
replacing  $\overline{A}(t_0)$  by $\overline{A}(t_1)$, and so forth.
\end{description}
\bigskip

\section{Sample calculation:
hard gluon evolution  with self-generated soft field}

\medskip

This Section is devoted to exemplify the practical applicability of the 
developed formalism by following the solution scheme of Sec. III C 
for the specific physics scenario advocated in the introduction and
schematically illustrated in Fig. 1.
I consider a high-energy beam current of hard gluons as it evolves
in space-time and momentum space, and eventually induces its soft mean field.
\smallskip

\subsection{The physics scenario}

\begin{description}
\item[(i)]
The initial state is modeled as an ensemble of a number ${\cal N}_0$ 
of uncorrelated hard gluons. 
The Lorentz-frame of reference is the one where the gluons move
with the speed of light in $+z$-direction.
The initial gluon beam is prepared at 
\begin{equation}
r_0^\mu \;=\; (t_0,\vec{r}_{\perp 0}, z_0)
\;,\;\;\;\;\;\;\;\;\;
t_0\;=\; z_0 \;=\; 0
\;,\;\;\;\;\;\;\;\;\;
0\;\le\;\, r_{\perp 0}\;\,\le\;\, R
\;,
\label{ps1}
\end{equation}
corresponding at $t_0$ to a sheet located with longitudinal position
$z_0$ with transverse extent up to a maximum $R$, specified later.
\item[(ii)]
The initial hard gluons are imagined to be produced at some very large 
momentum scale $Q^2 \gg\Lambda^2$, with their energies and longitudinal
momentum along the $z$-axis being  $\simeq Q$. 
These gluons are therefore strongly concentrated
around the lightcone  with  momenta
\begin{equation}
k^0 \;\simeq \;k^3 \approx\; Q
\;,
\;\;\;\;\;\;\;\;\;
\frac{k_{\perp}^2}{Q^2} \;\approx \; 0
\;,
\label{ps2}
\end{equation}
and hence have very small spatial extent $\Delta r \sim 1/Q$.
That is, the initial state gluons are taken as
{\it bare} quanta without any radiation field around them.
\item[(iii)]
The  subsequent time-like evolution of these bare gluons proceeds then by two
competing processes: a) the regeneration of the radiation field
by emission and re-absorpton of virtual quanta, and b) the brems-strahlung
emission of real gluonic off-spring.
As a consequence, phase-space will be populated with progressing time
by more and more gluons. The typical energies decrease, whereas the
average transverse momentum increases
(c.f. Fig. 1a), but yet within the hard momentum range
\begin{equation}
Q^2 \;\gg\; k_{\perp}^2 \;\ge \mu^2 \;\,\gg\;\,\Lambda^2
\label{ps3}
\;.
\end{equation}
Eventually, the evolving gluon system reaches the  point
at which the transverse momenta become of the order of the energies.
This point is defined to be characterized by the scale $\mu$ - 
the transition from
hard, perturbative to soft, non-perturbative regimes.
When $K_\perp^2 \,\lower3pt\hbox{$\buildrel < \over\sim$}\,\mu^2$,
the individual gluons cannot be resolved anymore,
and their coherent color current acts as the source
of the soft mean field.
\item[(iv)]
Because of the restricted kinematic region (\ref{ps3}) of
the hard gluon dynamics,  the coupling $\alpha_s = g^2/4\pi$ satisfies
\begin{equation}
\alpha_s(k_\perp^2) \;\ll\;1 
\;\;\;\;\;\;\;\;\;
\alpha_s(k_\perp^2) \ln\left(Q^2/k_\perp^2\right) \;\simeq\;1 
\;\;\;\;\;\;\;\;\;
\mbox{for all} \;\; k_\perp^2 \;\ge\;  \mu^2
\label{ps4}
\;,
\end{equation}
so that a perturbative evaluation of the hard gluon interactions is applicable,
provided $\mu  \,\lower3pt\hbox{$\buildrel > \over\sim$}\, 1$ GeV.
The perturbative analysis in the following subsections will be restricted 
to leading order:
the hard gluon interactions then includes  {\it only radiative self-interactions}
$\sim g^2$, but {\it no gluon-gluon scatterings} $\sim g^4$
\footnote{
Aside from $g^4 \ll g^2$, the neglect of scatterings 
is reasonable here, because for a beam of almost colinearly moving
gluons, the scattering cross-section $\sigma_{g_1g_2} \propto
\alpha_s\,|v_1-v_2| \,M_{g_1,g_2}(k_1,k_2)$ is negligible
for vanishing flux $|v_1-v_2|\approx 0$.
},
or other higher-loop contributions.
Hence, for the hard gluon propagator
$\widehat{\Delta} = \widehat{\Delta}_{[\overline{0}]} +
\delta\widehat{\Delta}_{[\overline{A}]}$ of (\ref{Dsep}), (\ref{DA}),
the required accuracy for the quantum contribution 
$\widehat{\Delta}_{[\overline{0}]}$ is,
\begin{equation}
\widehat{\Delta}_{[\overline{0}]} \;=\;
\Delta_0 \;+\; \;\widehat{C}(g^2;\Delta_0)
\;\;\;+\;\;\;O(g^4)
\label{ps5}
\;,
\end{equation}
where $\Delta_0$ is the free-field solution.
On top of this 
the interaction of the hard gluons
with the soft field is treated as a correction 
$\delta\widehat{\Delta}_{[\overline{A}]}$ as in (\ref{DA2}),
to leading order $\sim g\overline{A}$ to the solution 
$\widehat{\Delta}_{[\overline{0}]}$ of (\ref{ps5}),
\begin{equation}
\delta\widehat{\Delta}_{[\overline{A}]} \;=\;
\overline{C}(g\overline{A};\Delta_{[\overline{0}]})
\;\;\;+\;\;\;O(g^2\overline{A}^2)
\label{ps6}
\;.
\end{equation}
\end{description}
\medskip

Although this so defined physics scenario,
with an  initial state of 
bare gluons, being only stastistically correlated and incoherent,
may appear to be rather academic, it has 
in fact valuable physical relevance:
For example, it  may be viewed as
the idealized version of the initial density of materialized gluons
in the very early stage of a high-energy collision of two heavy nuclei.
In this example,
one expects the materialization of a large number ${\cal N}_0$ of
virtual gluons in the wavefunctions of the colliding nuclei, to occur
very shortly after the nuclear overlap by means of hard scatterings.
If one imagines  the time of nuclear overlap equal to $t_0 =0$, and assume
the average momentum transfer of initial hard scatterings $\approx Q^2$,
then the above idealistic scenario acquires a more realistic meaning.
\bigskip

\subsection{Choice of lightcone gauge and kinematics}
\medskip

For purpose of calculational convenience,
I will henceforth work in the {\it lightcone gauge} which is
a special case of the axial-type gauges (\ref{gauge100}). It is
defined by (\ref{gauge1})-(\ref{gauge1a}) of Appendix C, that is,
\begin{equation}
n\cdot{\cal A}^a \;=\; 0
\;,\;\;\;\;\;\;\; n^2 \;=\; 0
\;\;\;\;\;\;\;\;\;\;
({\cal A}_\mu^a = \overline{A}_\mu^a, a_\mu^a)
\;,
\label{gauge10}
\end{equation}
corresponding to the gauge fixing term in (\ref{Ieff}), 
\begin{equation}
I_{GF}\left[n\cdot{\cal A}\right]
\;=\;
\int_P d^4x \;\left( 
- \frac{1}{2\alpha}
\left[n\cdot {\cal A}^a(x)\right]^2
\right)
\;,\;\;\;\;\;\;\;\;
\mbox{with} \;\alpha \;\longrightarrow \;0
\;.
\label{gauge11}
\end{equation}
I choose the lightlike vector $n^\mu$ parallel to the direction
of motion of the gluon beam along the forward lightcone,
\begin{equation}
n^\mu\;=\; \left(n^0,\vec{n}_\perp, n^3\right) \;=\;  
(1,\vec{0}_\perp, -1)
\end{equation}
and employ {\it lightcone variables},
i.e., for any four-vector $v^\mu$,
\begin{eqnarray}
& &
v^\mu\;=\; \left( v^+, v^-, \vec{v}_\perp \right)
\;\;\;\;\;\;\;\;
v^2 \;=\; v^+v^- \;-\;v_\perp^2
\\
& &
v^\pm \;=\; v_\mp \;=\; v^0\pm v^3
\;\;\;\;\;\;\;
\vec{v}_\perp \;=\; \left(v^1,v^2\right)
\;\;\;\;\;\;\;
v_\perp \;=\; \sqrt{\vec{v}_\perp^{\;2}}
\\
& &
v_\mu w^\mu \;=\; \frac{1}{2}\,
\left( v^+ w^- + v^- w^+ \right) \;-\;\vec{v}_\perp\cdot\vec{w}_\perp
\label{lcv1}
\;.
\end{eqnarray}
Then
$n^\mu = \left(n^+, n^-, \vec{n}_\perp\right) =
(0, 1, \vec{0}_\perp)$, so that the gauge constraint (\ref{gauge10})
reads
\begin{equation}
n\cdot {\cal A} \;=\; {\cal A}^+ \;=\; {\cal A}_- \;=\; 0
\;,
\end{equation}
and the non-vanishing components of the gauge-field tensor 
$\overline{F}^{\mu\nu} = -\overline{F}^{\nu\mu}$ are
\begin{eqnarray}
& &\;\;\;\;\;\;\;
\overline{F}^{+-} \;=\; -\partial^+ \overline{A}^-
\;\;\;\;\;\;\;
\overline{F}^{+\,i} \;=\; \partial^+ \overline{A}^i
\nonumber \\
& &\;\;\;\;\;\;\;
\overline{F}^{-\,i} \;=\; \partial^- \overline{A}^i
-\partial^i \overline{A}^-
-ig \left[\overline{A}^-,\overline{A}^i\right]
\nonumber \\
& &\;\;\;\;\;\;\;
\overline{F}^{ij} \;=\; \partial^i \overline{A}^j
-\partial^j \overline{A}^i
-ig \left[\overline{A}^i,\overline{A}^j\right]
\;,
\label{FmunuLC}
\end{eqnarray}
where 
$\partial^\pm = \partial/\partial r^\pm$ and 
the index $ i = 1,2$ labels the transverse components.
\smallskip

Finally,
the kinematic imposition (\ref{ps2}) reads in terms
of lightcone variables 
\begin{equation}
K^+ K^- \;=\; K^2 \;+\; K_\perp^2 \;\;\ll\;\; \left(K^+\right)^2
\end{equation}
\begin{equation}
K^+ \;\simeq \; 2 K^0 \;\simeq\; 2 K^3
\;\;\;\;\;\;\;\;
K^- \;\simeq \; 0
\;.
\;\;\;\;\;\;\;\;
K_\perp^2\,\gg\, K^2 
\label{K+K-}
\end{equation}
Physically this implies that the hard gluons are
effectively on mass shell, i.e., their actual virtuality (degree of off-shellness)
$K^2$ is  small compared to $K_\perp^2$, the transverse momentum squared,
and negligibly small compared to the scale $(K^+)^2$.
Within this kinematic regime, I henceforth consider $K^2/(K^+)^2 \rightarrow 0$.
\bigskip

\subsection{Properties of $\widehat{\Delta}_{\mu\nu}$ and $\Pi_{\mu\nu}$ in
lightcone representation}
\medskip

The most general Lorentz decomposition of the polarization
tensor $\Pi= \widehat{\Pi}+ \overline{\Pi}$ in lightcone 
gauge can be written as
$\Pi^{ab}_{\mu\nu} (r,K) = \delta^{ab} \Pi_{\mu\nu} (r,K)$, with
\begin{equation}
\Pi_{\mu\nu} (r,K) \;=\;
\left(g_{\mu\nu} -\frac{K_\mu K_\nu}{K^2}\right) \,\Pi_\perp
\;+\;
\left(\frac{K_\mu K_\nu}{K^2}\right) \,\Pi_\parallel
\;+\;
\left(\frac{n_\mu K_\nu + K_\mu n_\nu}{n\cdot K}\right) \,\Pi_1
\;+\;
\left(\frac{K^2 n_\mu n_\nu}{(n\cdot K)^2}\right) \,\Pi_2
\label{decomp1}
\end{equation}
where $\Pi_\perp, \Pi_\parallel, \Pi_1, \Pi_2$ are scalar functions of dimension
mass squared and depend on the four-vectors $r^\mu$ and 
$K^\mu= k^\mu-g\overline{A}^\mu$.  In lightcone gauge, 
the Ward identity for the gluon propagator \cite{gaugereview}
\begin{equation}
\lim_{\alpha \rightarrow 0} 
\left\{
\frac{1}{\alpha} \,(n\cdot K) \,n^\mu \widehat{\Delta}_{\mu\nu} \;+\;
\frac{1}{(2\pi)^4} \,K_\nu \right\} \;\stackrel{!}{=} \;0
\end{equation}
enforces $\Pi_{\mu\nu}$ to be transverse with respect to $n_\mu$ and 
symmetric in its arguments and indices, 
\begin{equation}
n^\mu \Pi^{ab}_{\mu\nu}  \;=\; 0 \;=\; \Pi^{ab}_{\mu\nu} n^\nu
\;\;\;\;\;\;\;\;\;\;\;\;
\Pi^{ab}_{\mu\nu}  \;=\; \Pi^{ba}_{\nu\mu}
\;,
\end{equation}
which implies that
\begin{equation}
\Pi_\parallel  \;=\; - \Pi_1\;=\; + \Pi_2
\;.
\end{equation}
Therefore, with $n\cdot K = K^+$,
\begin{equation}
\Pi_{\mu\nu}^{ab} (r,K) \;=\;
\delta^{ab}\;
\left(g_{\mu\nu} -\frac{K_\mu K_\nu}{K^2}\right) \,\Pi_\perp
\;+\;
\delta^{ab}\;
\left(\frac{K_\mu K_\nu}{K^2}
+
\frac{n_\mu K_\nu + K_\mu n_\nu}{K^+}
+
\frac{K^2 n_\mu n_\nu}{(K^+)^2}\right) \,\Pi_\parallel
\label{decomp2}
\end{equation}
\begin{equation}
\Pi_\perp \;=\;
\frac{1}{2} \left(g_{\mu\nu} +\frac{K^2}{(K^+)^2} \,n_\mu n_\nu\right) \,\Pi^{\mu\nu}
\;\;\;\;\;\;\;\;\;\;
\Pi_\parallel \;=\;
3 \,\Pi_\perp \;-\; \Pi_\mu^\mu
\;.
\label{Piperp}
\end{equation}
The corresponding full gluon propagator is given by the inverse of
$(\widehat{\Delta})^{-1} =(\Delta_0)^{-1} - \Pi$. Using
the free-field form
\begin{equation}
\Delta_{0\;\mu\nu}^{ab} (r,K) \;=\;
\delta^{ab}\,d_{\mu\nu}(K)\; \Delta_0 (r,K)
\;,\;\;\;\;\;\;\;\;\;\;\;\;\;\;\;
d_{\mu\nu}(K)\;=\;
g_{\mu\nu} - \frac{n_\mu K_\nu + K_\mu n_\nu}{K^+} + 
\frac{K^2 n_\mu n_\nu}{(K^+)^2}
\;,
\label{decomp3a}
\end{equation}
with the scalar functions 
$\Delta_0\equiv(\Delta_0^{ret}, \Delta_0^{adv}, \Delta_0^{cor})$
 [c.f. eqs.(\ref{freesol1})-(\ref{freesol3})],
\begin{equation}
\Delta_0^\ra (K) \;=\;\frac{1}{K^2\pm i\epsilon}
\;,\;\;\;\;\;\;\;\;\;\;\;\;\;\;\;\; 
\Delta_0^{cor} (r,K) \;=\;- 2\pi\,i \delta(K^2)\;{\tt G}_0(r,K)
\;,
\label{freesol10}
\end{equation}
one finds
\begin{equation}
\widehat{\Delta}_{\mu\nu}^{ab} (r,K)\;=\;
\delta^{ab}\;
\Delta_0 (r,K)
\;\left(\frac{1}{1 - \Pi_\perp/K^2}\right)
\;
\left\{g_{\mu\nu} - \frac{n_\mu K_\nu + K_\mu n_\nu}{K^+} + 
\frac{K^2 n_\mu n_\nu}{(K^+)^2}
\;\left(\frac{\Pi_\parallel/K^2}{1 - (\Pi_\perp-\Pi_\parallel)/K^2}\right)
\right\}
\label{decomp3}
\;.
\end{equation}
\smallskip
Now, 
because of (\ref{K+K-}), 
the last term in (\ref{decomp3}) vanishes
for   $K^2/(K^+)^2 \rightarrow 0$, and the 
full propagator $\widehat{\Delta}$ can be expressed as
the free-field counterparts $\Delta_0$ times a scalar {\it formfactor function}
${\cal Z}$ whose momentum dependence contains only the 
Lorentz invariants $n\cdot K = K^+$ and $K_\perp^2$:
\begin{equation} 
\widehat{\Delta}_{\mu\nu}^{ab}(r,K) \;=\;
\delta^{ab} \,d_{\mu\nu}(K) \;\Delta_0 (r,K) \;\;{\cal Z}(r,K^+, K_\perp^2)
\label{decomp4}
\;,
\end{equation} 
where, because of $K^2/(K^+)^2 \rightarrow 0$, the function $d_{\mu\nu}$  reduces
now to
\begin{equation}
d_{\mu\nu}(K) \;=\;
g_{\mu\nu} - \frac{n_\mu K_\nu + K_\mu n_\nu}{K^+}
\end{equation}
and the formfactor ${\cal Z}$ is related to the polarization tensor by
\begin{equation}
{\cal Z}(r,K^+,K_\perp^2) \;=\; \frac{1}{1\,-\,\Pi_\perp/K^2}
\label{Zdef1}
\;,
\end{equation}
with boundary condition
\begin{equation}
\left. {\cal Z}(0, K^+,K_\perp^2)\right|_{K=Q} \;\;=\;\; 1
\label{Zdef2}
\;.
\end{equation}
Here $Q$ is the renormalization point, determined by the momentum scale of
the initial state hard gluons (which specified in the next subsection).

The great advantage of the lightcone gauge becomes evident now: the
solution of the full retarded, advanced and correlation
functions  (\ref{gensol1})-(\ref{gensol3})
boils down to calculating a single scalar function for each of them,
namely the formactor ${\cal Z}$,
which is simply multiplied to the free-field forms (\ref{freesol1})-(\ref{freesol3}).
For the retarded and advanced functions, 
with 
\begin{equation}
{\cal Z}^\ra\;=\; \frac{1}{1-\Pi_\perp^\ra/K^2}
\label{gensol10}
\end{equation}
one has
\begin{equation}
\widehat{\Delta}_{\mu\nu}^{ret}(K)\; =\;
\frac{d_{\mu\nu}(K)}{K^2 + i\epsilon}
\;{\cal Z}^{ret}(r,K^+,K_\perp^2)
\;\;\;\;\;\;\;\;\;\;\;
\widehat{\Delta}_{0\;\mu\nu}^{adv}(K)\; =\;
\frac{d_{\mu\nu}(K)}{K^2 - i\epsilon}
\;{\cal Z}^{adv}(r,K^+,K_\perp^2)
\label{gensol11}
\;,
\end{equation}
which satisfy the useful relations
$\widehat{\Delta}_{\mu\nu}^{ret}=
(\widehat{\Delta}_{\mu\nu}^{adv})^\ast$ and
$\widehat{\Delta}_{\mu\nu}^{ret}(K^0,\vec{K})=
\widehat{\Delta}_{\mu\nu}^{adv}(-K^0,\vec{K})$.
Defining
\begin{equation}
{\cal Z}^\rho\;\,\equiv\;\, {\cal Z}^{ret}-{\cal Z}^{adv}
\;,
\end{equation}
the spectral density follows immediately as
\begin{equation}
\widehat{\rho}_{\mu\nu}(r,K)\; =\;
i\left(\widehat{\Delta}^{ret} - \widehat{\Delta}^{adv}\right)_{\mu\nu}(r,K)
\;=\;
d_{\mu\nu}(K)\;(-2\pi\,i)\;\, {\cal Z}^\rho(r,K^+,K_\perp^2)
\label{gensol12}
\;,
\end{equation}
and  the correlation function is obtained as
\begin{equation}
\widehat{\Delta}^{cor}_{\mu\nu}(r,K)\;=\;
d_{\mu\nu}(K)\;(-2\pi\,i)\;\, {\cal Z}^\rho(r,K^+,K_\perp^2)
\;\left( 1 + 2 \,{\tt g}(r,K)\right)
\label{gensol13}
\;.
\end{equation}
\bigskip

\subsection{Specifying the initial state}
\medskip

To fix the initial conditions for the scenario described in Sec. VI A,
both $\widehat{\Delta}$ and $\overline{A}$ have to be provided
with initial values at $r^0\equiv t_0 = 0$.
The initial condition for the hard propagator is chosen as
\begin{equation}
\left.
\widehat{\Delta}_{\mu\nu}(r,K)\right|_{r^0=r^3=0} \;\;=\;\;
\Delta_{0\;\;\mu\nu}(0,\vec{r}_\perp,K)
\;,
\end{equation}
referring to a statistical ensemble of 
bare gluon states at time $r^0=0$,
which can be characterized by a single-particle density matrix of
the Gaussian form as given by eq. (\ref{rho}) of Appendix B.
This ansatz corresponds  to an initial state source term 
in (\ref{DSE2}) of the form 
\begin{equation}
\left.
\Delta_{0\;\mu\nu}^{cor}(r,K)\right|_{(r=(0,\vec{r}_\perp,0)} \;=\;
{\cal K}^{(2)}_{\mu\nu}(r,K) \,\delta(r^0)\delta(r^3) \;=\; 
\rho_{0\;\mu\nu}(K)\,{\tt G}_0(r,K)
\label{in1}
\;.
\end{equation}
As assumed in Sec. IV A,
the initial ensemble consists of
a total number ${\cal N}_0$ of bare gluons 
with total invariant mass ${\cal Q}^2$, all moving
with equal fractions of the total  momentum 
${\cal Q}^\mu = Q^\mu/{\cal N}_0$. That is, each gluon
moves initially with momentum 
$Q^\mu = (Q^+,0,0_\perp)$ colinearly to the others along the lightcone.
Throughout the ultra-relativistic limit is understood, 
$"Q^2 \rightarrow \infty"$, i.e. $Q^+ \gg \Lambda$, where 
$\Lambda  \approx 0.2-0.3$ GeV.
The spatial distribution of these ${\cal N}_0$ initial  gluons 
at $r^0  = 0$
is taken as a $\delta$-distribution
along the lightcone
at $r^3 \equiv z_0 = 0$, and a random distribution
transverse to the lightcone motion.
That is, the initial multi-gluon ensemble is prepared at
lightcone position $r^+ = t_0 + z_0 = 0$,
lightcone time $r^- =t_0 - z_0$ with a transverse smearing 
$0 \le r_\perp \le {\cal N}_0/\sqrt{Q^2}$, where the typical
transverse extent of each gluon is $\delta r_\perp \approx 1/Q^2 \ll 1$ fm.
Accordingly, 
the initial state spectral density $\rho_0$ in (\ref{in1}) is
taken as
\begin{equation}
\rho_0(K)
\;=\; \frac{(2\pi)^4}{K^+}
\;\delta\left(K^+-Q^+\right) \;\delta\left(K^-- \frac{K_\perp^2 + Q^2}{K^+}\right)
\;\delta^2\left(\vec{K}_\perp\right)
\;,\;\;\;\;\;\;\;\;\;\;\;\;\;\;\;\;\;
\int\frac{dK^+dK^-d^2K_\perp}{(2\pi)^4} \; \rho_0(K) \;=\;1
\label{in3}
\;.
\end{equation}
The corresponding retarded and advanced functions $\Delta_0^\ra$
are of the form (\ref{freesol1})
\begin{equation}
\Delta_0^\ra(K)
\;=\;		
\mbox{PV}\left(\frac{1}{K^2}\right)\;\,\mp \;\,\frac{i}{2} \,\rho_0(K)
\label{in3a}
\;.
\end{equation}
Finally, the initial state correlation function $\Delta^{cor}_0$
is the convolution of
$\rho_0$ with the  density of bare gluons at  
the scale $Q$ and lightcone time (position)
$r^- = r^+ = 0$,
\begin{equation}
\Delta_{0\;\mu\nu}^{cor}(r,K) \;=\; \int{d^4K'}{(2\pi)^4} \;
d_{\mu\nu}(K')\;\rho_0(K')\;{\tt G}_0(r,K) 
\;,
\label{in5}
\end{equation}
where 
$
{\tt G}_0(r,K) \equiv {\tt G}_0(r) {\tt G}_0(K)
$
with 
\begin{eqnarray}
{\tt G}_0(r) &=&
\frac{{\cal N}_0}{\pi} \;\delta(r^-)\;\delta(r^+)\;
\theta\left(1 - {\cal N}_0 r_\perp^2 Q^2\right) 
\nonumber \\
{\tt G}_0(K) &=&
\;=\; \frac{(2\pi)^4}{K^+}
\;\delta\left(K^+-Q^+\right) \;\delta\left(K^-- \frac{K_\perp^2 + Q^2}{K^+}\right)
\;\delta^2\left(\vec{K}_\perp\right)
\;.
\label{in6}
\end{eqnarray}
The visualization of the initial gluon density ${\tt G}_0$  
in (\ref{in6}) is  a  
{\it 2-dimensional color-charge density}: 
It is spread out in the two transversee
directions $\vec{r}_\perp$ in a disc with radius $R = 1/\sqrt{{\cal N}_0 Q^2} =
1/{\cal Q}$, and a delta-function in longitudinal direction at $r^+=0$ at 
time $r^-=0$. 
The normalization is  such that the total number ${\cal N}_0$ of initial bare gluons
is given by
\begin{equation}
\int dr^-dr^+d^2r_\perp \; {\tt G}_0(r,K) \;\equiv\;{\cal N}_0\;{\tt G}_0(K)
\label{in4}
\;.
\end{equation}

Finally, because of this statistical ensemble of almost pointlike, bare 
gluons, one does  not expect any collective mean-field behaviour 
at initial time $r^0 = t_0 = 0$ and at large $Q^2$, 
so that the  magnitude of the soft field 
is initially equal to zero which is consistent with (\ref{MFconstraint1}):
\begin{equation}
\overline{A}_\mu(r^+,r^-,\vec{r}_\perp)\vert_{r^-=r^+=0} \;=\;0
\label{in1a}
\;.
\end{equation}

This completes the construction of the initial state, starting from which, 
I now 
address the solution of the set of equations (\ref{X1})-(\ref{YME6}).
\bigskip

\subsection{Solving for the spectral density 
$\widehat{\rho}_{\mu\nu}$
}
\medskip

To find the spectral density $\widehat{\rho}_{\mu\nu}$, 
the solution of 
$\widehat{\Delta}^{ret}_{[\overline{0}]}$ 
and $\widehat{\Delta}^{adv}_{[\overline{0}]}$ 
is needed. 
The first correction to the free-field solution (\ref{freesol2})
arises from two contributions: a) from  one-loop hard gluon
self-interaction of order $g^2$ that is contained 
in the hard polarization tensor $\widehat{\Pi}$, and b) from
the coupling of the hard gluon propagator to the soft field $\overline{A}$
in $\overline{\Pi}$ which is of order $g\overline{A}$.
Within the perturbative scheme (\ref{ps5}) and (\ref{ps6}) 
the retarded and advanced propagators are to be
evaluated to order $g^2$ from eq. (\ref{gensol1}) with the internal
propagators in $\Pi^\ra$ taken as the free-field solutions,
\begin{equation}
\widehat{\Delta}^\ra_{[\overline{0}]}=\;
{\Delta}^\ra_{0}\;\;+\;\;
{\Delta}^\ra_{0}\;\;\widehat{\Pi}^\ra\left[ g^2; \Delta_0\right]\;
\;{\Delta}^\ra_0
\label{gensol21}
\;,
\end{equation}
with the subsidary condition
$
[ K\cdot \overline{D}_r ,  \widehat{\Delta}^\ra_{[\overline{0}]}
]
=
0+O(g^4)
$.
To order $g^2$, 
the gluon polarization tensor $\widehat{\Pi}$ 
as given by (\ref{RPiQU})-(\ref{RPid}),
reduces to the one-loop term $\widehat{\Pi}^{(2)}$, because the tadpole
term $\widehat{\Pi}^{(1)}$ vanishes as usual in the context of
dimensional regularization \cite{gaugereview}, and the two-loop terms
$\widehat{\Pi}^{(3)}$, $\widehat{\Pi}^{(4)}$ are of order $g^4$.
Hence,  
$\left(\widehat{\Pi}^{ret}-\widehat{\Pi}^{adv}\right)\left[ g^2; \Delta_0\right]$ in (\ref{gensol21}) reduces to
\begin{eqnarray} 
& &
\left(\widehat{\Pi}^{ret}-\widehat{\Pi}^{adv}
\right)^{ab}_{\mu\nu}(r,K)
\;=\;
-\frac{i g^2}{2}
\; \int \frac{d^4q}{(2\pi)^4}
\;\;V_{0\;\mu\lambda\sigma}^{\;\;\;\;acd}(K,-q,-K+q)\;\;
\widehat{V}_{0\;\sigma'\lambda'\nu}^{d'c'b}(r; K-q, q, -K)
\nonumber \\
& &
\nonumber \\
& &
\;\;\;\;\;\;\;\;\;\;\;\;\;
\times
\;\delta^{cc'}\delta^{dd'}
\; d^{\lambda\lambda'}(q)\; d^{\sigma\sigma'}(K-q)\;
 \;\left\{
\frac{}{}
{\Delta}_{0}^{adv}(r,q)\;
{\Delta}_{0}^{cor}(r,K-q)\;
\,-\,
{\Delta}_{0}^{cor}(r,q)\;
{\Delta}_{0}^{ret}(r,K-q)
\right\}
\label{e1}
\;,
\end{eqnarray}
where the
$\Delta_{0\;\,\mu\nu}^{ab}(r,K) = \Delta_0(r,K)\,d^{\mu\nu}(K) \,\delta^{ab}$
are the zeroth order solutions (\ref{freesol1}) and (\ref{freesol3}).

The mean-field contribution (\ref{RPiMF})-(\ref{PiMF6}) 
to the retarded and advanced components of 
$\overline{\Pi}$,  on the other hand,
vanishes, 
because  $\overline{F}_{\mu\nu} = T^a \overline{F}^a_{\mu\nu}$ 
is antismmetric and traceless
\footnote{
Note, however,  that this cancellation occurs only in the
lightcone gauge (\ref{gauge11}) with gauge parameter $\alpha = 0$.
In a general non-covariant gauge with $\alpha\ne 0$, one
encounters on the right hand side of (\ref{e2}) a finite
term $\alpha\,(n\cdot \partial_r)
\left(n_\nu \overline{A}_\mu(r) +n_\mu \overline{A}_\nu(r) \right)$.
}.
\begin{equation}
\left(
\overline{\Pi}^{\ra}\;\widehat{\Delta}^\ra
\right)^{ab}_{\mu\nu}(r,K)
\;=\;
- 2\,g \;\delta^{ab}\,d_{\mu\nu} \;\Delta^\ra_0(r,K) \; 
\left( \frac{1}{3} g^{\rho\lambda}\,\overline{F}_{\rho\lambda}(r)
\right)
\;\;=\;\;0
\label{e2}
\;.
\end{equation}
Hence, 
the dependence on the soft field $\overline{F}_{\mu\nu}$ or 
$\overline{A}_\mu$
is resident only implicitely in the kinetic momentum
$K_\mu = k_\mu -g\overline{A}_\mu$, so that eq. (\ref{X1}) 
becomes formally identical to the case of $\overline{A} = 0$, in
which $K_\mu = k_\mu$.
Exploiting this formal analogy, one can  evaluate explicitly
$\widehat{\Pi}_{\mu\nu}^{ret}-\widehat{\Pi}_{\mu\nu}^{adv}$
in the kinematic range 
$Q^2 \,\lower3pt\hbox{$\buildrel > \over\sim$}\,
(K^+)^2 \gg K_\perp^2 \ge \mu^2$
by using standard techniques of QCD evolution calculus
\cite{ms39,dok80}.
Inserting into (\ref{e1}) the free-field expressions for 
$\Delta_0^{ret}$, $\Delta_0^{adv}$, and  $\Delta_0^{cor}$, from
eqs. (\ref{freesol1}), (\ref{freesol3}), one finds that
to $O(g^2)$  the polarization tensor
$\widehat{\Pi}^\ra$ does not depend on $r$,
hence one may write
\begin{equation}
\widehat{\Pi}^\ra(r,K) \;\equiv \widehat{\Pi}^\ra(K) 
\;\;\;\;\;\;\;
{\cal Z}^\ra(r,K) \;\equiv {\cal Z}^\ra(K) 
\;.
\end{equation}
Using the lightcone variables (\ref{lcv1}), 
for the momenta, together with the lightcone phase-space element
\begin{equation}
d^4 q \;\stackrel{"q^2 \rightarrow 0"}{=}\; 
\frac{1}{2} \,dq^+dq^- d^2q_\perp\,\delta\left(q^+q^--q_\perp^2\right)
\;=\;
\frac{\pi}{2} \,\frac{dq^+}{q^+} dq_\perp^{2}
\;,
\label{e3}
\end{equation}
and using  (\ref{Piperp}) and  (\ref{Zdef1}),
$\widehat{\Pi}_\perp^{ret}-\widehat{\Pi}_\perp^{adv} = 
\frac{1}{2} (\widehat{\Pi}^\ra)_\mu^{\;\mu}$ 
one finds the formfactors 
${\cal Z}^\ra= (1\,-\,\Pi_\perp^\ra/K^2)^{-1)}$
to leading-log accuracy:
\begin{equation}
{\cal Z}^{ret} (K^{+},K_\perp^2)
\;=\;
\exp\left\{
-\,\frac{1}{2}\,\int_{K_\perp^2}^{K^{+\,2}=Q^2}
d q_\perp^{2}
\int_0^{K^+} \frac{dq^+}{q^+} \;
\frac{\alpha_s(q_\perp^{2})}{2\pi q_\perp^2}\,
\,\gamma\left(\frac{q^+}{K^+}\right)
\right\}
\;,
\label{e4}
\end{equation}
\begin{equation}
{\cal Z}^{adv} (K^{+},K_\perp^2) \;\theta(K^+) \;=\;
-\;{\cal Z}^{ret} (-K^{+},K_\perp^2) \;\theta(-K^+)
\label{e4a}
\;,
\end{equation}
where 
\begin{equation}
\alpha_s(q_\perp^2) \;=\; \frac{4\pi}{11\ln(q_\perp^2/\Lambda^2)}
\;\;\;\;\;\;\;\;\;\;\;\;\;
\gamma(z)\;=\;
2\,N_c\;\left(
 z ( 1 - z )  + \frac{1-z}{z} + \frac{z}{1-z}  \right)
\label{e5}
\;,
\end{equation}
and $z= q^{+}/{K^+}$, 
$1-z= q^{'\;+}/{K^+}$,
$ q^{'} = K - q$.
The effective formfactor function ${\cal Z}^\rho$ 
can be approximately evaluated,
\begin{equation}
{\cal Z}^\rho(K^+,K_\perp^2)
\;=\;\; {\cal Z}^{ret}-{\cal Z}^{adv}
\;\; \approx\;\;
\left\{
\begin{array}{l}
\exp\left\{
-\frac{3\alpha_s}{2\pi}\;
\ln^2\left(\frac{Q^2}{K_\perp^2}\right)
\right\}
\;\;\;\;\;\;\;\;\; \;\;\;\;\;\;\;
\;\;\;\;\;\;\;\;\; \;\;\;\;\;\;\;\;\;
\mbox{for} \;\; K_\perp^2 \,\ge \, \mu\,Q
\\
\exp\left\{
-\frac{3\alpha_s}{2\pi}\;
\left[ \frac{1}{2} \ln^2\left(\frac{Q^2}{\mu^2}\right)
-  \ln^2\left(\frac{K_\perp^2}{\mu^2}\right)
\right]
\right\}
\;\;\;\;\;\;\;\;\;
\mbox{for} \;\; K_\perp^2 \,< \, \mu\,Q
\end{array}
\right.
\label{e4b}
\end{equation}

Substituting ${\cal Z}^\rho$ into eq. (\ref{gensol11})
for $\widehat{\Delta}^{ret}_{\mu\nu}$ and 
$\widehat{\Delta}^{adv}_{\mu\nu}$, 
one obtains for the spectral density 
$\widehat{\rho}
=id_{\mu\nu}^{-1}\left(
\widehat{\Delta}_{\mu\nu}^{ret}-\widehat{\Delta}_{\mu\nu}^{adv}
\right)$, 
\begin{eqnarray}
\widehat{\rho}(K^+,K_\perp^2) &=&
{\cal Z}^\rho(K^+,K_\perp^2) 
\;\frac{(2\pi)^4}{K^+}\;
\;\left[
\frac{}{}
\delta\left(K^+ - Q^+\right) \;\delta\left(K_\perp^2\right)
\right.
\nonumber \\
& &
\;\;\;\;\;\;\;\;\;\;\;\;\;\;\;\;\;\;\;\;\;\;\;\;\;\;\;\;\;
\;+\;
\left.\frac{}{}
\int_{K_\perp^2}^{Q^2}\frac{d q_\perp^{2}}{q_\perp^{2}}
\;\frac{\alpha_s(q_\perp^2)}{2\pi}\;
\int_0^1 dz \;
\gamma(z)\;
\widehat{\rho}\left(\frac{K^+}{z},q_\perp^2\right)
\;{\cal Z}^{\rho\;\;\;-1}\left(K^+,\frac{q_\perp^{2}}{z}\right) 
\right]
\;.
\label{e6}
\end{eqnarray}
The previously advocated interpretation of the spectral density
$\widehat{\rho}$
of an initial state gluon
as the density of its `intrinsic' gluon fluctuations becomes clearer now:
$\widehat{\rho}$ represents the structure function of a gluon that
was initialized as a bare state at $Q^2$. Looking at this gluon state
with a resolution scale $K_\perp^2$, one sees at $K_\perp^2 = Q^2$
only the initial bare gluon itself, because ${\cal Z}^\rho(Q^+,Q^2)=1$,
eq. (\ref{Zdef2}), 
and the integral term in (\ref{e6}) vanishes.
For $K_\perp^2 < Q^2$, the formfactor ${\cal Z}^\rho(K^+,K_\perp^2)$ decreases
(c.f. eqs. (\ref{e4}), (\ref{e4b})), and so the first term, which
is the probablity that the
gluon remains in its bare initial state, is suppressed by ${\cal Z}^\rho$,
whereas the integral term, which is adjoint probability that the
gluon contains a distribution of intrinsic gluons, increases
with weight ${\cal Z}^\rho(K^+,K_\perp^2)/{\cal Z}^\rho(K^+,q_\perp^2/z)$.
Hence the evolution of the spectral density $\widehat{\rho}$ describes the change of structure of the initially bare gluon state due to real and virtual emission
and absorption of daughter gluons, corresponding to the generation of virtual
Coulomb field coat and real brems-strahlung, respectively.
\smallskip

Eq. (\ref{e6}) can be solved in closed form by using the following
trick to effectively eliminate ${\cal Z}^\rho$.
First, note that $\widehat{\rho}$ satisfies
the momentum sum rule \cite{chou},
\begin{equation}
\int_0^{K^+} dq^+ \,q^+\,
\widehat{\rho}\left(q^+,q_\perp^2\right) \;=\;(K^+)^2
\;\;\;\;\;\;\;\;\;\;\;\;\;\;\;
\int_0^{K^+} dq^+ \,q^+\,
\frac{\partial}{\partial q_\perp^2}\,
\widehat{\rho}\left(q^+,q_\perp^2\right) \;=\;0
\label{sumrule1}
\;,
\end{equation}
for any value of $q_\perp^2$.
Eq. (\ref{sumrule1}) is nothing but 
a manifestation of lightcone-momentum conservation, 
meaning that the  aggregat of $q^+$ momentum from intrinsic
gluons must add up to the total $K^+$ of the gluon state composed of those.
This is a general property, which is immediately evident in the free-field case.
Next,  multiply  (\ref{e6})
by $q^+/K^+$ and integrate over $q^+$ from 0 to $K^+$, which yields on account
of the sumrule (\ref{sumrule1}),
\begin{equation}
1 \;\;=\;\;
{\cal Z}^\rho(K^+,K_\perp^2) 
\;\left[
\frac{}{}
1\;\;+\;\;
\frac{}{}
\int_{K_\perp^2}^{Q^2}\frac{d q_\perp^{2}}{q_\perp^{2}}
\;\frac{\alpha_s(q_\perp^2)}{2\pi}\;
\int_0^1 dz \;
\gamma(z)\;
\;{\cal Z}^{\rho\;\;\;-1}\left(K^+,\frac{q_\perp^{2}}{z}\right) 
\right]
\;,
\end{equation}
which does not contain $\widehat{\rho}$.
Next, multiply this formula 
 with ${\cal Z}^{\rho\;\;-1}(K^+,K_\perp^2)$
from the left, and then differentiate with respect to $K_\perp^2$ by
applying $K_\perp^2\partial /\partial K_\perp^2$:
\begin{eqnarray}
& &
\left(
K_\perp^2 \frac{\partial}{\partial K_\perp^2}
{\cal Z}^{\rho\;\;\;-1}(K^+,K_\perp^2)
\right)
\;
\widehat{\rho}(K^+,K_\perp^2)
\;+\;
{\cal Z}^{\rho\;\;\;-1}(K^+,K_\perp^2)
\;
\left(
K_\perp^2 \frac{\partial}{\partial K_\perp^2}
\widehat{\rho}(K^+,K_\perp^2)
\right)
\nonumber \\
& &
\;\;\;\;=\;
-{\cal Z}^{\rho\;\;\;-1}(K^+,K_\perp^2)
\;\frac{\alpha_s(K_\perp^2)}{2\pi}\;
\int_0^1 dz \;
\gamma(z)\;
\frac{1}{z}
\widehat{\rho}\left(\frac{K^+}{z},zK_\perp^2\right)
\label{e7}
\;.
\end{eqnarray}
Using (\ref{e4}), 
the derivative $\partial {\cal Z}^\rho/\partial K_\perp^2$
on the left hand side can be rewritten as
\begin{equation}
K_\perp^2 \frac{\partial}{\partial K_\perp^2}
{\cal Z}^{\rho\;\;\;-1}(K^+,K_\perp^2)
\;=\;
- {\cal Z}^{\rho\;\;\;-1}(K^+,K_\perp^2)
\;
\frac{1}{2}
\;\frac{\alpha_s(K_\perp^2)}{2\pi}\;
\int_0^1 dz \;
\gamma(z)\;
\;.
\end{equation}
Substituing this into (\ref{e6}) and multiplying by ${\cal Z}^\rho$,
one obtains a differential evolution equation \'a la DGLAP
\cite{DGLAP,dok80} that involves only 
$\widehat{\rho}$, but not ${\cal Z}^\rho$ anymore
\footnote{
It should be noted that in obtaining (\ref{e9}),
the fact that
$\widehat{\rho}\left(\frac{K^+}{z},zK_\perp^2\right)
\simeq 
\widehat{\rho}\left(\frac{K^+}{z},K_\perp^2\right)$ was
used -- a property that is due to the very weak $z$-dependence of the 
the $K_\perp^2$-argument of $\widehat{\rho}$ \cite{dok80}.
}:
\begin{equation}
K_\perp^2 \frac{\partial}{\partial K_\perp^2}
\widehat{\rho}(K^+,K_\perp^2)
\;=\;
\;\frac{\alpha_s(K_\perp^2)}{2\pi}\;
\int_0^1 \frac{dz}{z} \;
\;\gamma(z)\;
\left[
\widehat{\rho} \left(\frac{K^+}{z},K_\perp^2\right)
-
\frac{z}{2}
\;
\widehat{\rho} \left(K^+,K_\perp^2\right)
\right]
\label{e9}
\;.
\end{equation}
The explicit solution of this equation is well known \cite{bassetto,ms36}:
\begin{equation}
\widehat{\rho}\left(K^+,K_\perp^2\right) 
\;=\;
\rho_0\left(K^+,K_\perp^2\right)
\;+\; 
\rho_1\left(K^+,K_\perp^2\right) 
\;\exp\left[-\frac{N_c}{12 \pi} g\left(K_\perp^2\right)\right]
\;\exp\left[
\sqrt{\frac{4 N_c}{11 \pi} 
\;g\left(K_\perp^2\right)\;h\left(K^+\right)
}
\right]
\label{e10}
\end{equation}
where 
\begin{eqnarray}
& &
\;\;\;\;\;\;\;\;
\rho_0\left(K^+,K_\perp^2\right)
\;=\;
\frac{(2\pi)^4}{K^+}
\;\delta(K^+-Q^+) \,\delta(K_\perp^2-Q^2)
\nonumber\\
& &
\;\;\;\;\;\;\;\;
\rho_1\left(K^+,K_\perp^2\right)
\;=\;
\frac{(2\pi)^4}{K^+}
\;\,
\frac{1}{\sqrt{4\pi}}\;\left[\frac{N_c}{11 \pi} \,g(K_\perp^2)\right]^{1/4}
\;\left[h(K^+)\right]^{-3/4}
\label{e10a}
\nonumber \\
& &
g(K_\perp^2)
\;=\;
\ln\left[\frac{\ln(Q^2/\Lambda^2)}{\ln(K_\perp^2/\Lambda^2)}\right]
\;\;\;\;\;\;\;\;\;\;\;\;\;\;
h(K^+)\;=\; \ln\left(\frac{Q^+}{K^+}\right)
\;.
\label{e10b}
\end{eqnarray}
\bigskip

\subsection{Solving for the correlation function
$\widehat{\Delta}^{cor}_{\mu\nu}$
}
\medskip

Within the perturbative scheme (\ref{ps5}) and (\ref{ps6}), 
the calculation of 
$\widehat{\Delta}^{cor} = \widehat{\Delta}_{[\overline{0}]}^{cor} 
+\delta\widehat{\Delta}_{[\overline{A}]}^{cor}$
is most conveniently split into two steps:
\begin{description}
\item[1.]
The quantum contribution
$\widehat{\Delta}_{[\overline{0}]}^{cor}$ is evaluated
to order $g^2$ from (\ref{gensol2}),
i.e., the hard polarization tensor $\widehat{\Pi}^{cor}[g^2,\Delta_0]$ 
is to be calculated again in one-loop approximation 
with free-field internal propagators.
The mean-field part polarization tensor, on the other hand,
is set to zero in this first step: $\overline{\Pi}^{cor}=0$.
\item[2.]
The mean-field induced correction 
$\delta\widehat{\Delta}_{[\overline{A}]}^{cor}$ 
in leading  order $g\overline{A}$ is then
added by calculating 
$\overline{\Pi}^{cor}[g; \Delta_0]$.
The quantum part now is set to zero in this second step:
$\widehat{\Pi}^{cor}=0$
(as it is already contained in 
$\widehat{\Delta}^{cor}_{[\overline{0}]}$ from Step 1).
\end{description}

\subsubsection{The contribution $\widehat{\Delta}_{[\overline{0}]}^{cor}$}

Since to order $g^2$ only the radiative self-interaction contributes to
to the hard propagator $\widehat{\Delta}_{[\overline{0}]}$,
and scattering processes that could alter the gluon trajectories  are 
absent, the transport equation 
for the part $\widehat{\Delta}_{[\overline{0}]}$ simplifies to,
\begin{equation}
\left[
K\cdot \overline{D}_r  \, ,\,  \widehat{\Delta}^{cor}_{[\overline{0}]}
\right]
\;=\;
\;0\;\;+\;\;O(g^4)
\;.
\label{e15}
\end{equation}
Therefore, with respect to the space-time variable $r$, 
(\ref{e15}) implies free-streaming behavior in the presence of
the soft mean field, as implicitly contained in $K=k-g\overline{A}$,
that is,
$\widehat{\Delta}^{cor}_{[\overline{0}]} (r,K) = 
\widehat{\Delta}^{cor}_{[\overline{0}]} (r'- V r^-)$ with
$V_\mu=K_\mu/K^+$ and $r' < r$.
Hence, it remains to consider eq. (\ref{gensol2}) with
$\widehat{\Delta}^{cor}_{[\overline{0}]} \rightarrow \Delta^{cor}_{0}$:
\begin{equation}
\widehat{\Delta}^{cor}_{[\overline{0}]}\;=\;
- 
{\Delta}^{ret}_0\; 
\left(\frac{}{}
{\Delta}^{cor\;\;-1}_{0}\;
\;-\;
\widehat{\Pi}^{cor}\left[g^2,\Delta_0\right]
\right)\;
{\Delta}^{adv}_0
\;.
\label{gensol22}
\end{equation}
The easiest way to obtain 
$\widehat{\Delta}_{[\overline{0}]}^{cor}$
is to use the formula (\ref{gensol3}) and simply convolute the
total number density of gluons ${\tt G} = 1 + 2 {\tt g}$ with 
the spectral density $\widehat{\rho}$ obtained in the preceeding subsection.
To prove that the relation  (\ref{gensol3}) is indeed consistent, one
calculates instead 
$\widehat{\Delta}_{[\overline{0}]}^{cor}$
from (\ref{gensol22}) directly.
The procedure is fully analogous to the previous subsection,
except that, instead of 
$\Pi^{ret} -\Pi^{adv}$, one  needs to evaluate 
$\Pi^{>} +\Pi^{<}$. The resulting form of 
$\widehat{\Pi}^{cor}\left[g^2,\Delta_0\right]$ in (\ref{gensol22}) is
\begin{eqnarray} 
& &
\left(\widehat{\Pi}^{cor}\right)^{ab}_{\mu\nu}(r,K)
\;\;=\;\;
 -\left(\widehat{\Pi}^{>}+\widehat{\Pi}^{<}
\right)^{ab}_{\mu\nu}(r,K)
\nonumber \\
& &
\;\;\;\;\;\;\;\;\;\;\;\;\;\;\;\;
\;=\;
\frac{i g^2}{2}
\; \int \frac{d^4q}{(2\pi)^4}
\;\;V_{0\;\mu\lambda\sigma}^{\;\;\;\;acd}(K,-q,-K+q)\;\;
\widehat{V}_{0\;\sigma'\lambda'\nu}^{d'c'b}(r; K-q, q, -K)
\nonumber \\
& &
\nonumber \\
& &
\;\;\;\;\;\;\;\;\;\;\;\;\;
\;\;\;\;\;\;\;\;\;\;\;\;\;
\times
\;\delta^{cc'}\delta^{dd'}
\; d^{\lambda\lambda'}(q)\; d^{\sigma\sigma'}(K-q)\;
 \;\left\{
\frac{}{}
{\Delta}_{0}^{>}(r,q)\;
{\Delta}_{0}^{<}(r,K-q)\;
\,+\,
{\Delta}_{0}^{<}(r,q)\;
{\Delta}_{0}^{>}(r,K-q)
\right\}
\label{e20}
\;,
\end{eqnarray}
where in the integral, the free-field forms 
$\Delta_{0\;\,\mu\nu}^\gl$ are given by (c.f. Appendix F):
\begin{equation}
\Delta_{0\;\,\mu\nu}^{\gl\;\;\;ab}(r,K)\; =\; 
\Delta_0^\gl(r,K)\,d^{\mu\nu}(K) \,\delta^{ab}
\;,\;\;\;\;\;\;\;
\Delta_0^\gl(r,K)\;=\;
  (-2\pi i) \;\delta(K^2) 
\left(\frac{}{}\theta(\pm K^+) \,+\, {\tt g}_0(r,\pm K)\right)
\;.
\label{e21}
\end{equation}
Inserting into (\ref{e20}) the expressions (\ref{e21}),
and observing that in (\ref{gensol2})  $\Pi^{cor}$ is sandwiched between
$\widehat{\Delta}^{ret}$ and  $\widehat{\Delta}^{adv}$, i.e.
appears only in the combination
\begin{equation}
\widehat{\Delta}_{\mu\lambda}^{ret} \;\widehat{\Pi}^{cor}_{\lambda\rho}
\;\widehat{\Delta}^{adv}_{\rho\nu} 
\;\propto\;
d_{\mu\lambda}(K) \,d_{\lambda\tau}(q)\,
d_{\tau\rho}(K-q) \,d_{\rho\nu}(K)
\;\propto\;
d_{\mu\nu}(K) \;\widehat{\Pi}_\perp^{cor}(K)
\;,
\end{equation}
where $\widehat{\Pi}^{cor}_\perp = g_{\rho\tau} \widehat{\Pi}^{cor}_{\rho\tau}$,
and $d_{\mu\lambda} d_{\lambda \nu} = d_{\mu\nu}$,
one finds after a calculation analogous as in the preceding subsection 
the following result for the $\widehat{\Delta}_{[\overline{0}]}$ part of the
correlation function:
\begin{eqnarray}
& &
\widehat{\Delta}_{[\overline{0}]}^{cor}\left(r,K^+,K_\perp^2\right) 
\;\;=\;\;
\mbox{Tr}\left[
d_{\mu\nu}^{-1}(K)\,
\widehat{\Delta}_{[\overline{0}]\;\mu\nu}^{cor}\left(r,K\right) 
\right]
\nonumber \\
&&
\;\;\;\;\;\;\;\;\;\;
\;=\;
\int_{K_\perp^2}^{Q^2} \frac{dK_\perp^{'\,2}}{K_\perp^{'\,2}}
\;\frac{\alpha_s(K_\perp^{'\,2})}{2\pi}\;
\int_0^{1} \frac{dz}{z} \;
\;\gamma(z)\;
\left[
\left\{\frac{}{}
\widehat{\rho} \left(1+2{\tt g}\right)
\right\}
\left(r,K^+/z,K_\perp^{'\,2}\right)
\;-\;
\frac{z}{2}
\left\{\frac{}{}
\widehat{\rho} \left(1+2{\tt g}\right)
\right\}
\left(r,K^+,K_\perp^{'\,2}\right)
\right]
\;.
\label{e22}
\end{eqnarray}
Comparison with (\ref{e9}) reveals that
$\widehat{\Delta}^{cor}_{[\overline{0}]}$
is indeed the convolution of the spectral density $\widehat{\rho}$ with the total
gluon density ${\tt G} = 1 + 2{\tt g}$, as advocated by eq. (\ref{gensol3}).
Hence,
\begin{equation}
{\tt G}_{[\overline{0}]}(r,K^+,K_\perp^2)
\;=\;
{\tt G}_{0}(r,Q^+,0_\perp)
\;+\;
\int_{K_\perp^2}^{Q^2} \frac{dK_\perp^{'\,2}}{K_\perp^{'\,2}}
\;\frac{\alpha_s(K_\perp^{'\,2})}{2\pi}\;
\int_0^{1} dz \;
\;\gamma(z)\;
\left[
\frac{1}{z}
{\tt G}_{[\overline{0}]}
\left(r,\frac{K^+}{z},K_\perp^{'\,2}\right)
\;-\;
\frac{1}{2}
{\tt G}_{[\overline{0}]}
\left(r,K^+,K_\perp^{'\,2}\right)
\right]
\;,
\label{e22a}
\end{equation}
where 
${\tt G}_{0}(r,Q^+,0_\perp)$ is the initial gluon density (\ref{in6}).
In the  limit $z\ll 1$ 
the integral (\ref{e22a}) can be approximately evaluated analytically
\cite{furmanski79,basetto80}.
This gives an estimate of the gluon multiplicity 
\cite{basetto83,mueller83}
as a function of $K_\perp^2$ at fixed space-time point $r$:
\begin{eqnarray}
{\tt G}_{[\overline{0}]}(r,K_\perp^2)
&=&
i \;\int_0^{Q^+} d K^+\;
{\tt G}_{[\overline{0}]}(r,K^+,K_\perp^2)
\nonumber \\
&=&
{\tt G}_0(r,Q^2)\;
\left(\frac{\ln(Q^2/\Lambda^2)}{\ln(K_\perp^2/\Lambda^2)}\right)^{-1/4}
\;
\exp\left[2\,\sqrt{\frac{N_c}{11\pi}}\;
\left(\frac{}{}
\sqrt{\ln\left(\frac{Q^2}{\Lambda^2}\right)} 
\;-\;\sqrt{\ln\left(\frac{K_\perp^2}{\Lambda^2}\right)}
\right)
\right]
\label{e23}
\;,
\end{eqnarray}
where ${\tt G}_0(r,Q^2)$ is given by (\ref{in6}).
It is evident that in the kinematic regime  $Q^2 \ge K_\perp^2 \ge \mu^2$, 
the hard gluon multiplicity is characterized by a rapid growth 
as the gap between the initial scale $Q^2$ and $K_\perp^2$ increases.
\smallskip

\subsubsection{The contribution $\delta\widehat{\Delta}_{[\overline{A}]}^{cor}$}
The leading-order mean-field contribution 
$\delta\widehat{\Delta}_{[\overline{A}]}^{cor}$ 
is now to be added to the result for
$\widehat{\Delta}_{[\overline{0}]}^{cor}$, eq. (\ref{e22}).
To do so, one needs to evaluate $\overline{\Pi}^{cor}$ to order $g\overline{A}$,
using the free-field solutions (\ref{freesol1})-(\ref{freesol3})
and set $\widehat{\Pi}^{cor} = 0$.
The analogon of (\ref{gensol22}) for
$\delta\widehat{\Delta}^{cor}_{[\overline{A}]}$ is
\begin{equation}
\delta\widehat{\Delta}^{cor}_{[\overline{A}]}\;=\;
- 
{\Delta}^{ret}_{[\overline{0}]}\;
\overline{\Pi}^{cor}\left[g,\Delta_0\right]
\; {\Delta}^{adv}_{[\overline{0}]}
\label{gensol23}
\;,
\end{equation}
and 
$\overline{\Pi}^{cor}\left[g,\Delta_0\right]$ can be read off from
(\ref{X2}), giving the contribution
\begin{eqnarray} 
& &
g \,K_\lambda  \overline{F}^{\lambda\sigma}(r) \; \partial^K_{\sigma}
\Delta_{[\overline{0}]\;\mu\nu}^{cor\;\;ab}(K)
\;+\; g \left(
\overline{F}_\mu^\lambda(r)\,{\Delta}_{[\overline{0}]\;\lambda\nu}^{cor}(K)
-
{\Delta}_{[\overline{0}]\;\mu}^{cor\;\;\lambda}(K)\, \overline{F}_{\lambda\nu}(r)
\right)^{ab}
\nonumber \\
& &
\;\;\;\;\;\;\;\;\;\;\;\;\;\;\;\;\;\;\;\;\;\;
\;\;\;\;\;\;\;\;\;\;\;\;\;\;\;\;\;\;\;\;\;\;
\;\;\;\;\;\;\;\;\;\;\;\;\;\;\;\;\;\;\;\;\;\;
\;\;\;\;\;\;\;\;\;\;\;\;\;\;\;\;\;\;\;\;\;\;
\;=\;
g \,\delta^{ab} \;d_{\mu\nu}(K)\;K_\lambda \overline{F}^{\lambda\sigma}(r)
 \partial^K_{\sigma} {\Delta}_{[\overline{0}]}^{cor\;\;ab}(K)
\label{e25}
\;.
\end{eqnarray}
The second term on the left side  cancels,
because
$\overline{F}_\mu^\lambda\,\widehat{\Delta}_{[\overline{0}]\;\lambda\nu}^{cor}
\propto
\overline{F}_\mu^\lambda\,g_{\lambda\nu}
-
\overline{F}_\mu^\lambda\,
(n_\lambda K_\nu+K_\lambda n_\nu)/K^+$, and 
$n^+=-n^-=1, \; n_\perp = 0$,  $K^-=0$.
Notice that this is a specific feature of the employed lightcone
representation, and does not hold in a general non-covariant gauge.
With (\ref{e25}), the function
$\delta\widehat{\Delta}_{[\overline{A}]}^{cor}$ satisfies the
transport equation
\begin{equation}
\left[
K\cdot \overline{D}_r  \, ,\,  \delta\widehat{\Delta}^{cor}_{[\overline{A}]}
\right]
\;=\;
-\;
g \;d_{\mu\nu}\;K_\lambda \overline{F}^{\lambda\sigma} \partial^K_{\sigma}
{\Delta}_{[\overline{0}]}^{cor}
\label{e26}
\;.
\end{equation}
To solve (\ref{e26}), it is convenient to express
$\delta\widehat{\Delta}_{[\overline{A}]\;\mu\nu}^{cor}$ in terms of a new
function $\Phi_\mu = T^a \Phi_\mu^a$ \cite{BI1}, defined by
\begin{equation}
\delta\widehat{\Delta}_{[\overline{A}]\;\mu\nu}^{cor}(r,K)
\;=\;
d_{\mu\nu}(K)\;
\;g\,\Phi^\lambda(r,K) \,\;\partial^K_\lambda 
\widehat{\Delta}_{[\overline{0}]}^{cor}(r,K)
\;,
\label{e26a}
\end{equation}
where 
$\widehat{\Delta}_{[\overline{0}]}^{cor}$ is the solution (\ref{e22}).
In terms of the function $\Phi^\mu$, the transport
equation (\ref{e26}) becomes now
\begin{equation}
\left[
K\cdot \overline{D}_r  \, ,\,  
\Phi_\mu(r,K)
\right]
\;=\;
\overline{F}_{\mu\nu}(r)\, K^\nu
\label{e27}
\;.
\end{equation}
The function $\Phi^\mu$ evidently satisfies
\begin{equation}
K_\mu \,\Phi^\mu(r,K) \;=\; 0
\;\;\;\;\;\;\;
\Longrightarrow
\;\;\;\;\;\;\;\;\;\;
\Phi^- \;=\; \frac{2}{K^+}\;\vec{K}_\perp\cdot \vec{\Phi}_\perp
\;\;\;\;\;\;\;\;\;
\Phi^+ \;=\; 0
\label{e29}
\;,
\end{equation}
i.e., $\Phi^-$ is not an independent variable, but is  expressable in terms
of the transverse components $\vec{\Phi}_\perp$, and
$\Phi^+$ is suppressed by $K^-/K^+$ and therefore may be set to zero.
The interpretation of the function  $\Phi_\mu$,
as was pointed out by Blaizot and Iancu \cite{BI1},
is the following:
The component $g \Phi_\mu$  corresponds to the kinetic momentum 
$K_\mu = k_\mu -g\overline{A}_\mu$
that is acquired by a gluon
propagating in the presence of the
soft field $\overline{A}_\mu$, or $\overline{F}_{\mu\nu}$.
The condition (\ref{e29}) reflects then the fact that the 
lightcone energy transferred by the soft field, namely
$g \Phi^-$, equals the mechanical work done by the Lorentz force
$g \vec{V}_\perp\cdot \Delta\vec{K}_\perp = g \vec{V}_\perp\cdot\vec{\Phi}_\perp$,
where $V^\mu = K^\mu/K^+ = (1,0,\vec{V}_\perp)$ is the velocity.
\smallskip

The transport equation (\ref{e27}) for $\Phi_\mu$ 
can be readily solved  \cite{BI1} with the help of
the  retarded and advanced functions
${\Delta}^\ra_{{0}\;\;\mu\nu}= 
d_{\mu\nu}\,{\Delta}^\ra_{{0}}$, 
\begin{equation}
\Phi_\mu(r,K) 
\;=\;
i \int d^4 r'\;{\Delta}^{ret}_{0}(r-r',K)
\;\overline{F}_{\mu\nu}(r')\;K^\nu
\;-\;
i \int d^4 r'\;{\Delta}^{adv}_{0}(r-r',K)
\;\overline{F}_{\mu\nu}(r')\; K^\nu
\label{e31}
\end{equation}
The free-field retarded and advanced functions
admit the space-time representation \cite{BI1},
\begin{eqnarray}
{\Delta}^{ret}_{{0}} (r-r',K)
&=& 
-i\; \theta(r^--r^{'-})\,
\delta\left({r}^+-{r}^{'+} -(r^--r^{'-})\right)
\;\delta^2\left(\vec{r}_\perp-\vec{r}^{\;'}_\perp-\frac{\vec{K_\perp}}{K^+}
(r^--r^{'-})\right)
\nonumber \\
{\Delta}^{adv}_{{0}} (r-r',K)
&=& 
+i \;\theta(r^{'\;-}-r)\,
\delta\left({r}^{'+} - r^+ -(r^{'-}-r^-)\right)
\;\delta^2\left(\vec{r}^{\;'}_\perp-\vec{r}_\perp-\frac{\vec{K_\perp}}{K^+}
(r^{'-}-r^-)\right)
\;,
\label{e32}
\end{eqnarray}
and therefore
$i\left({\Delta}^{ret}_{{0}}  - {\Delta}^{adv}_{{0}}\right) (r-r',K)
=
2\;\delta\left({r}^+-{r}^{'+} -(r^--r^{'-})\right)
\delta^2\left(\vec{r}_\perp-\vec{r}^{\;'}_\perp-(\vec{K_\perp}/K^+)
(r^--r^{'-})\right)
$.
Insertion into (\ref{e31}) then yields,
\begin{equation}
\Phi_\mu(r,K) 
\;=\;
 2\; K^\nu \;\int_0^{r^-} dr^{'-}\;
\;\overline{F}_{\mu\nu}\left(r - \frac{K}{K^+}\,r^{'-}\right)
\;\;\equiv\;\;
 2\; K^\nu \;
\;\overline{{\tt F}}_{\mu\nu}(r)
\;.
\label{e33}
\end{equation}
Substituting  this result into (\ref{e26a}) and using the lightcone components
of $\overline{{\tt F}}_{\mu\nu}$, eq. (\ref{FmunuLC}), the
result for the mean-field induced correction
$\delta\widehat{\Delta}^{cor}_{[\overline{A}]}$ is,
\begin{equation}
\delta\widehat{\Delta}^{cor}_{[\overline{A}]}(r,K) 
\;=\;
\mbox{Tr}\left[
d_{\mu\nu}^{-1}(K)\,
\delta\widehat{\Delta}_{[\overline{A}]\;\mu\nu}^{cor}\left(r,K\right) 
\right]
\;=\;
- g \;{\overline{\tt F}}_{\perp +}(r) \;
\left( K_\perp \,\frac{\partial}{\partial K^+} \;-\;
 K^+\,\frac{\partial}{\partial K_\perp} \right)
\; \widehat{\Delta}^{cor}_{[\overline{0}]}(r,K) 
\label{e34a}
\end{equation}
where $\perp$ denotes the transverse vector components $i=1,2$,
and $\Phi_\perp = \frac{1}{2} (\Phi^1 + \Phi^2)$,
$K_\perp = \frac{1}{\sqrt{2}} (K^1 + K^2)$.

With (\ref{e33}), the  addition 
$\delta{\tt G}_{[\overline{A}]}(r,K^+,K_\perp^2)$
to the gluon density ${\tt G}_{[\overline{0}]}(r,K^+,K_\perp^2)$
of eq. (\ref{e22a}) is,
\begin{equation}
\delta{\tt G}_{[\overline{A}]}(r,K^+,K_\perp^2)
\;=\;
- 2 g \;\frac{K^+}{K_\perp} \;{\overline{\tt F}}_{\perp +}(r) \;
\left( K_\perp^2 \,\frac{\partial}{\partial K_\perp^2}
\, {\tt G}_{[\overline{0}]}(r,K^+,K_\perp^2) 
\right)
\;\;+\;\;O\left(K_\perp^2/(K^+)^2\right)
\label{e34}
\;,
\end{equation}
where the explicit form of derivative term in brackets can be easily
read off the right-hand-side of (\ref{e22a}).
\bigskip
\bigskip

\subsection{Expansion in space-time of the hard gluon ensemble}
\medskip

As argued before in (\ref{e15}), the evolution of the hard gluon density
${\tt G}$, described by $\widehat{\Delta}^{cor}$, can in the
present context be viewed as a purely multiplicative
casade of gluon emissions , since to order $g^2$ and due to
the quasi-collinear motion of the gluons, statistical scatterings
between them do not contribute.
Therefore
the space-time development of ${\tt G}(r,K^+,K_\perp^2)$
with respect to $r = (r^-,r^+,\vec{r}_\perp)$ is of
`free-streaming nature'.
That is, the expansion with time of the ensemble of gluons as-a-whole
proceeds through a deterministic diffusion in momentum and space-time, as
qualitatively scetched in Fig. 1a.

To quantify this heuristic picture, one needs to
invoke the uncertainty principle to relate the development in
space and time to the evolution in momentum space, i.e. with respect to
$K^+$ and $K_\perp^2$.
Specifically, what is the characteristic time $r^-$ 
in the chosen Lorentz fame, that it takes to 
build up the density ${\tt G}(r,K^+,K_\perp^2)$ from the initial
form ${\tt G}_0(r_0, Q^+,0_\perp)$ at time $r^-_0 =0$.
Viewing the gluon evolution as a  cascade of successive 
branchings $K_{n-1}\rightarrow K_n + K_n'$, where $n$ labels the
generation in the cascade tree,
the life-time of gluon $K_{n-1}$ is given by the
time span $\Delta r^-_n$ that it takes to emit and form
the daughters $K_n$ and $K_n'$ as individual off-spring, that is,
by the formation time
\begin{equation}
\Delta r^-_n \;=\;
\frac{1}{2} 
\left(\frac{K_n^+}{K_{\perp n}^2}-\frac{K_n^{+\,'}}{K_{\perp n}^{'\;2}}
\right)
\;=\; \frac{K_{n-1}^+}{K_{\perp n}^2}
\;\;\equiv\;\; \tau_n\;\gamma_n
\label{time1}
\;,
\end{equation}
with $K_n^{+\,'}=K_{n-1}^+-K_n^{+}$, $\vec{K}_{\perp n}=-\vec{K}_{\perp n}'$,
and $K_{\perp n} \equiv \sqrt{\vec{K}_{\perp n}^2}$.
Here $\tau_n =1/K_{\perp n}$ and $\gamma_n = K^+_{n-1}/K_{\perp n}$
play the role of the proper time and the Lorentz gamma-factor, 
respectively , in agreement with the uncertainty principle.
Similarly, the average longitudinal 
and transverse distances travelled by the gluons
$K_n$ and $K_{n}'$ during the time span $\Delta r^-_n$ are
\begin{eqnarray}
\Delta r^+_n 
&=&
\frac{1}{2}\left( V_n^+ + V_n^{+\,'}\right) \;\Delta r^-_n
\;\simeq\; \frac{K_{n-1}^+}{K_{\perp n}^2}
\\
\Delta r_{\perp n} 
&=& 
\;\;\;\;\left| \vec{V}_{\perp n} - \vec{V}_{\perp n}'\right| 
\;\Delta r^-_n
\;\simeq \;\frac{2}{K_{\perp n}}
\;,
\label{time2}
\end{eqnarray}
where $V^\mu = K^\mu/K^+$ and $K_\perp \ll K^+$ is assumed as before.
The average total time $\langle \,r^-\,\rangle$ 
elapsed up to the $n$-th cascade generation
with mean gluon momentum $K^+$ and $K_\perp^2$, and the associated
spatial spread 
$\langle \,r^+\,\rangle$, $\langle \,r_\perp\,\rangle$, 
of the diffusing gluon ensemble, is then
obtained by weighting the  evolution 
of the gluon density ${\tt G}$, eq. (\ref{e22a}),
with $\Delta r(K^+,K_\perp^2) \equiv (\Delta r^-, \Delta r^+, \Delta r_\perp)$
from (\ref{time1})-(\ref{time2}). Taking the real emission
part of eq. (\ref{e22a}), differentiating it with respect to $K_\perp^2$,
convoluting it with the weight $\Delta r(K^+,K_\perp^2)$, integrating
over all possible branchings, and normalizing it to the density 
${\tt G}(r,K^+,K_\perp^2)$ itself, the desired average is
\begin{eqnarray}
\langle \;\,r(K^+,K_\perp^2)\;\,\rangle 
&=&
\frac{1}{{\tt G}(r,K^+,K_\perp^2)}
\;\;
\int_{K_\perp^2}^{Q^2} \frac{dK_\perp^{'\,2}}{K_\perp^{'\,2}}
\;\frac{\alpha_s(K_\perp^{'\,2})}{2\pi}\;
\left[\frac{}{}
\int_{K^+}^{Q^+}\frac{dK^{+\,'}}{K^{+\,'}}
\; {\tt G}(r,K^{+\,},K_\perp^{'\,2}) \;\,\Delta r(K^{+\,'},K_\perp^{'\,2})
\right]
\nonumber
\\
& &
\;\;\;\;\;\;\;\;\;\;\;\;\;\;\;\;\;\;
\;\;\;\;\;\;\;\;\;\;\;\;\;\;\;\;\;\;
\times\;
\int_0^{1} \frac{dz}{z} \;
\;\gamma(z)\;
{\tt G}
\left(r,K^+/z,K_\perp^{'\,2}\right)
\label{time3}
\;.
\end{eqnarray}
This complicated formula can be approximately evaluated from the known
behaviour of ${\tt G}$, as has been worked out in detail
in \cite{marchesini81}. For $K^+\ll Q^+$ the result is, up to
powers of $\ln(Q^+/K^+)$, the following estimate:
\begin{eqnarray}
\langle \;\,r^-(K^+,K_\perp^2)\;\,\rangle 
 &\simeq&
\langle \;\, r^+(K^+,K_\perp^2)\;\,\rangle 
\;=\;\frac{K^+}{K_\perp^2}\;{\cal T}(K^+,K_\perp^2)
\nonumber
\\
\langle \;\,r_\perp(K^+,K_\perp^2)\;\,\rangle 
&=&
\frac{2}{K_\perp}\;{\cal T}(K^+,K_\perp^2)
\;,
\label{time4}
\end{eqnarray}
where 
\begin{equation}
{\cal T}(K^+,K_\perp^2) \;=\ c_1(K_\perp^2) \;
\exp\left[ -\,c_2(K_\perp^2)\;\sqrt{\ln \left(\frac{Q^+}{K^+}\right)}\right]
\;,
\end{equation}
with $c_1, c_2 > 0$ very slowly varying functions of $K_\perp^2$.
This estimate shows that those gluons which are emitted either with
large $K_\perp^2$, or with small $K^+/Q^+$, 
appear the earliest in time $r^-$ and contribute the
quickest to the diffusion in $r^+$, $r_\perp$.
\bigskip
\bigskip

\subsection{Constructing the hard current  $\widehat{j}_\mu$ and the
induced soft field $\overline{A}_\mu$}
\medskip

The final task of the solution scheme Sec. III C is to solve for the soft
field $\overline{A}_\mu$, or $\overline{F}_{\mu\nu}$, which is
induced by the color current $\widehat{j}_\mu$, being
generated by the aggregat of initial plus emitted hard gluons
from the evolution of the gluon density (\ref{e22a}).
In the equation of motion for  $\overline{F}_{\mu\nu}$, recall (\ref{YME6}),
\begin{equation}
\left[\frac{}{}
\overline{D}_r^{\lambda}  , \; \overline{F}_{\lambda \mu}
\right]^a (r)
\;=\;
-\;\widehat{j}^a_\mu(r)
\;,
\label{YME6a}
\end{equation}
the current on the right-hand-side is determined by the  
hard gluon correlation function
$\widehat{\Delta}^{cor}_{\mu\nu}
=\widehat{\Delta}^{cor}_{[\overline{0}]\;\mu\nu}+
\delta\widehat{\Delta}^{cor}_{[\overline{A}]\;\mu\nu}$,
and therefore by the gluon density 
${\tt G}={\tt G}_{[\overline{0}]}+\delta{\tt G}_{[\overline{A}]}$,
as obtained in the previous subsection,
\begin{equation}
\widehat{j}_\mu(r)
\;=\;
T^a \,\widehat{j}^a_\mu(r)
\;=\;
- g
\;\int \frac{d^4k}{(2\pi)^2}
\;
\mbox{Tr} 
\left[ T^a\,
\left(
K_\mu \,\widehat{\Delta}_{\;\nu}^{cor\;\nu}(r,K)
\;-\;
K_\nu\,\widehat{\Delta}_{\;\mu}^{cor\;\nu}(r,K)
\right) 
\right] 
\label{j1}
\;.
\end{equation}
The first point to be made here is that, for the lightcone gauge 
condition $A^+ = 0$, the gauge-field tensor 
$\overline{F}_{\mu\nu}$ has only the non-vanishing components (\ref{FmunuLC}),
and if one requires in addition $A^-=0$, then in (\ref{YME6a}),
$
\overline{D}^{\lambda}  \, \overline{F}_{\lambda \mu}
=
\delta^\lambda_{\;\,\perp} \,\delta_{\mu +}\;\,
\overline{D}_\perp \, \overline{F}_{\perp +}
$.
The second observation is that the left-hand-side of (\ref{j1}) is essentially
the density ${\tt G}$ of hard gluons weighted with their momentum
$K^\mu$. Because the gluons evolve with the velocity $V^\mu=K^\mu/K^+$ 
along the lightcone,  at a given lightcone time $r^-$ and corresponding
coordinate $r^+ = r_0^++V^+r^- = r^- $,
these gluons appear as an extremely  thin Lorentz-contracted sheet, but
are spread out in transverse direction $r_\perp$ over a disc with radius 
$\sim 1/\langle K_\perp\rangle$.
As a consequence, the gluon current 
$\widehat{j}^\mu = ( \widehat{j}^+, \widehat{j}^-, \vec{\widehat{j}}_\perp)$
has only a component in $+$-direction, 
$
\widehat{j}^\mu= \delta^{\mu+} \; \widehat{j}^+
$
\cite{lmcl,kovchegov}.
Denoting as before the two transverse vector components
$i=1,2$ by $\perp$ with summation convention $a_\perp b_\perp \equiv \sum_{i=1,2} a_i b_i$,
eq. (\ref{YME6a}) now becomes,
\begin{equation}
\left[\frac{}{}
\overline{D}_\perp  , \; \overline{F}_{+\perp}
\right]^a (r)
\;=\;
\left[\frac{}{}
\delta^{ab}\partial_\perp-gf^{abc} \overline{A}_\perp^c(r)\, , \; \overline{F}_{+\perp}^b
\right] (r)
\;=\;
-\;\widehat{j}_+^a(r)
\;,
\label{YME6b}
\end{equation}
where $\widehat{j}_+^a$ is the color-charge density at $r^-=r^+$,
\begin{equation}
\widehat{j}_+^{a}(r)
\;=\; g\;T^a\,{\cal J}(\vec{r}_\perp)\;\delta(r^- - r^+)
\;,
\label{j0}
\end{equation}
where
\begin{equation}
{\cal J}(\vec{r}_\perp)
\;=\;
2\pi\;
\int_0^{Q^+} \frac{dK^+}{(2\pi)^3 \,2K^+}
\int_{\mu^2}^{Q^2} dK_\perp^2
\; \mbox{Tr} \left[ T^a\, K^+\widehat{\Delta}^{cor}(r,K^+,K_\perp^2)
\right]
\label{j2}
\end{equation}
and, 
$\widehat{\Delta}^{cor}
=
d_{\mu\nu}^{-1}\, (\widehat{\Delta}_{[\overline{0}]\;\mu\nu}^{cor}+
\delta\widehat{\Delta}_{[\overline{A}]\;\mu\nu}^{cor})$, 
using (\ref{e22a}), (\ref{e34}), 
\begin{equation}
\widehat{\Delta}^{cor}\left(r,K^+,K_\perp^2\right) 
\;=\;
2 \;\left(
\frac{}{} 1\;-\; 2 \,g\,\frac{K^+}{K_\perp} \;{\overline{\tt F}}_{\perp +}(r) \;
K_\perp^2 \,\frac{\partial}{\partial K_\perp^2}
\right)
\, {\tt G}_{[\overline{0}]}(r,K^+,K_\perp^2) 
\label{j3}
\;.
\end{equation}
Eqs. (\ref{j2}) and (\ref{j3}) follow from the fact that, on the 
left-hand-side of 
(\ref{j1}), the correlation function obeys the transversality condition 
$K^\mu \widehat{\Delta}^{cor}_{\mu\nu}= 
K^\mu d_{\mu\nu} \widehat{\Delta}^{cor} = 0$
and because
$K^\mu \widehat{\Delta}_\nu^{cor\;\,\nu}
= K^\mu d_{\nu}^\nu \widehat{\Delta}^{cor} =
2 K^\mu \widehat{\Delta}^{cor}$.
Notice that
in (\ref{j2}) the limits of the integration over $K^+$ and $K_\perp^2$
correspond to the average time $r^-$ and spatial extent $r_\perp$ of the 
gluon system, as estimated in (\ref{time4}) above, and hence,
${\cal J}$ accounts for the total gluon multiplicity
accumulated by the evolution between $Q$ and $\mu$.
\smallskip

Integrating both sides of (\ref{YME6b}) over $r^+$, $r^-$, and using
(\ref{j2}), (\ref{j3}), gives 
\begin{equation}
{\cal J}(\vec{r}_\perp)
\;=\;
\int_0^{r^+} dr^+ \int_0^{r^-} dr^-
\;
\left(\frac{}{}
\partial_\perp  \overline{F}_{\perp +}
\;-\;i g \left[
\overline{A}_\perp\, , \, \partial_\perp  \overline{F}_{\perp +}
\right]
\right)
(r^+,r^-,\vec{r}_\perp)
\label{YME6c}
\;.
\end{equation}
An approximate method to determine the soft field from (\ref{YME6c}), is to
adopt the approach of Kovchegov \cite{kovchegov}, who recently calculated
the lightcone gauge field induced by an ultra-relativistic current
of quarks with a uniform momentum distribution, using
the known form of the lightcone gauge potential of a single color charge
\cite{mueller88}.
Applying his concept
to the present case of gluons with a non-uniform distribution
${\tt G}(r,K)$, the first step is to write the color-charge density ${\cal J}$
of (\ref{j2}) in a `discretized version' as a superposition of
${\cal N}$ individual gluon charges,
\begin{equation}
\widehat{j}_+^{a}(r)
\;=\; \,g\;
\mbox{Tr} 
\left[\frac{}{}
\sum_{i=1}^{{\cal N}} \,T_i^a \,
\delta(r^+-r^+_i)\delta(r^--r^-_i) \delta^2(\vec{r}_\perp-\vec{r}_{\perp\,i})
\;\,{\cal J}(\vec{r}_{\perp\,i})
\right]
\;\delta(r^- - r^+)
\;,
\label{j5}
\end{equation}
where ${\cal N}$ is the total number of
gluons at a given $r^+=r^-$,
$
{\cal N}(r)
=
\pi \int_0^{Q^+} \frac{dK^+}{(2\pi)^3 \,2K^+}
\int_{\mu^2}^{Q^2} dK_\perp^2
\; {\tt G}(r,K)
$.
Now the approximate solution to (\ref{YME6c}) for the lightcone gauge-potential 
$\overline{A}_\mu$ at space-time point $r$ 
is obtained, 
by the superposition of contributions that are induced by the hard gluons 
at points $r_i$. 
Following Kovchegov \cite{kovchegov} in detail, 
the result is that $\overline{A}_\mu$ has only
non-vanishing transverse components,
\begin{eqnarray}
\overline{A}^+(r) &=& \overline{A}^{\,-}(r) \;=\; 0
\nonumber \\
\vec{\overline{A}}_\perp(r) &=&
2\pi g \,
\;\sum_{i=1}^{{\cal N}}
\,
\theta(r^+-r^+_i)\,\theta(r^--r^-_i) 
\;
\ln\left(
\frac{
\vec{r}_\perp-\vec{r}_{\perp\,i}
}{
|\vec{r}_\perp-\vec{r}_{\perp\,i}|^2
}
\right)
\;\,{\cal J}(\vec{r}_{\perp\,i})
\;\;
\mbox{Tr}\left[T_i^a \,S(r)\, T^a\,S^{-1}(r)\right]
\;,
\label{A1}
\end{eqnarray}
where
\begin{equation}
S(r) \;=\;
\prod_{i=1}^{{\cal N}}
\exp\left[
\;\frac{}{}
2\pi i \,g^2 \,
\;T^a\,T^a_i\;
\theta(r^+-r^+_i)\,\theta(r^--r^-_i) \;
\ln\left(
\frac{
\vec{r}_\perp-\vec{r}_{\perp\,i}
}{
|\vec{r}_\perp-\vec{r}_{\perp\,i}|^2
}
\right)
\right]
\;.
\label{A2}
\end{equation}
It is important to note that
eq. (\ref{A1}) is only 
an approximate solution of (\ref{YME6c}) for the 
induced soft field.
It is an {\it estimate}
of the classical equation of motion for the soft mean field 
$\vec{\overline{A}}_\perp$ that
is generated by the collective motion of a given configuration of hard
gluons with a distribution ${\tt G}(r,K)$.
In other words,
eq. (\ref{A1}) is the non-abelian Weizs\"acker-Williams field
due to the hard gluons.
\bigskip
\bigskip

\subsection{Resum\'e}
\medskip

Let me summarize the input and results of the preceding sample calculation
for the evolution of a high-energy gluon beam along the lightcone.
On the basis of the calculation scheme of Sec. III C,
the logic of application proceeded in the following steps:
\begin{description}
\item[1.]
Choice of lightcone gauge with gauge vector $n$ along the gluon
beam direction $K^+/K$ and gauge constraint $A^+=0$.
\item[2.]
Specification of the initial bare gluon ensemble at time $r^-=0$
with a momentum distribution of equal momenta $K^+ = Q^+$, $K_\perp =0$,
and a spatial distribution being uniform $r_\perp\le R$ in the transverse plane, but 
a delta-function sheet in longitudinal beam direction at $r^+=0$.
\item[3.]
Calculation of the retarded and advanced functions
$\widehat{\Delta}^\ra$ and the associated spectral density $\widehat{\rho}$
to order $g^2$ from the initial values of the hard gluon propagators.
The result is stated by eqs. (\ref{e10}), (\ref{e10b}).
\item[4.]
Evaluation of the quantum part
$\widehat{\Delta}^{cor}_{[\overline{0}]}$
of the correlation function,
involving the result for $\widehat{\Delta}^\ra$ of point 3.
The solution for 
$\widehat{\Delta}^{cor}_{[\overline{0}]}$ and the corresponding gluon
phase-space-density ${\tt G}_{[\overline{0}]}$ is given by
(\ref{e22a}) and (\ref{e23}),
respectively.
\item[5.]
Evaluation of the mean-field part
$\delta\widehat{\Delta}^{cor}_{[\overline{A}]}$,
involving the solution for
$\widehat{\Delta}^{cor}_{[\overline{0}]}$ of point 4.
The result for 
$\delta\widehat{\Delta}^{cor}_{[\overline{A}]}$
and the correction to the gluon density $\delta{\tt G}_{[\overline{A}]}$ 
is given by (\ref{e34a}) and (\ref{e34}), respectively.
\item[6.]
Construction of the hard gluon current $\widehat{j}$ from the solution 
$\widehat{\Delta}^{cor}=
\widehat{\Delta}^{cor}_{[\overline{0}]}+
\delta\widehat{\Delta}^{cor}_{[\overline{A}]}$
of points 4. and 5., with explcit form given by the formulae (\ref{j0})-(\ref{j3}).
Approximate evaluation of the soft mean field $\overline{A}$ from the 
classical Yang-Mills equation
(\ref{YME6b}) with resulting Weizs\"acker-Williams form (\ref{A1}).
\end{description}
With this procedure,  the  original master equations
(\ref{X1})-(\ref{YME6})
are solved in first iteration to order $g^2 (1+g\overline{A})$.
One could in principle now repeat this cycle, with the first order
solutions replacing the zeroth-order forms as input.
This, however, is another story. The story of this paper ends here. Cheers.

\newpage

\noindent {\bf ACKNOWLEDGEMENTS}
\medskip

Most sincere thanks to Edmond Iancu (for inspiration
and teaching of tricky details), to Berndt M\"uller (for his
guidance throughout this work and for careful proof-reading),
and to Rob Pisarski (for lots of confusion, clearification and enthusiasm).
Moreover I thank the INT at the University of Washington, Seattle,
for its hospitality during my time at the workshop on
`Ultrarelativistic nuclei: from structure functions to the quark-gluon plasma',
which was organized with fun by Larry McLerran and Miklos Gyulassy.

This work was supported in part by the D.O.E under contract no.
DE-AC02-76H00016.
\bigskip

\newpage
\appendix

\section{Conventions and notation}
\label{sec:appa}

Throughout the paper pure $SU(3)_c$ Yang-Mills theory for $N_c=3$ colors
is considered,
in the absence of quark degrees of freedom, with the
{\it gauge field tensor}
\begin{equation}
{\cal F}_{\mu\nu}^a\;=\;
\partial^x_\mu {\cal A}_{\nu}^a \;-\; \partial^x_\nu {\cal A}_{\mu}^a 
\;+\; g\, f^{abc} \,{\cal A}_\mu^b {\cal A}_\nu^c
\;,
\label{F}
\end{equation}
and the classical  {\it Yang-Mills Lagrangian}
\begin{eqnarray}
{\cal L}_{YM}(x) \;=\;
-\;\frac{1}{4} {\cal F}_{\mu\nu}^a(x) {\cal F}^{\mu\nu,\,a}(x)
& =&
-\;\frac{1}{2} 
\left\{
\frac{}{}
\left(\partial^x_\mu {\cal A}_{\nu}^a\right)^2 \;-\;
\left(\partial^x_\mu {\cal A}_{\nu}^a\right) 
\left(\partial_x^\nu {\cal A}^{\nu,\,a}\right)
\right\} (x)
\nonumber \\
& &
\;+\;
g\, f_{abc} 
\left\{
\frac{}{}
\left(\partial^x_\mu {\cal A}_\nu^a\right) 
{\cal A}^{\mu,\,b} {\cal A}^{\nu,\,c}
\right\}(x)
\;+\;
g^2\, f^{abc} f^{ab'c'}
\left\{
\frac{}{}
{\cal A}_\mu^b {\cal A}_\nu^c {\cal A}^{\mu,\,b'} {\cal A}^{\nu,\,c'}
\right\}(x)
\label{LYM}
\end{eqnarray}
Because only gluonic degrees of freedom are considered, only the 
{\it fundamental representation} of color space
is relevant, with the color indices $a, b, \ldots$ running from
$1$ to $N_c$.
The generators of the $SU(3)$ color group are the
traceless hermitian matrices $T_a$ with the structure constants $f^{abc}$,
as matrix elements, satisfying
\begin{equation}
\mbox{Tr}\left(T^a, T^b\right) \;=\; N_c\;\delta^{ab}
\;,\;\;\;\;\;\;\;\;
\left[T^a,T^b\right] \;= \; + i\, f^{abc} T_c
\;,\;\;\;\;\;\;\;\;
- i\,f^{abc}\;=\; \left(T^a\right)^{bc}
\;.
\label{Tab}
\end{equation}
In compact notation,
\begin{equation}
{\cal A}_\mu \;\equiv\; T^a\,{\cal A}_\mu^a
\;,\;\;\;\;\;\;\;\;\;\;\;\;
{\cal F}_{\mu\nu} \;\equiv\; T^a\,{\cal F}_{\mu\nu}^a
\;=\;
\partial^x_\mu {\cal A}_{\nu} \;-\; \partial^x_\nu {\cal A}_{\mu} 
\;-\;i\, g\, \left[{\cal A}_\mu , \, {\cal A}_\nu \right]
\;=\; \frac{1}{(-i g)} \,\left[ D_\mu\,, D_\nu \right]
\;,
\label{AF}
\end{equation}
where 
$\partial_\mu^x \equiv \partial /\partial x^\mu$ 
acting on the space-time argument $x^\mu = (x^0,\vec{x})$.
The {\it covariant derivative},  denoted by $D_\mu$, is defined as
\begin{equation}
D_\mu(x) \;\equiv\; \partial_\mu^x \,-\,ig\,T^a\,{\cal A}_\mu^a(x)
\;=\; \partial_\mu^x \,-\,ig\,{\cal A}_\mu(x)
\;,
\end{equation}
and its adjoint is
$D_\mu^\dagger(y)\equiv \partial_\mu^y \,+\,ig\,{\cal A}_\mu(y)$.
In components, using (\ref{Tab}),
\begin{equation}
D_\mu^{ab}(x) \;=\; \delta^{ab} \partial^x_\mu \,-\, 
g\,f^{abc}\,{\cal A}_\mu^c(x)
\;,
\label{covder}
\end{equation}
with the color coupling strength $g$ being related to the 
strong coupling $\alpha_s =g^2/(4\pi)$.
In general, for any color matrix $O$ with matrix elements $O_{ab}(x)$,
the action of the covariant derivative is 
\begin{equation}
\left[ D_\mu\,, O(x) \right]\;\equiv\;
\partial^x_\mu O(x) \;-\;i\,g\, \left[ {\cal A}_\mu(x), O(x) \right]
\;,
\end{equation}
and in particular, the covariant derivative of the field strength tensor reads
$
\left[ D_\mu\,, {\cal F}_{\nu\lambda} \right] = \partial^x_\mu {\cal F}_{\nu\lambda}
-ig \left[ {\cal A}_\mu, {\cal F}_{\nu\lambda}\right]
$.

The convention for placing indices and labels are such that
{\it color indices} $a, b, \ldots$ are always written as superscripts, whereas
all other labels may be subscripts or superscripts. In particular,
the {\it Lorentz vector indices} $\mu,\nu,\ldots$ may be raised
or lowered according to the Minkowski metric $g_{\mu\nu} = \mbox{diag}(1,-1,-1,-1)$, 
and the usual convention for summation over repeated indices is understood.
Finally, some shorthand notations are employed, namely
\begin{eqnarray}
A\cdot B &\equiv& A_\mu g^{\mu \nu} B_\nu
\;,\;\;\;\;\;\;\;\;\;\;\;\;\;\;\;\;
\;\;\;\;\;\;\;\;\;\;\;\;\;\;\;\;
K \cdot \left( A \,B \right) \;\equiv \; 
K_{\mu\nu} \,A^\mu B^\nu
\\
A\circ B &\equiv&  \int_P d^4x \,A(x)\cdot B(x)
\;,\;\;\;\;\;\;\;\;\;\;\;\;\;\;\;\;
K \circ \left( A \, B \right) \;\equiv \;
\int_P d^4x d^4 y \, K(x,y)\cdot\left(\frac{}{} A(x)\, B(y)\right)
\;,
\label{conv5}
\end{eqnarray}
where the label $P$ under the integral sign refers to the integration
of the time components $x^0$ ($y^0$) along a closed path in the complex
time plane.
\bigskip

\section{Basics of the Closed-Time-Path formalism}
\label{sec:appb}
\medskip

\subsection{The {\it in-in} amplitude  $Z_P$}

The key problem  in this paper is 
to describe the dynamical development
of a multi-particle system (here gluons), that evolves from an initially
prepared quantum state, e.g., produced
by a  high-energy particle collision.
There is a  crucial difference between the evolution of the system in
{\it in vacuum}  (which means, free space in the absence of surrounding matter), 
and {\it in medium} (which could be either an external
matter distribution, or an internal particle density induced by
the gluons themselves). 
As illustrated by Fig. 2 in the introduction, this difference arises from
the interactions, and hence,
non-trivial statistical correlations between the gluons
and the particles of the environment.

In the case of {\it vacuum}, the usual quantum field theory desrcibes
the time evolution of the system by
the vacuum-vacuum transition amplitude, also called {\it in-out} amplitude
(see Fig. 2a, left panel).  
That is, one 
starts at time $t_0$ in the remote past with appropriate asymptotic
{\it in}-state and evolves it to $t_\infty$ in the asymptotic future, 
by means of the time evolution operator $U(t_\infty,t_0)$.
Multiplication with the hermitian conjugate counterpart, which corresponds
to a backward evolution from $t_\infty$ to $t_0$ under the action of
$U^\dagger(t_0,t_\infty)$.
The resulting {\it in-out} amplitude may be interpreted 
as the sum over all $n$-point Green functions
for space-time points
along a path in the complex $t$ plane, 
exclusively on the upper (lower) branch for the forward (backward) evolution.
In vacuum there is no correlation between the two time branches, 
and so, 
for instance, the 2-point Greenfunctions are the usual time-ordered Feynman 
$\Delta^{F}$  (anti-time-ordered $\Delta^{\overline{F}}$) propagator
(see Fig. 2a, right panel).  
Because $U^\dagger(t_0,t_\infty) = U(t_\infty,t_0)$, one has 
$\Delta^{F}(t_1,t_2) = - \Delta^{\overline{F}}(t_2,t_1)$.

In the case of a {\it medium}, the above concept fails, because of
the a-priori-presence of medium particles described by the density 
matrix $\hat{\rho}(t_0)$.
Instead one has to construct a
generalized transition amplitude, called {\it in-in} amplitude, which
accounts for the non-trivial initial state at $t_0$ embodied
in the density matrix $\hat{\rho}(t_0)$, and evolves
the system in the presence of the medium from $t_0$
to $t_\infty$ in the future, 
by means of the time evolution operator $U(t_0,t_\infty)$
(see Fig. 2b, left panel).  
Because now
$U^\dagger(t_0,t_\infty) \ne U(t_\infty,t_0) \hat{\rho}(t_0)$,
forward and backward contributions are not merely conjugate to each other,
but interfere, giving rise to statistical correlations between
upper and lower time branch of the contour in the $t$-plane.
As a consequence the space of Green functions is enlarged by non-time-ordered
correlation functions. For example, the 2-point functions are now
$\Delta^F$, $\Delta^{\overline{F}}$ plus the new functions $\Delta^{<}$ and 
$\Delta^{>}$
(see Fig. 2b, right panel).  
\smallskip

The fundamental quantity of interest is
the {\it in-in amplitude} $Z_P$ 
for the evolution of
the initial quantum state $\vert in\rangle$ forward in time
into the remote future, starting from a specified initial state
that could be either the vacuum or a medium.
Within the CTP formalism the amplitude $Z_P$ can be evaluated by time integration
over the {\it closed-time-path} $P$ in the complex $t$-plane.
As illustrated in Fig. 6,  
this closed contour extends from $t=t_0$ to $t=t_\infty$ in the remote future
along the positive ($+$) branch and back to $t=t_0$ along the negative ($-$) branch.
where any point on the $+$ branch is understood at an earlier instant
than any point on the $-$ branch:
\begin{equation}
Z_P[{\cal J},\hat{\rho}] 
\;\equiv\;
Z_P[{\cal J}^+,{\cal J}^-,\,\hat{\rho}] 
\;=\;
\mbox{Tr}\left\{
U^\dagger_{{\cal J}^-}(t_0 ,t)\, U(t,t_0)_{{\cal J}^+}\,\; \hat{\rho}(t_0)\right\}
\;,
\label{Z11}
\end{equation}
where the  $\hat{\rho}(t_0)$ is the initial state density matrix, and
$U$ and $U^\dagger$ are the time evolution operator
and its adjoint, 
\begin{eqnarray}
U_{{\cal J}^+}(t,t_0) &=& T \,\exp\left\{ -i 
\left[\int_{t_0}^{t} dt'd^3x'
\,{\cal J}^+(x')\cdot {\cal A}^+(x')
\right]
\right\}
\nonumber \\
U_{{\cal J}^+}^\dagger(t_0,t) &=& T^\dagger \,\exp\left\{ +i 
\left[\int_{t_0}^{t} dt'd^3x'
\,{\cal J}^-(x')\cdot {\cal A}^-(x')
\right]
\right\}
\label{U}
\;,
\end{eqnarray}
with
$T$ ($T^\dagger$) denoting the time (anti-time) ordering operator.
Note that
$J_+$ ($J_-$) is the source along the positive (negative) branch 
of the closed-time-path of Fig. 6, and 
in general $J_+\ne J_-$, so that $Z_P$ depends on two different sources. If these
are set equal, one has
$Z_P(J,J,\rho)=\mbox{Tr} \hat{\rho}$, which is equal to unity in the
absence of initial correlations, being a statement of unitarity.
\smallskip

$Z_P$ contains the full information
about the development of the initial state via the creation, interaction, and
destruction of quanta, through the agency of
the sources: the quanta are initially created (e.g., by particle collision),
they evolve by further creation and annihilation (real and virtual
emission/absorption as well as scattering), and are finally destroyed (e.g., by
detection in a calorimeter).
Both the act of initial creation and final destruction represent the external sources
${\cal J}$ in the sense of a probing aparatus,
whereas the intermediate dynamics is governed by the underlying
quantum theory. 
Hence, in order to describe the time evolution of the initially prepared quantum system,
to the final detected state, the knowledge of $Z_P$ allows to extract objectively the
self-contained development of the system, when the external influence removed
(i.e. the sources are switched off).
\bigskip

\begin{figure}
\epsfxsize=400pt
\centerline{ \epsfbox{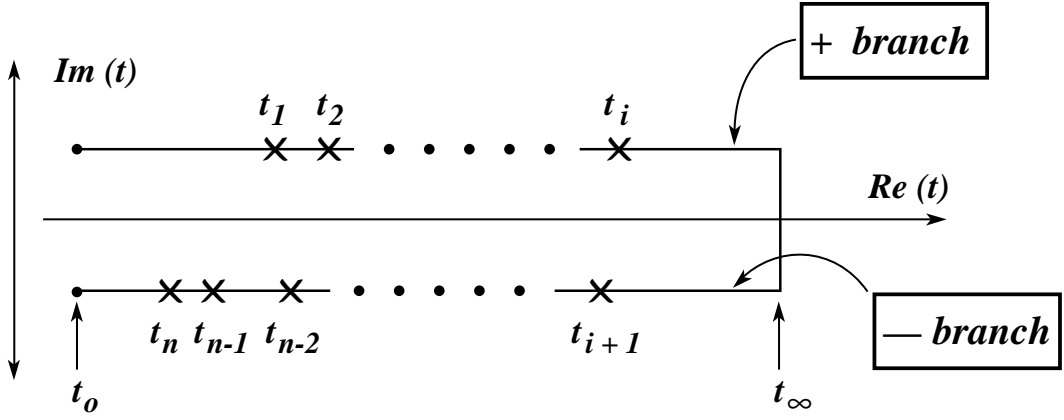} }
\bigskip
\bigskip
\caption{
The close-time-path in the complex $t$-plane 
for the evolution of operator expectation
values in an arbitrary initial state. 
Any point on the  forward, positive branch $t_0\rightarrow t_\infty$ 
is understood at an earlier instant than any point on the  
backward, negative branch $t_\infty\rightarrow t_0$. 
\label{fig:fig6}
}
\end{figure}
\bigskip

The functional $Z_P$ can be represented as a path integral by 
imposing boundary conditions
in terms of complete sets of eigenstates of
the gauge fields ${\cal A}_\mu$ at initial time $t=t_0$,
\begin{equation}
{\cal A}(t_0,\vec x) \,|\,{\cal A}^+ (t_0)\,\rangle \;=\;
{\cal A}^+(\vec x) \,|\,{\cal A}^+ (t_0)\,\rangle
\;,\;\;\;\;\;\;\;\;\;\;\;\;\;
{\cal A}(t_0,\vec x) \,|\,{\cal A}^-(t_0)\,\rangle \;=\;
{\cal A}^-(\vec x) \,|\,{\cal A}^- (t_0)\,\rangle
\label{bc1}
\;,
\end{equation}
and in the remote future at $t=t_{\infty}$,
\begin{equation}
{\cal A}'(t_\infty,\vec x) \,|\,{\cal A}'(t_{\infty})\,\rangle 
\;=\;
{\cal A}'(\vec x) \,|\,{\cal A}' (t_\infty)\,\rangle
\;.
\label{bc2}
\end{equation}
Then, making use of the  completeness of the eigenstates, 
one obtains from (\ref{Z11}) the following functional integral representation for $Z_P$,
\begin{eqnarray}	
Z_P[{\cal J}^+,{\cal J}^-,\hat{\rho}] &=&
\int \,{\cal D} {\cal A}^+ {\cal D} {\cal A}^-
{\cal D} {\cal A}'
\;
\langle \,{\cal A}^-(t_0)\, | \,U^\dagger_{{\cal J}^-}(t_0,t_\infty)\,|\,{\cal A} (t_\infty)\,\rangle
\nonumber \\
& & \;\;\;\;\;\;\;\;\;
\times \;
\langle \,{\cal A}(t_\infty)\, | \,U_{{\cal J}^+}(t_\infty,t_0)\,|\,{\cal A}^+ (t_0)\,\rangle
\;\times \;
\langle\,{\cal A}^+(t_0)\, | \,\hat{\rho} \,|\,{\cal A}^+(t_0)\,\rangle
\;.
\end{eqnarray}
The first two amplitudes are  the  transition amplitudes
in the presence of ${\cal J}^+$ and ${\cal J}^-$, whereas the density matrix element incorporates
the initial state correlations at $t_0$ at the endpoints of the closed-time path $P$.
Hence,
one obtains the path integral representation for $Z_P$ in analogy
to usual field theory \cite{calzetta,jordan}
\begin{equation}
Z_P[{\cal J}^+,{\cal J}^-,\hat{\rho}]
\;=\;
\int \,{\cal D} {\cal A}^+{\cal D} {\cal A}^-
\;\, \exp \left[i \left( \frac{}{} I[{\cal A}^+] \,+\, {\cal J}^+\circ{\cal A}^+
\right)
\;-\;  
i \left( \frac{}{} I^\ast[{\cal A}^-] \,+\, {\cal J}^-\circ{\cal A}^-
\right)
\right]
\;{\cal M}[\hat{\rho}]
\label{Z14}
\;,
\end{equation}
where
\begin{equation}
{\cal M}(\hat{\rho}) 
\;=\;
\langle \,{\cal A}^+ (t_0) \vert \,{\hat \rho}
\,\vert\,{\cal A}^-(t_0)\,\rangle
\;.
\end{equation}
The generalized classical  action $I[{\cal A}]$ 
accounts for all four field orderings on the closed-time path $P$:
\begin{equation}
I[{\cal A}] \;\equiv\; 
I[{\cal A}^+] \,- I^\ast[{\cal A}^-] \;=\;
I^{(0)}[u_{\alpha\beta} {\cal A}^\alpha_\mu {\cal A}^\beta_\nu] 
\;+\;
I^{(1)}[ g \,v_{\alpha\beta\gamma}(\partial_\mu {\cal A}_{\nu}^\alpha) 
{\cal A}_\mu^\beta {\cal A}_\nu^\gamma ]
\;+\;
I^{(2)}[g^2\,w_{\alpha\beta\gamma\delta} 
{\cal A}_{\mu}^\alpha {\cal A}_\nu^\beta {\cal A}_\mu^\gamma {\cal A}_\nu^\delta ]
\;,
\label{I14}
\end{equation}
where the correspondance with the terms with the ones of (\ref{LYM}) is obvious
(the color indices are suppressed here), and where
$\alpha, \beta,\gamma,\delta = +, -$,
\begin{equation}
u_{\alpha\beta}\;=\; u^{\alpha\beta} \;=\; \mbox{diag} (1, -1)
\;\;\;\;\;\;\;\;\;
v_{\alpha\beta\gamma}\;=\; \delta_{\alpha\beta} u_{\beta\gamma} 
\;\;,\;\;\;\;\;\;
w_{\alpha\beta\gamma\delta}\;=\; \mbox{sign}(\alpha)\, \delta_{\alpha\beta} \delta_{\beta\gamma} 
\delta_{\gamma\delta}
\;,
\label{sigma2}
\end{equation}
with the usual summation convention over repeated greek indices $\alpha, \beta,..$.
Eqs. (\ref{Z14})-(\ref{I14}) represent the detailed version of the
compact form (\ref{Z2}) used in Sec. 2 as the starting point, except
for the Fadeev-Popov determinant and the gauge fixing constraint, 
which is omitted here, and which is addressed in Appendix C.
\medskip

\subsection{The density matrix and the initial state}

Turning to the properties of the initial state incorporated
in  the functional ${\cal M}(\hat{\rho})$ 
with
the density matrix $\hat{\rho}(t_0)$, we
denote by $t_0$ the initial point of time from which on the
evolution of the multi-gluon state is followed, and assume that
all the dynamics prior to $t_0$ is contained in the form 
of the {\it initial state} 
\begin{equation}
|\;in\;\rangle \;\equiv\;
|\;{\cal A}(t_0)\;\rangle
\;=\; \prod_{\mu, a}\,|\,{\cal A}_\mu^a (t_0) \,\rangle
\;.
\end{equation}
The initial state at $t=t_0$ can be constructed by expanding
the gauge field operator ${\cal A}$ in the Heisenberg representation
in terms of a Fock basis of non-interacting single-gluon states, 
the {\it in}-basis,
\begin{equation}
{\cal A}_\mu^a(t_0,\vec{x})\;=\;
\int \frac{d^3k}{(2\pi)^3}\,\theta(k^0)\,(2\pi)\delta(k^2)\;
\sum_s
\left(e^{-ik\cdot x}\,\hat{c}_\mu^a(k,s)\;+\;e^{ik\cdot x}\,\hat{c}_\mu^{a\;\dagger}(k,s)\right)
\end{equation}
so that a particular Fock state is given by (supressing color and Lorentz indices)
\begin{equation}
|\,n^{(1)}, n^{(2)}, \ldots , n^{(\infty)}\,\rangle 
\;=\;
\prod_i\,
\frac{1}{\sqrt{n^{(i)}!}}\;
\left(\frac{}{}\hat{c}^\dagger (k_i,s_i)\right)^{n^{(i)}}
\,\;|\;0\;\rangle
\;.
\end{equation}
Here
$\hat{c}^\dagger$ ($\hat{c}$) 
is the creation (annihilation) operator for a single-gluon state
with definite 4-momentum $k$ and spin $s$, satisfying
$\hat{c}^\dagger\,|\,0\,\rangle=1$
($\hat{c}\,|\,0\,\rangle=0$).
and the $n^{(i)}$ are the occupation numbers of the different gluon states,
$n^{(i)}\equiv \langle\,n^{(i)}| \hat{c}(k_i,s_i) \hat{c}^\dagger (k_i,s_i)|\,n^{(i)}\, \rangle$.
Finally,  $\vert\, 0\, \rangle$ denotes the vacuum state or 
a ground state different from vacuum (e.g. a hadron).
Thus, a general multi-gluon state $|\phi\rangle$ at time $t_0$ is given by a superposition
of such states,
\begin{equation}
|\,{\cal A} (t_0) \,\rangle\;=\;
\prod_{\mu, a}\, \sum_{n^{(i)}} \,
C_\mu^a(n^{(1)}, n^{(2)}, \ldots , n^{(\infty)})\;
|\,n^{(1)}, n^{(2)}, \ldots , n^{(\infty)}\,\rangle 
\;,
\end{equation}
with real-valued coefficients $C$.
Alternatively,
the initial state of the system at  $t_0$ can be characterized by 
the {\it density matrix},
\begin{equation}
\hat{\rho}(t_0) \;\equiv\; |\, {\cal A}(t_0)\,\rangle \,\langle \,{\cal A} (t_0)\,|
\;\;\;\;\;\;\;\;\;
\left(\,\hat{\rho}_0\,\right)_{ij} \;\equiv\; 
\langle \,n^{(i)}\,|\, \hat{\rho}(t_0) \,|\,n^{(j)} \,\rangle
\label{rho}
\;.
\end{equation}
For instance, 
the case of empty vacuum corresponds to a diagonal density matrix
$\hat{\rho}(t_0)=|0\rangle \langle 0|$ with
$\left(\,\hat{\rho}_0\,\right)_{ij}\propto \delta_{ij}$,
whereas a general density matrix that describes any form
of a single-particle density distribution at $t_0$ is
\begin{equation}
\hat{\rho}(t_0)\;=\; N\;
\exp\left[ \sum_s\,\int_\Omega d^3x \int \frac{d^3 k}{(2\pi)^3 \,2 k_0} 
\,\theta(k_0) \,F(t_0,\vec x,k)\; \hat{c}_\mu^{a\;\dagger} (k,s) \hat{c}_\mu^a (k,s) 
\;\right]
\;,
\label{rho1}
\end{equation}
where $\Omega$ denotes the hypersurface of the initial values and $F$ is
a $c$-number function related to the single-particle phase-space density of
gluons around $\vec x+d\vec{x}$ with four-momentum within $k^\mu+dk^0d\vec{k}$, 
and $N$ a normalization factor.
The form (\ref{rho1}) describes  a large class of intreresting
non-equilibrium systems \cite{chou}, and contains as a special case
the thermal equilibrium distribution, namely when  
$t\rightarrow -i /T$ and
$F(t_0,\vec x,k)  \rightarrow k_0\,\delta(k^2) \,T^{-1}$, so that
$\hat{\rho}(T_0)\rightarrow  N\;\exp\left[ - \hat{\cal H}_{YM}/T\right]$.
\medskip

\subsection{Perturbation theory and Feynman rules}

The convenient feature of the CTP  formalism 
is that it is formally completely analogous to standard quantum field theory,
except for the fact that the fields have contributions from both
time branches.
In particular, one obtains as in usual field theory, from
the path-integral representation (\ref{Z3}) 
the  $n$-point Green functions $G^{(n)}(x_1,\ldots, x_n)$,
which however now include all correlations between
points on either positive and negative time branches,
\begin{equation}
G^{(n)}_{\alpha_1\alpha_2 \ldots\,\alpha_n}(x_1,\ldots, x_n)
\;\;=\;\;
\left.
\frac{1}{Z_P[0]}\;
\frac{\delta}{i \,\delta {\cal K}^{(n)}}
Z_P[{\cal K}]\right|_{{\cal K}=0}
\;\;, \;\;\;\;\;\;\;\;\;\;\;\;
\alpha_i \;= \pm
\;,
\label{Green0}
\end{equation}
depending on whether the space-time points
$x_i$ lie on the $+$ or $-$ time branch.
One can then  construct
a perturbative expansion of the non-equilibrium Green functions in terms
of modified Feynman rules (as compared to standard field theory)
\cite{lifshitz,chou,rammer}.
\begin{description}
\item[(i)]
All local 1-point functions $G^{(1)}_\alpha(x)$, such as the 
gauge-field or the color current, are `vectors' with 2 components,
\begin{equation}
{\cal A}(x) \;\equiv\;
\left( \begin{array}{c}
{\cal A}^{+} \\ {\cal A}^{-}
\end{array}\right) 
\;\;\;\;\;\;\;\;\;\;\;\;\;\;\;\;
{\cal J}(x) \;\equiv\;
\left( \begin{array}{c}
{\cal J}^{+} \\ {\cal J}^{-}
\end{array}\right) 
\end{equation}
Similarly, 
all 2-point functions
$G^{(2)}_{\alpha\beta}(x,y)$, such as the 
as the  gluon propagator $i\Delta_{\mu\nu}$ and the polarization tensor 
$\Pi_{\mu\nu}$, are  2$\times$2 matrices with components,
\begin{equation}
\Delta(x_1,x_2) \;\equiv\;
\left(\begin{array}{cc}
\Delta^{++} & \;\Delta^{+-} \\
\Delta^{-+} & \;\Delta^{--}
\end{array} \right) 
\;\;\;\;\;\;\;\;\;\;\;\;\;\;\;\;
\Pi(x_1,x_2) \;\equiv\;
\left(\begin{array}{cc}
\Pi^{++} &  \;\Pi^{+-} \\
\Pi^{-+} &  \;\Pi^{--}
\end{array} \right) 
\;.
\nonumber
\end{equation}
Explicitely, the components of the propagator are 
\begin{eqnarray}
\Delta_{\mu\nu}^{F}(x,y)\;&\equiv&
\Delta_{\mu\nu}^{++}(x,y)\;=\;
-i\,
\langle \;T\, {\cal A}_\mu^+(x)\, {\cal A}_\nu^+(y) \;\rangle
\;\;\;\;\;\;\;\;\;\;\;\;
\Delta_{\mu\nu}^{<}(x,y) \;\;\equiv\;
\Delta_{\mu\nu}^{+-}(x,y)\;=\;
-i\,
\langle \; {\cal A}_\nu^+(y)\, {\cal A}_\mu^-(x) \;\rangle
\nonumber \\
\Delta_{\mu\nu}^{>}(x,y)\;&\equiv&
\Delta_{\mu\nu}^{-+}(x,y)
\;=\;
-i\,
\langle \; {\cal A}_\mu^-(x)\, {\cal A}_\nu^+(y) \;\rangle
\;\;\;\;\;\;\;\;\;\;\;\;\;\;
\Delta_{\mu\nu}^{\overline{F}}(x,y) \;\;\equiv\;
\Delta_{\mu\nu}^{--}(x,y)\;=\;
-i\,
\langle \;\overline{T}\, {\cal A}_\mu^-(x)\, {\cal A}_\nu^-(y) \;\rangle
\label{D22}
\;,
\end{eqnarray}
where
$\Delta^F$ is the usual time-ordered Feynman propagator, $\Delta^{\overline{F}}$
is the corresponding anti-time-ordered propagator, and $\Delta^>$ ($\Delta^<$) is
the unordered correlation function for $x_0 > y_0$ ($x_0 < y_0$).
In compact  notation,
\begin{equation}
\Delta_{\mu\nu}(x,y)\;=\;
-i\,\langle \,T_P {\cal A}(x) {\cal A}(y) (y) \,\rangle
\; ,
\end{equation}
where
the generalized time-ordering operator $T_P$ is defined as
\begin{equation}
 T_P\,A(x)B(y)
\;:=\;
\theta_P (x_0,y_0)\, A(x) B(y) \;+ \;\theta_P(y_0,x_0) \,B(y) A(x)
\label{gto1}
\;, 
\end{equation}
with the $\theta_P$-function  defined as 
\begin{equation}
\theta_P(x_0,y_0)\;=\;
\left\{
\begin{array}{ll}
1 & \;\; \mbox {if $x_0$ {\it succeeds} $y_0$ on the contour $P$}
\\
0 & \;\;  \mbox{if $x_0$ {\it precedes} $y_0$ on  the contour $P$}
\end{array}
\right. 
\;.
\label{gto2}
\end{equation}
Higher order products $A(x)B(y)C(z)\ldots$ are ordered analogously.
Finally, 
the generalized $\delta_P$-function  on the closed-time path $P$ is defined as:
\begin{equation}
\delta^4_P(x,y) \;\,:=\;\,
\left\{
\begin{array}{ll}
+\delta^4(x-y) & \;\; \mbox {if} \; x_0 \;\mbox{and} \;y_0 \; \mbox{from positive
branch} \\
-\delta^4(x-y) & \;\; \mbox {if} \; x_0 \;\mbox{and} \;y_0 \; \mbox{from negative
branch} \\
0 & \;\; \mbox {otherwise}
\end{array}
\right. 
\;.
\label{gto3}
\end{equation}
\item[(ii)]
The number of elementary vertices is doubled,
because each propagator line of a Feynman diagram can be
either of the four components of the Green functions.
The  interaction  vertices in which 
all the fields are on the $+$ branch are the usual ones,
while the vertices in which the fields are on 
the $-$ branch have the opposite sign. 
On the other hand, combinatoric factors, rules for loop integrals, etc.,
remain  the same as in usual field theory.
Specifically, the
3-gluon and 4-gluon vertices,
\begin{eqnarray}
G^{(3)}_{\alpha\beta\gamma}(x_1,x_2,x_3) &\equiv&
\int_P d^4x\,
G^{(2)}_{\alpha'\alpha}(x_1,x)
G^{(2)}_{\beta'\beta}(x_2,x)
G^{(2)}_{\gamma'\gamma}(x_3,x)
\;\;
\Gamma^{(3)}_{\alpha'\beta'\gamma'}(x) 
\nonumber \\
G^{(4)}_{\alpha\beta\gamma\delta}(x_1,x_2,x_3,x_4) &\equiv&
\int_P d^4x\,
G^{(2)}_{\alpha'\alpha}(x_1,x)
G^{(2)}_{\beta'\beta}(x_2,x)
G^{(2)}_{\gamma'\gamma}(x_3,x)
G^{(2)}_{\delta'\delta}(x_4,x)
\;\;
\Gamma^{(4)}_{\alpha'\beta'\gamma'\delta'}(x) 
\;,
\nonumber
\end{eqnarray}
with $\Gamma^{(3)}(x)$ and $\Gamma^{(4)}(x)$ denoting the elementary, amputated 
vertices (with the external legs removed),
have, for fixed $\alpha, \beta,\gamma, \delta$ two components.
For instance, as in Fig. 7, for the external points on the $+$-branch,
\begin{eqnarray}
\mbox{for $\alpha=\beta=\gamma=\,+$}: 
& &
\;\;\;\;\;\;
\Gamma^{(3)}_{\alpha'\beta'\gamma'}(x) 
\;\;=\;\; \left(\Gamma^{(3)}_{+++}, \;-\, \Gamma^{(3)}_{---} \right) 
\nonumber \\
\mbox{for $\alpha=\beta=\gamma=\delta=\,+$} : & & 
\;\;\;\;
\Gamma^{(4)}_{\alpha'\beta'\gamma'\delta'}(x) 
\;=\; \left(\Gamma^{(4)}_{++++}, -\, \Gamma^{(4)}_{----} \right)
\nonumber 
\end{eqnarray}
\end{description}
\bigskip

\begin{figure}
\epsfxsize=400pt
\centerline{ \epsfbox{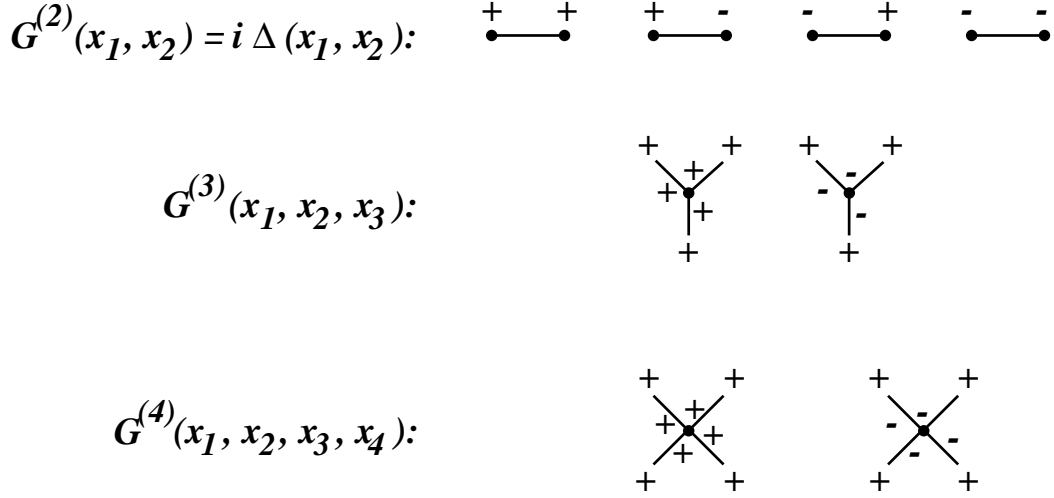} }
\bigskip
\bigskip
\caption{
Example for  the appearance of additional contributions to the $n$-point
functions $G^{(n)}$ for  the propagator
$G^{(2)}$, the 3-vertex $G^{(3)}$, and the 4-vertex $G^{(4)}$ 
In usual quantum field theory 
referring to free space or `vacuum', only the $+$-graphs are non-zero.
In the CTP formalism, accounting for the presence of surrounding 
matter or `medium', new diagrams arise that correspond to statistical 
correlations
between the field living on the $+$ and $-$ time branches of Fig. 6.
\label{fig:fig7} 
}
\end{figure}

\section{The {\it in-in} amplitude $Z_P$ for QCD in non-covariant gauges}
\label{sec:appc}

For the case of QCD, 
path-integral representation of
the {\it in-in} amplitude $Z_P$ is obtained along the lines of Appendix B1,
except that one has to extend the generic
formula (\ref{Z14}) to account for eliminating the
spurious gauge degrees of
freedom by the usual Fadeev-Popov procedure \cite{FP}.
The gauge theory version of (\ref{Z14}) for the class of non-covariant gauges 
(\ref{gauge100}) reads therefore:
\begin{equation}
Z_P[{\cal J},\hat{\rho}] \;=\;
{\cal N}\;\int \,{\cal D}{\cal A} \; \mbox{det}{\cal F}\;\delta \left( f[{\cal A}]\right)
\;\,
\exp \left\{i \left( \frac{}{} I \left[ {\cal A}, {\cal J}\right]
\right)\right\}
\;\; {\cal M}(\hat{\rho})
\label{Z2}
\;,
\end{equation}
where ${\cal A} = ({\cal A}^+,{\cal A}^-)$ and  
${\cal J} = ({\cal J}^+,{\cal J}^-)$ 
have two components, living on the $+$ and $-$ time branches of Fig. 2.
Physical expectation values are defined as 
functional averages over 
$T_P$-ordered 
products of $n$ field operators ($n\ge 1$), weighted by $Z_P$,
\begin{equation}
\langle\; {\cal O}_1(x_1)\ldots {\cal O}_n(x_n)\;\rangle_P
\;\equiv\;
\frac{1}{Z_P[0,\hat{\rho}]}
\;\int \,{\cal D}{\cal A} \; \mbox{det}{\cal F}\;\delta \left( f[{\cal A}]\right)
\exp \left\{i \left( \frac{}{} I \left[ {\cal A}, {\cal J}\right]
\right)\right\}
\,\, {\cal M}(\hat{\rho})
\;\;\;T_P\left[{\cal O}_1(x_1)\ldots {\cal O}_n(x_n)\right]
\label{Z2a}
\;.
\end{equation}
\medskip

\noindent
The structure of the functional $Z_P$ in (\ref{Z2}) and (\ref{Z2a}) 
is the following:

\begin{description}
\item[(i)]
The functional integral (with normalization ${\cal N}$) 
is over all gauge field configurations
with measure ${\cal D}{\cal A} \equiv \prod_{\mu,a} {\cal D}{\cal A}_\mu^a$,
subject to the condition of gauge fixing, here  for the 
{\it class of non-covariant gauges} defined by
\begin{equation}
f^a[{\cal A}] \;:=\; n\cdot {\cal A}^a(x) \;-\; B^a(x)
\;\;\;\;\;\;\;\;\;\;\;
\Longrightarrow
\;\;\;\;\;\;\;\;\;\;
\langle \;n^\mu\,{\cal A}_\mu^a(x)\;\rangle \;=\;0
\;, \;\;\;\;\;\;\;\;\;\;\;\;\; 
n^\mu\,\equiv \,\frac{n^\mu}{\sqrt{\vert n^2\vert }}
\label{gauge1}
\;,
\end{equation}
where
$n^\mu$ is a constant 4-vector, being either space-like ($n^2 < 0$), 
time-like ($n^2 > 0$), or light-like ($n^2=0$).
With this choice of gauge class the {\it local gauge constraint}
on the fields ${\cal A}^a_\mu(x)$ in
the path-integral (\ref{Z2}) becomes,
\begin{equation}
\mbox{det}{\cal F}\;\delta \left( n\cdot {\cal A}^a-B^a\right)
\;=\; 
\mbox{const}\,\times\,
\exp\left\{i\,
I_{GF}\left[n\cdot{\cal A}\right]
\right\}
\end{equation}
\begin{equation}
I_{GF}\left[n\cdot{\cal A}\right]
\;=\;
- \frac{1}{2\alpha}
\int_P d^4x \, \left[n\cdot {\cal A}^a(x)\right]^2
\;,
\label{gauge2}
\end{equation}
where $\mbox{det}{\cal F}$ is the Fadeev-Popov determinant (which in the case
of the non-covariant gauges turns out to be a constant factor, c.f. Appendix C), 
and where $\delta (n\cdot{\cal A}) \equiv \prod_{a}\delta (n\cdot{\cal A}^a)$.
The right side translates this constraint into a
the {\it gauge fixing} functional $I_{GF}$.
The particular choice of 
the vector $n^\mu$ and of 
the real-valued parameter $\alpha$ is dictated by
the physics or computational convenience, and distinguishes 
further within the class
of non-covariant gauges \cite{gaugereview,gaugebook}:
\footnote{
The analogy with the class of covariant gauges defined by
$f^a[{\cal A}]:= \partial_x\cdot {\cal A}^a - B^a$, instead of (\ref{gauge1}),
is evident: in place of (\ref{gauge2}), it results in the familiar gauge-fixing functional
$\exp\left\{- i/2\alpha \int_P d^4x \left(\partial\cdot {\cal A}^a\right)^2) \right\}$,
where $\alpha = 1$ gives the {\it Feynman gauge} and $\alpha = 0$ the {\it Landau
gauge}.
}
\begin{eqnarray}
\mbox{homogenous axial gauge}: \;& & \;\;\;\;\; n^2 \;< \; 0 \;\;\;\;\;\;\alpha \;=\;0 
\nonumber \\
\mbox{inhomogenous axial gauge}: \;& & \;\;\;\;\; n^2 \;< \; 0 \;\;\;\;\;\;\alpha \;=\;1 
\nonumber \\
\mbox{temporal axial gauge}: \;& & \;\;\;\;\; n^2 \;> \; 0 \;\;\;\;\;\;\alpha \;=\;0 
\nonumber \\
\mbox{lightcone gauge}: \;& & \;\;\;\;\; n^2 \;= \; 0 \;\;\;\;\;\;\alpha \;=\;0 
\label{gauge1a}
\;.
\end{eqnarray}

\item[(ii)]
The exponential $I$ is the {\it effective classical action} 
with respect to both the $+$ and the $-$ time contour, 
$
I\left[ {\cal A}, {\cal J}\right]
\equiv I\left[ {\cal A}^+, {\cal J}^+\right]
\;-\;
I^\ast\left[ {\cal A}^-, {\cal J}^-\right]
$,
including
the usual Yang-Mills action $I_{YM}=\int d^4x {\cal L}_{YM}$, plus the source
${\cal J}$ coupled to the gauge field ${\cal A}$:
\begin{eqnarray}
I\left[ {\cal A}, {\cal J}\right]
& =&
-\frac{1}{4} \int_P d^4x \,{\cal F}_{\mu\nu}^a(x) {\cal F}^{\mu\nu, \,a}(x) 
\;+\;
\int_P d^4x \,{\cal J}_{\mu}^a(x) {\cal A}^{\mu, \,a}(x) 
\nonumber \\
& &\nonumber \\
&\equiv&
\;\;\;\;\;\;\;\;
I_{YM}\left[{\cal A}\right] \;\;\;+\;\;\; {\cal J}\circ {\cal A}
\label{I1}
\;.
\end{eqnarray}

\item[(iii)]
The form of the initial state at $t=t_0$ as
described by the density matrix $\hat{\rho}$
(an example is given in Appendix B2, eq. (\ref{rho})) is
embodied in the function      
${\cal M}(\hat{\rho})$ which is the density-matrix element of the gauge fields
at initial time $t_0$, 
\begin{equation}
{\cal M}(\hat{\rho}) 
\;=\;
\langle \,{\cal A}^+ (t_0) \vert \,{\hat \rho}
\,\vert\,{\cal A}^-(t_0)\,\rangle
\;\equiv\;\,
\exp\left( i\; {\cal K}[{\cal A}] \right)
\label{K}
\;,
\end{equation}
where ${\cal A}^\pm$ refers to the $+$ and $-$ time branch {\it at} $t_0$,
respectively (c.f. Fig. 2).
The functional ${\cal K}$ may be expanded in a series of
non-local kernels corresponding to multi-point correlations concentrated
at $t=t_0$,
\begin{eqnarray} 
{\cal K}[{\cal A}]
&=&
{\cal K}^{(0)} \;+\;
\int_P d^4x \;{\cal K}^{(1)\;a}_{\;\;\;\,\mu}(x) \;{\cal A}^{\mu, \,a}(x) 
\;+\;
\frac{1}{2}
\int_P d^4xd^4y \;{\cal K}^{(2) \;ab}_{\;\;\;\,\mu\nu}(x,y)\;{\cal A}^{\mu, \,a} (x)
\,{\cal A}^{\nu, \,b}(y) 
\,\ldots
\nonumber
\\
& &
\;\;\;\;\;\;\;\;\;\;\;\;\;\;
\equiv\;
\sum_{n=0}^\infty \;\,\frac{1}{n !} 
\;{\cal K}^{(n)}\circ \left(\frac{}{}{\cal A}(1) \,{\cal A}(2)\, \ldots \,{\cal A}(n)\right)
\label{Kexpansion}
\;.
\end{eqnarray}
Clearly, the sequence of kernels ${\cal K}^{(n)}$ contains as much information
as the original density matrix.
In the special case that $\hat{\rho}$ is diagonal,
the kernels  ${\cal K}^{(n)}=0$ for all $n$, and 
the  usual `vacuum field theory' is recovered.
\end{description}
\medskip

The path-integral representation (\ref{Z2}) can be rewritten in a 
more convenient form:
First,  the gauge-fixing functional $I_{GF}[n\cdot{\cal A}]$ is implemented, 
using (\ref{gauge2}).
Second,   the series representation (\ref{Kexpansion}) is inserted into
the initial state functional ${\cal M}(\hat{\rho})$.
Third, 
${\cal K}^{(0)}$ is absorbed in the overall normalization ${\cal N}$ of $Z_P$ 
(henceforth set to unity),
and  the external source ${\cal J}$ in the 1-point kernel ${\cal K}^{(1)}$:
\begin{equation}
{\cal K}^{(0)}\;:=\; i\, \ln {\cal N}
\;,\;\;\;\;\;\;\;\;\;\;\;\;\;\;
{\cal K}^{(1)} \;:=\; {\cal K}^{(1)}\;+\; {\cal J} 
\label{NJ}
\;.
\end{equation}
Then  (\ref{Z2}) becomes,
\begin{equation}
Z_P[{\cal J},\hat{\rho}] 
\;\;\Longrightarrow\;\;
Z_P[{\cal K}] \;=\;
\int \,{\cal D}{\cal A} \; 
\exp \left\{i \left( \frac{}{} I \left[ {\cal A}, {\cal K}\right]
\right)\right\}
\label{Z3}
\;,
\end{equation}
where, instead of (\ref{I1}),
\begin{equation}
I \left[ {\cal A}, {\cal K}\right]
\;\equiv\;
I_{YM}\left[{\cal A}\right] 
\;+\;
I_{GF}\left[n\cdot{\cal A}\right]
\;+\; {\cal K}^{(1)}\circ {\cal A} \;+\; 
\frac{1}{2}\,{\cal K}^{(2)}\circ \left({\cal A}\,{\cal A}\right)
\;+\;
\frac{1}{6}\,{\cal K}^{(3)}\circ \left({\cal A}\,{\cal A} \,{\cal A}\right)
\;+\;\ldots
\label{I2}
\;.
\end{equation}
\medskip

The  objects of physical relevance are the
$n$-point {\it Green functions} $G^{(n)}$, defined as the
coefficients in a functional expansion of $Z_P$,
\begin{equation}
Z_P[{\cal K}] 
\;=\;
Z_P[0] \;\,\sum_{n=1}^\infty \,\frac{i^n}{n !} 
\int \prod_{i=1}^n d^4x_i \;G^{(n)}(x_1,\ldots, x_n)
\,\;{\cal K}^{(n)}(x_1,\ldots, x_n)
\label{Zexpansion}
\;,
\end{equation}
that is, the $G^{(n)}$ are 
functional averages in the sense of (\ref{Z2a}),
\begin{equation}
G_{\;\;\;\mu_1\ldots \mu_n}^{(n)\; a_1\ldots a_n}(x_1,\ldots, x_n)
\;\;\equiv\;\;
\langle\; {\cal A}_{\mu_1}^{a_1}(x_1)\ldots {\cal A}_{\mu_n}^{a_n}(x_n)\;\rangle_P
\;\;=\;\;
\left.
\frac{1}{Z_P[0]}\;
\frac{\delta}{i \,\delta {\cal K}^{(n)}}
Z_P[{\cal K}]\right|_{{\cal K}=0}
\label{Green1}
\;.
\end{equation}
\smallskip

The practical evaluation of $Z_P$ amounts 
therefore to calculating the $G^{(n)}$ in the expansion (\ref{Zexpansion}) up to
the order of desired accuracy.
For instance, the 1-, 2-, and 3-point Green functions according to (\ref{Green1}) are
\begin{eqnarray}
G_{\;\;\;\;\mu}^{(1)\;a} (x) 
&=&
\langle\; {\cal A}_{\mu}^{a}(x)\;\rangle_P
\nonumber \\
G_{\;\;\;\;\mu\nu}^{(2)\;a b}(x,y)
&=&
\langle\; {\cal A}_{\mu}^{a}(x){\cal A}_{\nu}^{b}(y)\;\rangle_P
\;=\;
\frac{\delta}{i \,\delta{\cal K}_{\;\;\;\mu}^{(1)\;a}(x)}
\,\langle\; {\cal A}_{\nu}^{b}(y)\;\rangle_P
\;+\;
\langle\; {\cal A}_{\mu}^{a}(x)\;\rangle_P \langle\; {\cal A}_{\nu}^{b}(y)\;\rangle_P
\nonumber \\
G_{\;\;\;\;\mu\nu\lambda}^{(3)\;a b c}(x,y,z)
&=&
\langle\; {\cal A}_{\mu}^{a}(x){\cal A}_{\nu}^{b}(y){\cal A}_{\lambda}^{c}(z)\;\rangle_P
\;=\;
\frac{1}{2} \left(
\frac{\delta}{i\,\delta {\cal K}_{\;\;\;\mu\nu}^{(2)\;ab}(x,y)} 
\;+\;
\frac{\delta}{i\,\delta {\cal K}_{\;\;\;\mu}^{(1)\;a}(x)} 
\frac{\delta}{i\,\delta {\cal K}_{\;\;\;\nu}^{(1)\;b}(y)}
\right)
\,\langle\; {\cal A}_{\lambda}^{c}(z)\;\rangle_P
\nonumber \\
& &
\;\;\;\;\;\;\;\;
\;+\;
\frac{\delta}{i\,\delta {\cal K}_{\;\;\;\mu}^{(1)\;a}(x)}
\left(
\frac{}{}
\langle\; {\cal A}_{\nu}^{b}(y)\;\rangle_P \langle \; {\cal A}_{\lambda}^{c}(z)\;\rangle_P
\right) 
\;+\;
\langle\; {\cal A}_{\mu}^{a}(x)\;\rangle_P 
\langle\; {\cal A}_{\nu}^{b}(y)\;\rangle_P \langle \; {\cal A}_{\lambda}^{c}(z)\;\rangle_P
\label{G123}
\;.
\end{eqnarray}
Higher order Green functions are generated similarly from ((\ref{Green1}).
\bigskip

\section{Non-covariant gauges and the absence of ghosts}
\label{sec:appd}

In this appendix the standard procedure of gauge field quantization
is applied to the class of non-covariant gauges (\ref{gauge1a}), and it is shown that
ghost degrees of freedom are indeed absent, reducing the
general non-linear dynamics in of QCD essentially to a linear QED type 
dynamics.
For an excellent review and bibliography, see Ref. \cite{gaugereview}.
Recall that
under local gauge transformations
\begin{equation} 
g[\theta^a]\;\equiv\; \exp\left( - i\,\theta^a(x)\, T^a\right)
\;,
\label{gtheta}
\end{equation} 
the gauge fields transform as
\begin{equation}
{\cal A}_\mu^a \;\longrightarrow\;
{\cal A}_\mu^{(\theta)\;a} \;=\;
g[\theta^a]\; {\cal A}_\mu^a \;g^{ -1}[\theta^a]
\;,
\end{equation}
implying that 
${\cal F}_{\mu\nu}^a {\cal F}_{\mu\nu}^a \;=\;
{\cal F}_{\mu\nu}^{(\theta)\;a} {\cal F}_{\mu\nu}^{(\theta)\;a}$,
that is the gauge invariance of the Yang-Mills action $I_{YM}[{\cal A}]$.
However, the source term ${\cal J}\circ {\cal A}$
in the generating functional $Z_P$ of (\ref{Z2}) is not gauge invariant
under the transformations (\ref{gtheta}).
Consequently, the {\it naive} functional
\begin{equation}
Z_P^{(naive)}
\;=\;
\int \,{\cal D}{\cal A} \;
\exp \left\{i \left( \frac{}{} I_{YM}\left[{\cal A}\right]+{\cal J}\circ {\cal A}
\right)\right\}
\,\times \, {\cal M}(\hat{\rho})
\label{Znaive}
\end{equation}
is also not a gauge invariant quantity.
As is well known, this can be remedied by applying the formal Fadeev-Popov 
\cite{FP} procedure and integrate in the path-integral
$Z_P$ over all possible gauge transformations $g(\theta^a)$ subject to
the linear subsidiary condition 
\begin{equation}
\phi^a [{\cal A}_\mu^{(\theta)}] \;\equiv\;
n^\mu\,{\cal A}_\mu^{(\theta)\;a}(x) \;-\; \beta^a(x)
\;\stackrel{!}{=}\;0
\label{phiA}
\end{equation}
with normalized space-like vector $n^\mu$ and $\beta^a(x)$ an arbitrary
weight function.
The Fadeev-Popov trick to implement  
the constraint (\ref{phiA}) in the non-invariant functional $Z_P^{(naive)}$
by multiplying with
\begin{equation}
1\;\,=\;\,
\int {\cal D} \theta\;\prod_a\delta\left(\phi^a [{\cal A}_\mu^{(\theta)}]\right)
\; \mbox{det}{\cal F}
\;,
\end{equation}
where the determinant is the Jacobian for the change of variables
$\phi^a\rightarrow \theta^a$,
\begin{equation}
\left(\frac{}{}\mbox{det} {\cal F}\right)^{ab}
\;=\; \mbox{det}\left(\frac{\delta \phi^a[{\cal A}_\mu^{(\theta)}]
}{
\delta \theta^b}\right)_{\phi^a [{\cal A}_\mu^{(\theta)}] = 0}
\;=\;
\left\{
\int {\cal D} \theta\;\prod_a\delta\left(\phi^a [{\cal A}_\mu^{(\theta)}]\right)
\right\}^{-1}
\;.
\label{detF}
\end{equation}
Following this procedure one arrives at 
\begin{equation}
Z_P^{(naive)}\;\longrightarrow\;
Z_P \;=\;
\int \,{\cal D}{\cal A} \; \mbox{det}{\cal F}\;
\prod_a \delta \left( \phi^a[{\cal A}_\mu]\right)
\; \exp \left\{i \left( \frac{}{} I_{YM}\left[{\cal A}\right]+{\cal J}\circ {\cal A}
\right)\right\}
\,\times \, {\cal M}(\hat{\rho})
\label{Znew}
\;,
\end{equation}
which is now a gauge invariant expression due to the proper account 
of the subsidiary condition (\ref{phiA}) that guarantees the correct 
transformation properties of the gauge fields in the presence of the 
sources ${\cal J}$.

To obtain the final form of $Z_P$ as quoted in (\ref{Z2}), one
integrates functionally over the arbitrary funtions $\beta^a(x)$
introduced in (\ref{phiA}), by choosing, e.g., a Gaussian weight functional
\begin{equation}
w[\beta^a]\;=\;
\exp\left\{- \frac{i}{2\alpha}
\,\int_P d^4x \,\left[\beta^a(x)\right]^2
\right\}
\;,
\end{equation}
with the real valued parameter $0\le\alpha\le 1$, upon which
the Fadeev-Popov determinant $\mbox{det}{\cal F}$ can be rewritten 
in a more suitable way:
\begin{equation}
\mbox{det}{\cal F}\;=\;
\int {\cal D}\beta \;\prod_a\;\exp\left\{-\frac{i}{2\alpha}
\,\int_P d^4x \,\left[\beta^a(x)\right]^2 
\right\}
\;\delta\left(
n^\mu\,{\cal A}_\mu^{(\theta)\;a}(x) \;-\; \beta^a(x)
\right)
\;.
\label{detFnew}
\end{equation}
In order to calculate the determinant, it is sufficient to integrate over 
$\theta^a$ in a small vicinity where the argument of the $\delta$-function 
passes through zero at given ${\cal A}^{(\theta)\;a}$ and $\beta^a$.
For {\it infinitesimal gauge transformations}
\begin{equation}
g[\theta^a] \;\longrightarrow\; 
\delta g[\theta^a] \;=\; 1\;-\;i \theta^a(x) \,T^a
\;,
\end{equation}
the gauge fields transform as
\begin{equation}
{\cal A}_\mu^{a} \;\longrightarrow\;
{\cal A}_\mu^{a} \;+\; \delta {\cal A}_\mu^{a} 
\;\;,\;\;\;\;\;\;\;\;\;\;\;\;\;\;
\delta {\cal A}_\mu^{a} 
\;=\;  f^{abc} \theta^b{\cal A}_\mu^c \;-\;\frac{1}{g}\,\partial_\mu^x \theta^a
\;,
\end{equation}
so that one obtains
\begin{eqnarray}
\;\delta\left( n^\mu\,{\cal A}_\mu^{(\theta)\;a}(x) \;-\; \beta^a(x) \right)
&=&
\;\delta\left( n^\mu\,{\cal A}_\mu^{(\theta)\;a}(x) 
\;+\; f^{abc}\theta^b\,n^\mu {\cal A}_\mu^{(\theta)\;c}\;-\; 
\frac{1}{g} \,n^\mu \partial_\mu^x \,\theta^a
\;-\;\beta^a \right)
\nonumber \\
&=&
\;\delta\left(
f^{abc}\theta^b\,\beta^c\;-\; 
\frac{1}{g} \,n^\mu \partial_\mu^x \,\theta^a
\right)
\;,
\end{eqnarray}
because $n^\mu{\cal A}_\mu^{(\theta)\;a} = \beta^a$.
This latter expression is evidently independent of
${\cal A}_\mu^a$.
Therefore, when substituted into (\ref{detFnew}), $\mbox{det}{\cal F}$ is
also explicitely independent of the gauge fields, and hence can be pulled out of
the path-integral $Z_P$ and absorbed in an (irrelevant) normalization, which may be set
equal to unity.  The final result is then:
\begin{equation}
Z_P[{\cal J},\hat{\rho}] \;=\;
\int \,{\cal D}{\cal A} \;
\;
\exp \left\{i \left( \frac{}{} I_{YM} \left[{\cal A} \right]
\;+\; I_{GF}\left[n\cdot{\cal A}\right]
\;+\;{\cal J}\circ {\cal A}
\right)
\right\}
\,\times \, {\cal M}(\hat{\rho})
\;,
\end{equation}
where, from (\ref{detFnew}),
\begin{equation}
I_{GF}\left[n\cdot{\cal A}\right]
\;\equiv\;
\exp\left\{- \frac{i}{2\alpha}
\,\int_P d^4x \,
\left[n\cdot {\cal A}^a(x)\right]^2\right\}
\;.
\end{equation}
In {\it conclusion}: 
the property of gauge field independence of the Fadeev-Popov determinant
proves that there are indeed
no ghost fields coupling to the gluon fields, hence the formulation is
{\it ghost-free}!
\bigskip

\section{The truncated effective action $\Gamma_P\left[\overline{A}, \widehat{\Delta}\right]$}
\label{sec:appe}

The generating functional for the 
{\it connected} Green functions, denoted by ${\cal G}^{(n)}$,
is defined as usual:
\begin{equation}
W_P\left[{\cal K}\right] \;\;=\;\;
- i \,\ln\, Z_P\left[ {\cal K}\right]
\label{W0}
\;.
\end{equation}
From $W_P$ one obtains
the {\it connected} Green functions ${\cal G}^{(n)}$
by functional differentiation analogous to (\ref{Green1}),
in terms of mixed products of $a_\mu$ and $A_\mu$ fields
\begin{equation}
(-i)\, {\cal G}_{\;\;\;\;\mu_1\ldots \mu_n}^{(n)\;a_1\ldots a_n}(x_1,\ldots, x_n)
\;\equiv\;
\left.
\frac{\delta}{i \,\delta{\cal K}^{(n)}}
W_P[{\cal K}]\right|_{{\cal K}=0}
\;\;=\;\;
\langle\; a_{\mu_1}^{a_1}(x_1)\ldots a_{\mu_k}^{a_k}(x_k)\,
A_{\mu_{k+1}}^{a_{k+1}}(x_{k+1})\ldots A_{\mu_n}^{a_n}(x_n)\;\rangle_P^{(c)}
\label{Green2}
\;,
\end{equation}
where the superscript $(c)$ indicates the `connected parts'.
It follows then that
\begin{eqnarray}
\frac{\delta W_P}{\delta {\cal K}^{(1)\; \mu,a}(x)} 
&=& 
\;\;\;\;
{\cal G}_{\;\;\;\;\mu}^{(1)\;a}(x)
\nonumber \\
\frac{\delta W_P}{\delta {\cal K}^{(2)\;\mu\nu, ab}(x,y)}&=&
\frac{1}{2}\,\left( 
{\cal G}_{\;\;\;\;\mu\nu}^{(2)\;ab}(x,y)
\;+\;{\cal G}_{\;\;\;\mu}^{(1)\;a}(x) {\cal G}_{\;\;\;\nu}^{(1)\;b}(y) \right)
\nonumber \\
\frac{\delta W_P}{\delta {\cal K}^{(3)\;\mu\nu\lambda, abc}(x,y,z)}
&=&
\frac{1}{6}\,\left( 
{\cal G}_{\;\;\;\;\mu\nu\lambda}^{(3)\;abc}(x,y,z)
\;+\;
3\;{\cal G}_{\;\;\;\;\mu\nu}^{(2)\;ab}(x,y)\,{\cal G}_{\;\;\;\;\lambda}^{(1)\;c}(z)
\;+\;{\cal G}_{\;\;\;\mu}^{(1)\;a}(x) {\cal G}_{\;\;\;\nu}^{(1)\;b}(y)
{\cal G}_{\;\;\;\lambda}^{(1)\;c}(z) 
\right)
\;,
\label{Green3a}
\end{eqnarray}
where, for example,
\begin{eqnarray}
{\cal G}_{\;\;\;\;\mu}^{(1)\;a}(x)
&=&
\langle \;A_\mu^a(x) \;\ \rangle_P^{(c)}
\;+\;
\langle \;a_\mu^a(x) \rangle_P^{(c)}
\nonumber \\
{\cal G}_{\;\;\;\;\mu\nu}^{(2)\;ab}(x,y)
&=&
\langle\;A_\mu^a(x) A_\nu^b(y)\;\rangle_P^{(c)}
\;+\;
\langle \;a_\mu^a(x) a_\nu^b(y)\;\rangle_P^{(c)}
\nonumber \\
{\cal G}_{\;\;\;\;\mu\nu\lambda}^{(3)\;abc}(x,y,z)
&=&
\langle\;A_\mu^a(x) A_\nu^b(y)A_\lambda^c(z) \;\rangle_P^{(c)}
\;+\;
\langle \;a_\mu^a(x) a_\nu^b(y) a_\lambda^c(z) \;\rangle_P^{(c)}
\;,
\label{Green3}
\end{eqnarray}
and similarly expressions higher order 
Green functions which  involve  4, 5 ... space-time points.
\smallskip

$W_P$ of (\ref{W0}) involves
the sources ${\cal K}$ that do not have any immediate physical interpretation, it is
more convenient to work
with the corresponding  effective action $\Gamma_P$, the
generating functional for the proper vertex functions, which determines
the equations of motion for 
the physically relevant Green functions.
The {\it effective action} $\Gamma_P$ is defined as the multiple Legendre transform,
and is obtained by
eliminating the  source variables ${\cal K}$ in favor of the 
connected Green functions ${\cal G}$:
\begin{eqnarray}
\Gamma_P\left[{\cal G}\right]
&=&
W_P\left[{\cal K}\right] 
\;-\;{\cal K}^{(1)}\circ {\cal G}^{(1)}
\;-\;\frac{1}{2}{\cal K}^{(2)}\circ
\left( {\cal G}^{(2)} + {\cal G}^{(1)}{\cal G}^{(1)}\right)
\nonumber \\
& &
\;\;\;\;\;\;\;\;\;\;\;\;\;\;\;
\;-\;\frac{1}{6}{\cal K}^{(3)}\circ\left( {\cal G}^{(3)}
+3\,{\cal G}^{(2)}{\cal G}^{(1)}
+{\cal G}^{(1)}{\cal G}^{(1)}{\cal G}^{(1)}\right)
\;-\;\ldots
\label{Gamma00}
\;.
\end{eqnarray}
So far no approximations have been made. The variation of $\Gamma_P$ with respect
to the Green functions ${\cal G}^{(n)}$ would yield an infinite set of 
coupled equations,
the analogue of the BBGKY hierarchy \cite{BBGKY}:
\begin{eqnarray}
\frac{\delta \Gamma_P}{{\cal G}^{(1)}} &=&
- {\cal K}^{(1)} \;-\; {\cal K}^{(2)} \circ{\cal G}^{(1)} \;-\; 
- \frac{1}{2} {\cal K}^{(3)}\circ\left({\cal G}^{(2)}+ {\cal G}^{(1)}{\cal G}^{(1)} \right)
\;-\;\ldots
\nonumber \\
\frac{\delta \Gamma_P}{{\cal G}^{(2)}} &=&
- \frac{1}{2} {\cal K}^{(2)}
- \frac{1}{2} {\cal K}^{(3)}\circ{\cal G}^{(1)}
\;-\;\ldots
\nonumber \\
\frac{\delta \Gamma_P}{{\cal G}^{(3)}} &=&
- \frac{1}{6} {\cal K}^{(3)}
\;-\;\ldots
\;,\;\;\;\;
\;\;\;\;\;\;
\mbox{etc.}
\label{EOM}
\end{eqnarray}
At this point approximation 1 and 2 of Section II A are invoked.
It is assumed that the initial state is
a (dilute) ensemble of hard gluons of very small spatial extent
$\ll \lambda$, corresponding to transverse momenta $k_\perp^2 \gg \mu^2$.
By definition of $\lambda$, or $\mu$, the short-range character of
these quantum fluctuations implies that the expectation value $\langle a_\mu\rangle$
vanishes at all times. However, the long-range correlations of the eventually
populated soft modes with very small momenta $k_\perp^2\ll\mu^2$ may lead
to a collective mean field with non-vanishing $\langle A\rangle$.
Accordingly,  the following condition
is imposed on the expectation values of the fields:
\begin{equation}
\langle \; A_\mu^a(x)\;\rangle \;
\left\{
\begin{array}{ll}
\;=\;0 \;\;\;\mbox{for} \;t\,\le\,t_0 \\
\;\ge\;0 \;\;\;\mbox{for} \;t\,>\,t_0
\end{array}
\right.
\;\;\;\;\;\;\;\;\;\;\;\;\;\;\;\;\;\;\;\;
\langle \; a_\mu^a(x)\;\rangle \;\stackrel{!}{=}\; 0
\;\;\;\mbox{for all} \;t
\;.
\label{MFconstraint}
\end{equation}
Furthermore, the quantum fluctuations of the soft field are ignored, assuming
any multi-point correlations of soft fields to be small,
\begin{equation}
\langle \; A_{\mu_1}^{a_1}(x_1)\; \ldots A_{\mu_n}^{a_n}(x_n) \;\rangle
\ll\;\langle \;  A_{\mu_1}^{a_1}(x_1)\;\rangle
\;\ldots \;\langle \;  A_{\mu_1}^{a_n}(x_n)\;\rangle
\;\;\;\;\;\;\mbox{for all $n\ge 2$}
\;,
\label{MFconstraint2}
\end{equation}
i.e.,   $A_\mu$ is treated as a non-propagating and non-fluctuating, 
classical field.
Hence, the set of Green functions (\ref{Green3}) reduces to
\begin{eqnarray}
{\cal G}_{\;\;\;\;\mu}^{(1)\;a}(x)
\;=\;\langle\;A_\mu^a(x)\;\rangle_P^{(c)}
\;\;\equiv\; \;\overline{A}_\mu^a(x)
\;\;\;\;\;\;\;\;\;
{\cal G}_{\;\;\;\;\mu\nu}^{(2)\;ab}(x,y)
\;=\;
\langle \;a_\mu^a(x) a_\nu^b(y)\;\rangle_P^{(c)}
\;\;\equiv\;\; 
i \widehat{\Delta}_{\mu\nu}^{ab}(x,y)
\;.
\label{Green3b}
\end{eqnarray}
These relations 
define the soft, classical  mean field $\overline{A}$, and the hard  
quantum propagators $\widehat{\Delta}$.
\smallskip

Now  the hierarchy is truncated for $n\ge 3$. However,
to perform this truncation properly, one must eliminate 
all the ${\cal G}_\mu^{(3)}$, ${\cal G}_\mu^{(4)}$, etc., as dynamical variables
by introducing \cite{calzetta}
\begin{equation}
\Gamma_P\left[{\cal G}^{(1)}, {\cal G}^{(2)}\right]
\;\equiv\;
\Gamma_P\left[
{\cal G}^{(1)}, {\cal G}^{(2)}, \widetilde{{\cal G}}^{(3)}, \widetilde{{\cal G}}^{(4)}, 
\ldots \right\}
\;\;\;\;\;\;\;
\end{equation}
where 
$\widetilde{{\cal G}}^{(n)}$ for all $n\ge 3$ are functionals of of the
1- and 2-point functions alone, and are determined by the implicit equations
\begin{equation}
\widetilde{{\cal G}}^{(n)}\;:=\; \widetilde{{\cal G}}^{(n)}
\left[{\cal G}^{(1)},{\cal G}^{(2)}\right]
\;,
\;\;\;\;\;\;\;\;\;\;\;\;\;
\frac{\delta \Gamma_P\left[{\cal G}^{(1)}, {\cal G}^{(2)}\right]}
{\delta \widetilde{{\cal G}}^{(n)}\left[{\cal G}^{(1)},{\cal G}^{(2)}\right]}
\;=\;0
\;\;\;\;\;\;\;\;\;\;\;
\mbox{ for all $n\ge 3$}
\;.
\end{equation}
From (\ref{Green3a}) and (\ref{Green3}) one sees that then
the infinite  set of Green functions reduces to involve only 
${\cal G}_\mu^{(1)} = \overline{A}_\mu$
and 
${\cal G}_{\mu\nu}^{(2)}= i\widehat{\Delta}_{\mu\nu}$,
so that   $\Gamma_P$ becomes a functional of only
the soft mean field $\overline{A}_\mu$
and the hard propagator
$\widehat{\Delta}_{\mu\nu}$:
\begin{equation}
\Gamma_P\left[{\cal G}\right]
\;\;\approx\;\;
\Gamma_P\left[\overline{A}, \widehat{\Delta}\right]
\;\;=\;\;
W_P\left[{\cal K}^{(1)},{\cal K}^{(2)}\right] 
\;-\;{\cal K}^{(1)}\circ\overline{A}
\;-\;\frac{1}{2} 
\;{\cal K}^{(2)}\circ \left(\,i\widehat{\Delta}\;+\;\overline{A}\;\overline{A}
\right)
\label{Gamma0}
\;.
\end{equation}
\smallskip

The  {\it equations of motion} for the 
mean field $\overline{A}$ and for the hard propagator $\widehat{\Delta}$ 
in the presence of external sources, follow now from (\ref{Green3}), (\ref{EOM}) 
and (\ref{Gamma0}):
\begin{eqnarray}
\frac{\delta \Gamma_P}{\delta \overline{A}_\mu^a(x)} &=&
- {\cal K}^{(1) \;\mu,a}(x) \;-\; 
\int_P d^4y \,K^{(2)\;\mu\nu,ab}(x,y) \;\overline{A}^{\nu,\,b}(y)
\;,
\label{eom0a}
\\
\frac{\delta \Gamma_P}{\delta \widehat{\Delta}_{\mu\nu}^{ab}(x,y)} 
&=& \frac{1}{2i}\;{\cal K}^{(2)\;\mu\nu,ab}(x,y)
\label{eom0b}
\;.
\end{eqnarray}
\bigskip

\section{Analytic properties of the free-field propagators} 
\label{sec:appf}

\noindent
The components of the free-field propagator 
$\Delta_{0\;\mu\nu}^{\;\;\;\;ab}$
are defined as in (\ref{D22}), i.e.,
\begin{eqnarray}
\Delta_{0\;\mu\nu}^{F\;ab}(x,y)
&=& -i \langle \;T\, a_\mu^a(x)a_\nu^b(y)\;\rangle 
\;\;\;\;\;\;\;\;\;\;
\Delta_{0\;\mu\nu}^{\overline{F}\;ab}(x,y)
\;=\; -i \langle \;\overline{T}\, a_\mu^a(x)a_\nu^b(y)\;\rangle 
\nonumber \\
\Delta_{0\;\mu\nu}^{>\;ab}(x,y)
&=& -i \langle \; a_\mu^a(x)a_\nu^b(y)\;\rangle 
\;\;\;\;\;\;\;\;\;\;\;\;\;\;
\Delta_{0\;\mu\nu}^{<\;ab}(x,y)
\;=\; -i \langle \; a_\nu^b(y)a_\mu^a(y)\;\rangle 
\;.
\label{C0}
\end{eqnarray}
For free fields, one may write 	
\begin{equation}
\Delta_{0\;\mu\nu}^{\;\;\;\;ab}(x,y) \;=\;
\delta^{ab}\;d_{\mu\nu}(\partial_x) \; \Delta_{0}(x,y) 
\;\;\;\;\;\;\;\;\;
(\Delta \equiv \Delta^F, \Delta^{\overline{F}}, \Delta^>, \Delta^<)
\;,
\end{equation}
where $d_{\mu\nu}(\partial_x)$ is defined by (\ref{Box}), and
the functions $\Delta_{0}$ on the right side are the {\it scalar}
parts of the propagators.
The $F,\overline{F},>,<$ components of the latter
obey the following free-field equations with 
different boundary conditions:
\begin{eqnarray}
\stackrel{\rightarrow}{\partial}_x^2 \,\Delta_{0}^F(x,y) 
&=&
\Delta_{0}^F(x,y)\,\stackrel{\leftarrow}{\partial}_y^2
\;\;=\;\;+\delta^4(x,y)
\nonumber\\
\stackrel{\rightarrow}{\partial}_x^2 \,\Delta_{0}^{\overline{F}}(x,y) 
&=&
\Delta_{0}^{\overline{F}}(x,y)\,\stackrel{\leftarrow}{\partial}_y^2
\;\;=\;\;-\delta^4(x,y)
\nonumber\\
\stackrel{\rightarrow}{\partial}_x^2 \,\Delta_{0}^{>}(x,y) 
&=&
\Delta_{0}^{>}(x,y)\,\stackrel{\leftarrow}{\partial}_y^2
\;\;=\;\;0
\nonumber\\
\stackrel{\rightarrow}{\partial}_x^2 \,\Delta_{0}^{<}(x,y) 
&=&
\Delta_{0}^{<}(x,y)\,\stackrel{\leftarrow}{\partial}_y^2
\;\;=\;\;0
\label{C1}
\;,
\end{eqnarray}
and the identities 
\begin{eqnarray}
\Delta_{0}^F(x,y) 
&=&\theta(x_0,y_0)\,\Delta_{0}^>(x,y)\;+\;
\theta(y_0,x_0)\,\Delta_{0}^<(x,y)
\nonumber \\
\Delta_{0}^{\overline{F}}(x,y) 
&=&\theta(x_0,y_0)\,\Delta_{0}^<(x,y)\;+\;
\theta(y_0,x_0)\,\Delta_{0}^>(x,y)
\;.
\label{C2}
\end{eqnarray}
Because of the relations (\ref{C2}), the set of equations (\ref{C1})
can be solved by only two independent functions, namely, (i) a purely
imaginary and odd function $i\Delta^-$, and (ii) a purely real and even
function $\Delta^+$:
\begin{eqnarray}
i\,\Delta^-(x,y) &\equiv& 
\delta^{ab}\;d_{\mu\nu}(\partial_x) \; 
\langle\;\left[a_\mu(x)\, ,\, a_\nu^b(y)\right]\;\rangle
\;=\;
i \left(\Delta_{0}^> - \Delta_{0}^<\right)(x,y)
\nonumber \\
\;\,\Delta^+(x,y) &\equiv& 
\delta^{ab}\;d_{\mu\nu}(\partial_x) \; 
\langle\;\left\{a_\mu(x)\, ,\, a_\nu^b(y)\right\}\;\rangle
\;=\;
i \left(\Delta_{0}^> + \Delta_{0}^<\right)(x,y)
\label{C3}
\end{eqnarray}
From (\ref{C1}) it follows that these functions obey
\begin{eqnarray}
\stackrel{\rightarrow}{\partial}_x^2 \,\Delta^-(x,y) 
&=&
\Delta^-(x,y)\,\stackrel{\leftarrow}{\partial}_y^2
\;\;=\;\;0
\;\;\;\;\;\;\;\;\;
\Delta^-(x,y)\;=\; - \Delta^-(y,x)
\nonumber\\
\stackrel{\rightarrow}{\partial}_x^2 \,\Delta^+(x,y) 
&=&
\Delta^+(x,y)\,\stackrel{\leftarrow}{\partial}_y^2
\;\;=\;\;0
\;\;\;\;\;\;\;\;\;
\Delta^+(x,y)\;=\; + \Delta^+(y,x)
\label{C4}
\;,
\end{eqnarray}
with the general solutions
\begin{eqnarray}
\Delta^-(x,y) &=& - i \,\int \frac{d^4k}{(2\pi)^4}
\;e^{-i k\cdot(x-y)}\;
2\pi \delta(k^2)\;\left( g_1(k) \;-\;g_2(-k)\right)
\nonumber \\
\Delta^+(x,y) &=& \;\;\;\,\int \frac{d^4k}{(2\pi)^4}
\;e^{-i k\cdot(x-y)}\;
2\pi \delta(k^2)\;\left( g_1(k) \;+\;g_2(-k)\right)
\;,
\label{C5}
\end{eqnarray}
where the functions
\begin{equation}
g_1(k) \;\equiv\; \theta(k_0) \,+\, f_1(k)
\;\;\;\;\;\;\;\;
g_2(-k) \;\equiv\; \theta(-k_0) \,+\, f_2(-k)
\end{equation}
contain the positive and negative frequency modes, respectively.
Here $\theta(\pm k_0)$ is the vacuum contribution, while
$f_{1,2}(\pm k)$ are the additional contributions from a medium.
\medskip

From (\ref{C3})-(\ref{C5}), one can now infer immediately the analytic
properties of $f_1$, $f_2$, corresponding to those of
$g_1$, $g_2$.
\begin{description}
\item[1.]
First, one observes, because $\Delta^-$ is purely imaginary
and $\Delta^+$ is purely real, that it must hold
$$
f_1(k), f_2(k) \;\;=\;\;\mbox{real}
\;.
$$
\item[2.]
Second, because the commutator of free fields, i.e. 
the imaginary function $\Delta^-$,
must be independent of the state of the medium, 
$$
\Delta^-(x,y) \stackrel{!}{=}\; - i \,\int \frac{d^4k}{(2\pi)^4}
\;e^{-i k\cdot(x-y)}\;
2\pi \delta(k^2)\;\left( \theta(k_0)-\theta(-k_0)\right)
\;,
$$
it follows that
$$
f_1(k)\;=\;f_2(-k) \;\;\equiv\;\;f(k)
\;.
$$
\item[3.]
Finally, because the anticommutator, i.e. the real function
$\Delta^+$ must satisfy
$$
\int d^4x d^4y \;h^\ast(x) \;\Delta^+(x,y)\;h(y)\;\stackrel{!}{\ge}\;0
$$
for any smooth, but in general complex-valued function $h$,
it follows that
$$
f(\vec{k})\;\equiv\; 
\int dk^0 f(k^0,\vec{k}) \;\ge\;\;0\;\;\;\;\;\;\;\;\mbox{for all $\vec{k}$}
\;,
$$
and so the `on-shell' function $f(\vec{k})$ is positive definite
may indeed be identified with the positive definite 
phase-space density $dN/d^3k$.
\end{description}

The free-field solutions of 
$\Delta^F, \Delta^{\overline{F}}, \Delta^>, \Delta^<$ can now
easily reconstructed using  the following identities implied by
(\ref{C0}) and (\ref{C2}):
\begin{eqnarray}
2 \Delta^F(x,y) &=& -i \Delta^+(x,y) 
\;+\; \left(2\theta(x^0,y^0) - 1\right) \Delta^-(x,y)
\nonumber \\
2 \Delta^{\overline{F}}(x,y) &=& -i \Delta^+(x,y) 
\;+\; \left(2\theta(y^0,x^0) - 1 \right) \Delta^-(x,y)
\end{eqnarray}
\begin{eqnarray}
2 \Delta^>(x,y) &=& -i \Delta^+(x,y) \;+\;  \Delta^-(x,y)
\nonumber \\
2 \Delta^<(x,y) &=& -i \Delta^+(x,y) \;-\;  \Delta^-(x,y)
\;,
\end{eqnarray}
from which, upon Fourier transformation, one obtains
\begin{eqnarray}
\Delta_{0}^F(x,y) &=&
\int \frac{d^4k}{(2\pi)^4}\,e^{-ik\cdot(x-y)}\;
\left( \frac{+1}{k^2 +i\epsilon} \;-\; i \,2\pi \delta(k^2) \;f(k)\right)
\nonumber \\
\Delta_{0}^{\overline{F}}(x,y) &=&
\int \frac{d^4k}{(2\pi)^4}\,e^{-ik\cdot(x-y)}\;
\left( \frac{- 1}{k^2 -i\epsilon} \;-\; i \,2\pi \delta(k^2) \;f(k)\right)
\nonumber \\
\Delta_{0}^>(x,y) &=&
\int \frac{d^4k}{(2\pi)^4}\,e^{-ik\cdot(x-y)}\;
\left( \frac{}{} \;-\; i \,2\pi \delta(k^2) \;
\left(\theta(k_0) + f(k)\right)\right)
\nonumber \\
\Delta_{0}^<(x,y) &=&
\int \frac{d^4k}{(2\pi)^4}\,e^{-ik\cdot(x-y)}\;
\left( \frac{}{} \;-\; i \,2\pi \delta(k^2) \;
\left(\theta(-k_0) + f(k)\right)\right)
\;.
\end{eqnarray}
Finally, it is straight forward to infer the corresponding
free-field forms of the retarded, advanced and correlation functions,
\begin{eqnarray}
\Delta^{ret}_{0}(x,y) &=&
+\theta(x_0,y_0)\,
\Delta^-(x,y) \;=\; \left(\Delta_{0}^F - \Delta_{0}^<\right)(x,y)
\;=\;
\int \frac{d^4k}{(2\pi)^4}\,e^{-ik\cdot(x-y)}\;
\left( \frac{1}{k^2 +2i\epsilon}\right)
\nonumber \\
\Delta^{adv}_{0}(x,y) &=&
- \theta(y_0,x_0)\,
\Delta^-(x,y) \;=\; \left(\Delta_{0}^F - \Delta_{0}^>\right)(x,y)
\;=\;
\int \frac{d^4k}{(2\pi)^4}\,e^{-ik\cdot(x-y)}\;
\left( \frac{1}{k^2 -2i\epsilon}\right)
\nonumber \\
\Delta^{cor}_{0}(x,y) &=&
-i\,\Delta^+(x,y) \;=\; \left(\Delta_{0}^> + \Delta_{0}^<\right)(x,y)
\nonumber \\
& &
\;\;\;\;\;\;\;\;\;\;\;\;\;\;\;\;\;\;
\;\;\;\;\;\;\;\;\;\;\;\;\;\;\;\;\;\;
\;=\;
\int \frac{d^4k}{(2\pi)^4}\,e^{-ik\cdot(x-y)}\;
\left(\frac{}{} -i \,2\pi \delta(k^2) \;\left(1 + 2 f(k)\right)\right)
\;.
\end{eqnarray}


\begin{thebibliography}{99}
\bibitem{remark1}
It should be remarked that the Wigner function can be transformed
from the here employed class of  non-covariant gauges to any other gauge 
by utilizing the general form of (\ref{W}) with
$$
{\cal G}_{\mu\nu}^{ab}\left(r,s\right)
\;\equiv\;
U^{aa'}\left(r,r+\frac{s}{2}\right)
\;\,
{\cal G}_{\mu\nu}^{a'b'}\left(r+\frac{s}{2}, r-\frac{s}{2}\right)
\,\;
U^{bb'}\left(r-\frac{s}{2},r \right)
\;,
$$
and $U(x,y)$ is the `parallel transporter' or `link operator'
\cite{BI1,EH}
that guarantees that the Wigner function transforms
covariantly under local gauge transformations, when moving
on an oriented straight line $z(\xi):=x+\xi (y-x)$, $0\le \xi \le 1$
from space-time point $x$ to point $y$:
$$
U^{ab}\left(x,y\right)
\;=\;\delta^{ab}\;P \exp\left\{\frac{}{} i g\int_0^\infty dz^\mu 
{\cal A}_\mu(z)\right\}
\;=\;
\delta^{ab}\;\left(\frac{}{}1\;+\; 
P\sum_{n=1}^\infty \prod_{i=1}^n \,ig \,z_i\cdot{\cal A}(z_i)
\right)
\;.
$$
The symbol $P$ denotes the path-ordering of the color matrices in the exponential
and ${\cal A} = T^a {\cal A}_\mu^a$ where ${\cal A}$ stands for $A$ or $a$.
Choosing the gauge vector $n^\mu$
such that it is parallel to the direction of propagation, $n \parallel z$,
the gauge constraint $n\cdot{\cal A} = 0$ implies
$z\cdot {\cal A}= 0$, and as a consequence one has
then
$
U^{ab}\left(x,y\right) \stackrel{n\cdot{\cal A}=0}{=}\delta^{ab}
$,
so that
$$
\widehat{{\cal G}}_{\mu\nu}^{ab}(r,k) \;
\stackrel{n\cdot{\cal A}=0}{=}\;
\int d^4 s \, e^{i\,k\,\cdot\, s}\;\, 
{\cal G}_{\mu\nu}^{ab}\left(r+\frac{s}{2}, r-\frac{s}{2}\right)
\;.
$$
This property is another
very advantagous feature of the non-covariant gauges: 
the link operator becomes  a unit matrix in color space, and so
the complicated  non-linear dependence (in the gauge fields) 
of the Wigner functions reduces to a form identical to 
those in classical field theory.
This is of course not possible in a general gauge, in particular, 
in covariant gauges one has to account for the infinite number of 
non-linear terms in the series representation of the path-ordered 
exponential.
\medskip
\end{thebibliography}
\end{document}